\documentclass[longauth]{aa}   

\usepackage{graphicx}
\usepackage{txfonts}
\usepackage{lipsum}
\usepackage{subcaption}         
\usepackage{lscape}             
\usepackage{placeins}           
                                
\usepackage[colorlinks=true,citecolor=blue,linkcolor=blue,urlcolor=blue,filecolor=blue]{hyperref}

\usepackage{changes}

\newcommand{\HI}{\mbox{$\mathrm{H}\,{\scriptstyle\mathrm{I}}$}}
\newcommand{\NHI}{\mbox{$N_{\scriptsize\HI}$}}
\graphicspath{{./}{figures/}}

\usepackage{natbib}
\bibpunct{(}{)}{;}{a}{}{,} 

\begin{document}


    \title{EP250226a: A fast X-ray transient associated with GRB 250226A and its multiwavelength afterglow properties}

   
    \titlerunning{EP250226a / GRB 250226A}
    
%
%
%

   \author{Rui-Zhi Li\inst{1,2,3}\email{liruizhi@ynao.ac.cn}
        \and Jirong Mao\inst{1,3,4}\corrauth{jirongmao@mail.ynao.ac.cn}
        \and Aishwarya Linesh Thakur\inst{5}\email{aishwarya.thakur@inaf.it}
        \and Chen-Wei Wang\inst{6}\email{cwwang@ihep.ac.cn}
        \and Jie An\inst{2,7}\email{anjie@nao.cas.cn}
        \and Fei-Fan Song\inst{1,2,3}\email{songfeifan@ynao.ac.cn}
        \and Yodgor Rajabov\inst{8}\email{rajabov@astrin.uz}
        \and Dong Xu\inst{7}\corrauth{dxu@nao.cas.cn}
        \and Shuai-Qing Jiang\inst{2,7}\email{sqjiang@bao.ac.cn}
        \and Bo-Ting Wang\inst{1,2,3}\email{wangbaiting@ynao.ac.cn}
        \and Tong Zhao\inst{2,9}\email{zhaotong@bao.ac.cn}
        \and Yujia Song\inst{2,9}\email{songyj@bao.ac.cn}
        \and Hong-Zhou Wu\inst{10}\email{wuhongzhou@hust.edu.cn}
        \and Yuan Liu\inst{7}\email{liuyuan@bao.ac.cn}
        \and Samaporn Tinyanont\inst{11}\email{samaporn@narit.or.th}
        \and Rungrit Anutarawiramkul\inst{11}\email{rungrit@narit.or.th}
        \and Pathompong Butpan\inst{11}\email{pathompong@narit.or.th}
        \and Zhou Fan\inst{7}\email{zfan@nao.cas.cn}
        \and Shaoyu Fu\inst{10}\email{syfu@nao.cas.cn}
        \and Xing Liu\inst{7}\email{liuxing@nao.cas.cn}
        \and Ning-Chen Sun\inst{2,7,12}\email{sunnc@ucas.ac.cn}
        \and Zi-Pei Zhu\inst{7}\email{zpzhu@nao.cas.cn}
        \and Benjamin Schneider\inst{13}\email{benjamin.schneider@lam.fr}
        \and Nial R. Tanvir\inst{14}\email{nrt3@leicester.ac.uk}
        \and Daniele B. Malesani\inst{15,16,17}\email{daniele.malesani@nbi.ku.dk}
        \and Susanna D. Vergani\inst{18}\email{susanna.vergani@obspm.fr}
        \and Andrew Levan\inst{17,19}\email{a.levan@astro.ru.nl}
        \and Giovanna Pugliese\inst{20}\email{pugliese@astroduo.org}
        \and Andrea Saccardi\inst{21,22}\email{andrea.saccardi@cea.fr}
        \and Luca Izzo\inst{23,24}\email{luca.izzo@inaf.it}
        \and Jesse T. Palmerio\inst{21}\email{jesse.palmerio@obspm.fr}
        \and Patrizia Barria\inst{5,25}\email{patrizia.barria@inaf.it}
        \and James Craig Rodi\inst{5}\email{james.rodi@inaf.it}
        \and Giulia Gianfagna\inst{5}\email{giulia.gianfagna@inaf.it}
        \and Lorenzo Natalucci\inst{5,25}\email{lorenzo.natalucci@inaf.it}
        \and Luigi Piro\inst{5}\email{luigi.piro@inaf.it}
        \and Nidhal Guessoum\inst{26}\email{nguessoum@aus.edu}
        \and Thomas Hussenot-Desenonges\inst{27}\email{thomhus+astro@gmail.com}
        \and Ernazar Abdikamalov\inst{28,29}\email{ernazar.abdikamalov@nu.edu.kz}
        \and D. Turpin\inst{30}\email{damien.turpin@cea.fr}
        \and Nelio Sasaki\inst{31,32}\email{neliosasaki@gmail.com}
        \and A. Takey\inst{33}\email{ali.takey@nriag.sci.eg}
        \and E. G. Elhosseiny\inst{33}\email{eslam_elhosseiny@nriag.sci.eg}
        \and Y.H.M. Hendy\inst{33}\email{yasserhendy@nriag.sci.eg}
        \and Ting-Wan Chen\inst{34}\email{twchen@astro.ncu.edu.tw}
        \and Amar Aryan\inst{34}\email{amar@astro.ncu.edu.tw}
        \and Albert K.H. Kong\inst{35}\email{akong@gapp.nthu.edu.tw}
        \and D. Andrew Howell\inst{36,37}\email{ahowell@lco.global}
        \and Curtis McCully\inst{36}\email{cmccully@lco.global}
        \and Joseph R. Farah\inst{36,37}\email{jfarah@lco.global}
        \and Moira Andrews\inst{36,37}\email{mandrews@lco.global}
        \and Kathryn Wynn\inst{36,37}\email{kwynn@lco.global}
        \and Yuan Qi Ni\inst{38,36}\email{christopher.of.ni@gmail.com}
        \and Wen-Xiong Li\inst{7}\email{liwx@bao.ac.cn}
        \and Hu Zou\inst{7}\email{zouhu@bao.ac.cn}
        \and Yun-Ao Xiao\inst{7}\email{xiaoya@nao.cas.cn}
        \and Wei-Jian Guo\inst{7}\email{guowj@bao.ac.cn}
        \and Niu Li\inst{7}\email{liniu@bao.ac.cn}
        \and A.S. Pozanenko\inst{39,40}\email{apozanen@iki.rssi.ru}
        \and A. S. Moskvitin\inst{41}\email{mosk@sao.ru}
        \and O. I. Spiridonova\inst{41}\email{ospir@sao.ru}
        \and Xingzhu Zou\inst{42,43}\email{xingzhuzou@mail.ynu.edu.cn}
        \and Brajesh Kumar\inst{42,43}\email{brajesh@ynu.edu.cn}
        \and Xiaowei Liu\inst{42,43}\email{x.liu@ynu.edu.cn}
        \and Hui Sun\inst{7}\email{hsun@nao.cas.cn}
        \and Xue-Feng Wu\inst{44}\email{xfwu@pmo.ac.cn}
        \and Weimin Yuan\inst{7}\email{wmy@nao.cas.cn}}

   \institute{Yunnan Observatories, Chinese Academy of Sciences, Kunming 650216, People's Republic of China
\and School of Astronomy and Space Science, University of Chinese Academy of Sciences, Beijing 101408, People's Republic of China
\and Center for Astronomical Mega-Science, Chinese Academy of Sciences, Beijing 100012, People's Republic of China
\and Key Laboratory for the Structure and Evolution of Celestial Objects, Chinese Academy of Sciences, Kunming 650216, People's Republic
of China 
\and INAF, Istituto di Astrofisica e Planetologia Spaziali, Via del Fosso del Cavaliere, 100, I-00133, Roma (RM), Italy
\and State Key Laboratory of Particle Astrophysics, Institute of High Energy Physics, Chinese Academy of Sciences, Beijing 100049, People’s
Republic of China
\and National Astronomical Observatories, Chinese Academy of Sciences, Beijing 100101, People's Republic of China
\and Ulugh Beg Astronomical Institute, Uzbekistan Academy of Sciences, Astronomy str. 33, Tashkent 100052, Uzbekistan
\and Key Laboratory for Computational Astrophysics, National Astronomical Observatories, Chinese Academy of Sciences, Datun Road A20,
Beijing 100012, People's Republic of China
\and Department of Astronomy, School of Physics, Huazhong University of Science and Technology, Wuhan 430074, People’s Republic of
China
\and National Astronomical Research Institute of Thailand, 260 Moo 4, Donkaew, Maerim, Chiang Mai 50180, Thailand
\and Institute for Frontiers in Astronomy and Astrophysics, Beijing Normal University, Beijing 102206, People's Republic of China
\and Aix Marseille Univ., CNRS, CNES, LAM, Marseille, France
\and School of Physics and Astronomy, University of Leicester, University Road, Leicester, LE1 7RH, United Kingdom
\and Cosmic Dawn Center (DAWN), Denmark
\and Niels Bohr Institute, University of Copenhagen, Jagtvej 128, 2200 Copenhagen N, Denmark
\and Department of Astrophysics/IMAPP, Radboud University, 6525 AJ Nijmegen, The Netherlands
\and LUX, Observatoire de Paris, Universit\'e PSL, CNRS, Sorbonne Universit\'e, Meudon 92190, France
\and Department of Physics, University of Warwick, Coventry, CV4 7AL, UK
\and Anton Pannekoek Institute of Astronomy, University of Amsterdam, Science Park 904, 1098 XH Amsterdam, The Netherlands
\and Universit\'e Paris-Saclay, Universit\'e Paris Cit\'e, CEA, CNRS, AIM, 91191, Gif-sur-Yvette, France
\and Centre national d'\'etudes spatiales (CNES), Paris, France
\and INAF, Osservatorio Astronomico di Capodimonte, Salita Moiariello 16, I-80131, Napoli, Italy
\and DARK, Niels Bohr Institute, University of Copenhagen, Jagtvej 155A, 2200, Copenhagen N, Denmark
\and Italian Research Center on High Performance Computing, Big Data and Quantum Computing, Via Magnanelli, 2, Casalecchio di Reno,
I-40033, Italy
\and Physics Department, American University of Sharjah, PO Box 26666, Sharjah, United Arab Emirates
\and Universit\'e Paris-Saclay, CNRS/IN2P3, IJCLab, F-91405 Orsay, France
\and Physics Department, Nazarbayev University, 53 Kabanbay Batyr Ave., Astana 010000, Kazakhstan
\and Energetic Cosmos Laboratory, Nazarbayev University, 53 Kabanbay Batyr Ave., Astana 010000, Kazakhstan
\and CEA Paris-Saclay, IRFU/DAp-AIM, 91191 Gif-sur-Yvette, France
\and Laboratorio Nacional de Astrofsica, R. dos Estados Unidos, 154 - Nacoes, Itajuba - MG, 37504-364, Brazil
\and NEPA, Universidade do Estado do Amazonas, Rd. Odovaldo Novo, Parintins - AM, 69152-470, Brazil
\and National Research Institute of Astronomy and Geophysics (NRIAG), 11421 Helwan, Cairo, Egypt
\and Graduate Institute of Astronomy, National Central University, 300 Jhongda Road, 32001 Jhongli, Taiwan
\and Institute of Astronomy, National Tsing Hua University, Hsinchu 300044, Taiwan
\and Las Cumbres Observatory, 6740 Cortona Drive Suite 102, Goleta, CA 93117-5575, USA
\and Department of Physics, University of California, Santa Barbara, CA 93106-9530, USA
\and Kavli Institute for Theoretical Physics, University of California, Santa Barbara, 552 University Road, Goleta, CA 93106-4030, USA
\and Space Research Institute, Russian Academy of Sciences, Moscow 117997, Russia
\and Faculty of Physics, Higher School of Economics, Moscow 101000, Russia
\and Special Astrophysical Observatory, Russian Academy of Sciences, Nizhnij Arkhyz 369167, Russia
\and South-Western Institute for Astronomy Research, Yunnan University, Kunming, Yunnan 650504, People's Republic of China
\and Yunnan Key Laboratory of Survey Science, Yunnan University, Kunming, Yunnan 650500, People's Republic of China
\and Purple Mountain Observatory, Chinese Academy of Sciences, Nanjing, 210023, People's Republic of China}

   \date{Received 18 August 2026 / Revised 12 November 2026
   }


  \abstract
   {Recent advances in the study of fast X-ray transients have been driven by the wide-field, sensitive soft X-ray monitoring capabilities of the Einstein Probe (EP) mission, enabling the prompt detection and precise localization of previously underexplored fast X-ray transients. EP observations provide new opportunities to investigate the physical origins, emission mechanisms, and population properties of newly discovered fast X-ray transients.}
   {As a representative example, we present the multiwavelength observations of the EP-detected extragalactic fast X-ray transient EP250226a, which is associated with the gamma-ray burst (GRB) 250226A. We report comprehensive observational results covering the multiwavelength emission from the prompt phase to the afterglow phase.}
   {We perform a joint spectral analysis of the prompt emission of the burst from soft X-rays to gamma rays, together with a spectral analysis of the X-ray afterglow. We collect optical photometric data and perform both empirical and afterglow model-based analyses. We present optical-to-infrared spectroscopic observations of the transient and measure the equivalent widths of typical absorption lines commonly identified in GRB afterglow spectra.}
   {The spectroscopic redshift of the transient is $z = 3.314$. The isotropic-equivalent gamma-ray energy of the burst is measured to be $E_\mathrm{iso}\simeq7.32\times10^{53}$ erg. The location of the transient in the Amati diagram suggests that the burst likely originated from the core collapse of a massive star (i.e., a Type II GRB). Furthermore, based on the estimated fluence ratio of $\rm{S}(25-50\,\rm{keV})/\rm{S}(50-100\,\rm{keV})=0.57\pm0.06$, the burst is classified as a classical GRB. 
   The X-ray and optical afterglows exhibit achromatic behavior.
   We model the multiwavelength afterglow emission using several afterglow models and find that a top-hat jet propagating into a uniform interstellar medium provides the best description of the observations.
   The preferred model suggests a highly collimated, nearly on-axis jet, with an isotropic-equivalent kinetic energy of $E_\mathrm{k,iso}\simeq4.07\times10^{55}$ erg, a low gamma-ray radiative efficiency of $\eta_{\gamma}\simeq 1.77\%$, and a small fraction of shock energy in the magnetic field ($\log \epsilon_\mathrm{B,FS} \simeq -6.52$).}
   {In the context of GRB population studies, EP250226a shows typical characteristics of a long-duration GRB, with the first peak of its soft X-ray emission coinciding with the gamma-ray prompt emission, while the soft X-ray emission extends significantly beyond the duration of the gamma-ray emission.}

   
   \maketitle

\nolinenumbers
\section{Introduction} \label{sec:intro}
Fast X-ray transients are a heterogeneous class of high-energy transients characterized by soft X-ray emission on timescales ranging from seconds to thousands of seconds \citep{2004NuPhS.132..263H,2010csxs.book..267H,2020ApJ...896...39A,2026arXiv260708090Q}. Their physical origins span a wide variety of Galactic and extragalactic phenomena, making them an important population for understanding the dynamic high-energy universe \citep{2004NuPhS.132..263H,2013ApJ...779...14J,2017MNRAS.467.4841B,2023A&A...675A..44Q,2026arXiv260708090Q,2026MNRAS.545f2064Q,2026arXiv260627048V}. To date, a substantial fraction of extragalactic fast X-ray transients have been discovered through systematic searches of historical X-ray archives \citep[e.g.,][]{2017MNRAS.467.4841B,2022A&A...663A.168Q,2023A&A...675A..44Q}. Although a small number of extragalactic fast X-ray transients were previously detected by several soft X-ray missions \citep[e.g.,][]{2001grba.conf...16H,2005ApJ...629..311S,2016PASJ...68S...1N}, the lack of rapid localization and timely multiwavelength follow-up observations has prevented detailed characterization of their physical properties and progenitor systems.

The Einstein Probe (EP) mission was recently launched to survey the soft X-ray sky in the 0.5--4 keV band and to provide rapid X-ray follow-up observations in the 0.5--10 keV energy range \citep{10.1007/s11433-024-2600-3}. The Wide-field X-ray Telescope \citep[WXT;][]{10.1007/978-981-16-4544-0_151-1} provides an instantaneous field of view of approximately 3,600 square degrees and enables near real-time detection and arcminute localization of X-ray transients. Autonomous follow-up observations with the Follow-up X-ray Telescope \citep[FXT;][]{10.1117/12.2562311} further improve the localization accuracy to the arcsecond level. Together with rapid-response observations from both space-based and ground-based facilities, EP has fundamentally expanded the discovery space of extragalactic fast X-ray transients. 

An increasing number of EP-discovered fast X-ray transients have now been found to be temporally and spatially associated with gamma-ray bursts (GRBs) \citep[e.g.,][]{2026arXiv260421624Z,2025ApJ...989L..39Y,2025NatAs...9.1375L,2025NatAs...9..564L,2025ApJ...988L..34J,2024ApJ...975L..27Y}. Likewise, the Space-based multi-band astronomical Variable Objects Monitor \citep[SVOM;][]{2016arXiv161006892W,2026arXiv260424257C}, equipped with the wide-field instrument ECLAIRs with a field of view of approximately 2 sr and a low-energy threshold down to 4 keV \citep{2014SPIE.9144E..24G,2026RAA....26j4007G}, is also discovering soft and faint GRBs that further bridge the observational connection between fast X-ray transients and the classical GRB population \citep{2026A&A...714A..82S}. 
For some subclasses of extragalactic fast X-ray transients, such as `CDF-S XT1'-like events, the inferred intrinsic event rate can be comparable to those of orphan and low-luminosity GRBs \citep{2017MNRAS.467.4841B,2022A&A...663A.168Q,2023A&A...675A..44Q}.


Among the diverse fast X-ray transient population, one important subclass is physically associated with long-duration GRBs. In these events, the prompt emission is generally attributed to internal energy dissipation processes within relativistic jets \citep{1994ApJ...430L..93R,1994ApJ...427..708P,1997MNRAS.287..110S,2004RvMP...76.1143P,2006ApJ...642..354Z}. 
The prompt emission is generally detected by GRB-triggering instruments, such as the Fermi Gamma-ray Burst Monitor \citep[Fermi/GBM;][]{2009ApJ...702..791M}, which covers an energy range of 8 keV-40 MeV, and the Swift Burst Alert Telescope \citep[Swift/BAT;][]{2004ApJ...611.1005G,2004SPIE.5165..175B,2005SSRv..120..143B}, which operates over 15--150 keV. 
In contrast, EP provides sensitive triggering in the soft X-ray band (0.5--4 keV), extending the observational window to lower energies than those accessible to previous GRB-triggering instruments.
This unique capability enables the detection and characterization of soft X-ray emission associated with GRBs, offering new insights into the early emission mechanisms. A distinguishing feature of EP-detected GRBs is that the soft X-ray emission lasts longer than the gamma-ray emission \citep{2025NatAs...9..564L,2025Galax..13...62W,2026arXiv260714317C,2026arXiv260627048V}. Following the prompt phase, broadband afterglow emission is produced as the relativistic ejecta interact with the circumburst medium, where shock-accelerated electrons radiate predominantly through synchrotron processes \citep{1997MNRAS.287..110S,1998ApJ...497L..17S,1997ApJ...476..232M,10.1016/S0370-1573(98)00127-6,2013NewAR..57..141G,2015PhR...561....1K,2018pgrb.book.....Z}. Multiwavelength observations of the prompt and afterglow emission therefore provide powerful diagnostics of the jet energetics, geometry, radiation mechanisms, and the circumburst environment. Consequently, establishing whether a newly discovered fast X-ray transient belongs to the GRB population is essential for understanding its physical origin.

EP250226a/GRB~250226A was discovered by EP as a bright fast X-ray transient and subsequently triggered extensive follow-up observations across the electromagnetic spectrum. The event was also independently detected by several gamma-ray instruments (see Sections~\ref{sec:Fermi/GBM_Observations}, \ref{sec:SPI-ACS_Observations}, and \ref{sec:GECAM_Observation} for details of the gamma-ray observations), providing further evidence for a connection with the GRB population. 
We investigate whether EP250226a can be robustly classified as a long-duration GRB and explore its physical nature through comprehensive multiwavelength analyses. In this paper, we present X-ray and gamma-ray observations of the prompt emission, X-ray and optical observations of the afterglow, together with optical spectroscopic measurements of the transient. We perform detailed spectral analyses of the prompt emission and broadband modeling of the afterglow to constrain the properties of the relativistic jet  and its surrounding environment. This event serves as a representative example of the growing population of EP-discovered fast X-ray transients that can be physically linked to classical GRBs, illustrating the new discovery space opened by sensitive soft X-ray transient surveys.

This paper is organized as follows. Section~\ref{sec:Observations_and_Data_Reduction} describes the multiwavelength observations and data reduction procedures. Section~\ref{sec:data_analysis} presents the prompt emission spectral analysis and the empirical analysis of the afterglow light curves. Section~\ref{sec:Results_and_Discussions} discusses the GRB classification, broadband afterglow modeling, and the corresponding physical implications. Finally, Section~\ref{sec:Summary} summarizes our main conclusions. Throughout the paper, the Lambda Cold Dark Matter ($\mathrm{\Lambda CDM}$) cosmological model is adopted with the Hubble constant $H_{0} = 67.66\mathrm{\,km\,s^{-1}\,Mpc^{-1}}$, the baryon and cold dark matter density $\Omega_\mathrm{m} = 0.3103$, and the dark energy density $\Omega_{\Lambda} = 0.6897$ \citep{2020A&A...641A...6P}.
\section{Observations and data reduction} \label{sec:Observations_and_Data_Reduction}

\subsection{Fermi/GBM observations}\label{sec:Fermi/GBM_Observations}
At 06:34:57 UT on 26 February 2025 (referred to as $T_0$), 
the Fermi/GBM triggered on and localized a burst designated GRB 250226A. The burst position was determined to be at R.A. = $222.1^\circ$ and Decl. = $23.8^\circ$ (J2000) with a statistical uncertainty of $3.9^\circ$ \citep{2025GCN.39479....1F}. 

The Fermi/GBM is equipped with 12 sodium iodide (NaI) detectors covering the 8--1000 keV energy range and two bismuth germanium oxide (BGO) detectors covering 0.2--40 MeV \citep{2009ApJ...702..791M}. For our analysis, we selected the NaI detector (na) and the BGO detector (b1) that had the smallest viewing angles to the burst among the NaI and BGO detectors, respectively. We used the temporally unbinned time-tagged event (TTE) data\footnote{\url{https://heasarc.gsfc.nasa.gov/FTP/fermi}}. The TTE data have a temporal resolution of 2 $\mu s$ and contain 128 energy channels. Using the Fermi GBM Data Tools \citep{GbmDataTools}, we extracted the light curves of GRB 250226A from the TTE data in three energy ranges: 44--300 keV, 8--1000 keV, and 0.2--40 MeV. The time binning for all light curves was set to 1 s. 
The Fermi/GBM light curve shows several distinct spikes within a single emission episode, with a duration ($T_{90}$) of about 61 s in the 50--300 keV band \citep{2025GCN.39530....1P}.
Time-averaged energy spectra were extracted over the interval $T_0 - 2.6\ \mathrm{s}$ to $T_0 + 62 \ \mathrm{s}$. The 8--1000 keV spectrum was obtained from the na detector, while the 0.2--40 MeV spectrum was extracted from the b1 detector. To estimate the background, we fitted a third-order polynomial to each energy channel, using two background intervals defined as $T_0 - 100 \ \mathrm{s}$ to $T_0 - 20 \ \mathrm{s}$ and  $T_0 + 75 \ \mathrm{s}$ to $T_0 + 150 \ \mathrm{s}$. The resulting background models were integrated over the 44--300 keV, 8--1000 keV, and 0.2--40 MeV energy bands to obtain the corresponding background light curves. The background models were integrated over the time interval $T_0 - 2.6\ \mathrm{s}$  to $T_0 + 62 \ \mathrm{s}$ to obtain the background spectra for the 8--1000 keV and 0.2--40 MeV energy bands.
Since the burst occurred near the South Atlantic Anomaly (SAA) entry, the Fermi/GBM spectral results (particularly the peak energy $E_\mathrm{p}$) are likely affected by contamination, and therefore may not accurately represent the intrinsic spectral properties of the burst \citep{2025GCN.39530....1P}. Therefore, the Fermi/GBM spectra were not used in the spectral analysis.




The light curves observed by Fermi/GBM are shown in Figure~\ref{fig:HE-lc}. We define the Fermi/GBM trigger time as $T_0$. All times presented in figures and tables are measured relative to $T_0$, unless stated otherwise.

\subsection{INTEGRAL observations}\label{sec:SPI-ACS_Observations}

At 06:34:59 UTC on 26 February 2025, the Anti-Coincidence Shield of the Spectrometer on board the International Gamma-ray Astrophysics Laboratory \citep[INTEGRAL/SPI-ACS;][]{2003A&A...411L..63V,SPI-ACS} recorded a transient event designated as GRB~250226A \citep{2025GCN.39518....1L}. SPI-ACS provided continuous all-sky monitoring of high-energy transients with a temporal resolution of 50~ms and sensitivity to photons with energies $\gtrsim 80$~keV. Although SPI-ACS did not provide spectral or positional information, its large effective area and sky coverage made it particularly effective for detecting short-duration gamma-ray events.

The SPI-ACS data were analyzed using an automated, multi-stage transient detection pipeline designed to identify statistically significant excesses above background across multiple timescales. The pipeline extracted ACS count-rate data and onboard times from standard FITS files, converted all timestamps into common astronomical systems (UTC, JD, MJD, and  IJD), and produced background-subtracted light curves. To account for the diversity of transient durations, the data were rebinned over timescales ranging from 50~ms to 1~s.

Candidate excesses were identified by computing signal-to-noise ratios at each timescale and evaluated using statistical false-alarm metrics, including the False Alarm Rate (FAR) and False Alarm Probability (FAP). Events were classified according to their statistical significance, and high-confidence detections were further characterized by measuring temporal properties such as burst duration and peak count rate.

In the SPI-ACS light curve above 80~keV, we identified a statistically significant signal temporally coincident with the detections reported by Fermi/GBM and EP/WXT \citep{2025GCN.39479....1F,2025GCN.39482....1J}. The emission consisted of a single pulse with a duration of $T_{90} = 33 \pm 1$ s above 80~keV. A temporally consistent, though marginal, excess was also observed in the Pixellated Imaging CaeSium Iodide Telescope \citep[PICsIT;][]{PICsIT} of the Imager on Board the INTEGRAL Satellite \citep[IBIS;][]{2003A&A...411L.131U} data.

The peak count rate recorded by SPI-ACS reached $\sim 7.4 \times 10^{4}$~cts~s$^{-1}$ above 80~keV, compared to a median background level of $\sim 6.76 \times 10^{4}$~cts~s$^{-1}$. The temporal coincidence and consistent duration across multiple instruments supported the association of the SPI-ACS signal with GRB~250226A. The light curves observed by INTEGRAL are shown in Figure~\ref{fig:HE-lc}.

\subsection{GECAM-B observations} \label{sec:GECAM_Observation}
The Gravitational wave high-energy Electromagnetic Counterpart All-sky Monitor \citep[GECAM;][]{2022RDTM....6...12L} mission was triggered both in-flight and on-ground by GRB 250226A  \citep{2025GCN.39492....1Z}. 
All 25 gamma-ray detectors (GRDs) of GECAM-B operate in two readout channels: high gain (HG) and low gain \citep[LG;][]{GEC_INS_An2022}. For spectral analysis, the energy range is 40--300\,keV for HG and 0.7--6\,MeV for LG. For temporal analysis, the energy range is 40--300\,keV for HG and 0.3--6\,MeV for LG. Detectors with an incident angle less than 60$^\circ$ (that is, GRD11, GRD12, and GRD21 of GECAM-B for this burst) were selected for spectral and temporal analysis with GECAMTools\footnote{\url{https://gecam.ihep.ac.cn/grbDataAnalysisSoftware.jhtml}}. 
According to the GECAM-B light curves in the energy range of approximately 40--6000 keV, this burst is mainly composed of a single pulse, with a total duration of about $63.5 \pm 11.7$ s ($T_{90} = T_0 + [-7, +57]$ s) as shown in Figure~\ref{fig:HE-lc}.




\subsection{Einstein Probe observations}\label{sec:Einstein_Probe_Observation}

At 06:35:16 on 26 February 2025 (UTC), the WXT onboard the EP mission \citep{10.1007/s11433-024-2600-3} triggered on a fast X-ray transient via the onboard processing unit. The fast X-ray transient was assigned as EP250226a and localized by EP/WXT at coordinates (J2000) R.A. = 224.273 degrees, Decl. = 20.973 degrees, with a typical error radius of 3 arcminutes \citep[90\% C.L. statistical and systematic;][]{2025GCN.39482....1J}. The refined EP/WXT trigger time was determined using the Bayesian block algorithm \citep{2013ApJ...764..167S} as 06:34:56.2 on 26 February 2025 (UTC). Upon inspection of the trigger times and localizations reported by Fermi/GBM and EP/WXT, EP250226a is found to be temporally and spatially consistent with GRB 250226A. The light curve observed by EP/WXT is presented in Figure~\ref{fig:HE-lc}. The EP/WXT light curve lasted $\sim250$ s, and the EP/WXT observation was interrupted by the EP/FXT autonomous follow-up observation \citep{2025GCN.39482....1J,2025GCN.39513....1J}. 

The autonomous follow-up observation was carried out by the EP/FXT, with an exposure time of 2.7 ks at a mid-time of 1.1 hours post-trigger, after an initial delay caused by the Earth occultation \citep{2025GCN.39513....1J}. The EP/FXT detected an uncatalogued X-ray source at coordinates (J2000) R.A. = 224.2641 deg, Decl. = 20.9754 deg, with a typical error radius of 10 arcsec \citep[90\% C.L. statistical and systematic;][]{2025GCN.39513....1J}. The second, third, and fourth X-ray follow-up observations were conducted by EP/FXT, with exposure times of 3.0 ks, 2.8 ks, and 3.0 ks at mid-times of 12.5, 31.6, and 82.8 hours post-trigger, respectively.

The EP/WXT data were preprocessed using the WXT Data Analysis Software (WXTDAS v2.10; Y. Liu et al., in preparation) and the WXT calibration database \citep[CALDB v1.0;][]{2026arXiv260628101C,2025arXiv250620369C}. The EP/FXT data were pre-processed with the FXT Data Analysis Software \citep[FXTDAS v1.30;][]{2025RDTM....9..215Z} using the FXT calibration database (CALDB v1.30).

\subsection{Optical photometric observations}\label{sec:optical_obs}
\subsubsection{TRT detection of the optical counterpart}
We observed the field of EP250226a detected by EP/WXT, using the 0.7-m telescope of the Thai Robotic Telescope network (TRT), located at Cerro Tololo Inter-American Observatory (CTIO), Chile. The TRT first conducted observations at a mid-time of 836 s after $T_0$ and obtained several frames in the Sloan-$r$ filter. An uncatalogued, fading optical source was detected within the EP/FXT error circle \citep{2025GCN.39482....1J} at coordinates (J2000): R.A. = 14:57:03.63 and Decl. = +20:58:32.31, with an uncertainty of about 1.0 arcsec \citep{2025GCN.39486....1A}. We inferred that this optical source was the optical counterpart of EP250226a (Figure~\ref{fig:fc}). The magnitudes of the optical counterpart were calibrated against Pan-STARRS DR2 stars \citep{2020ApJS..251....7F} and are listed in Table~\ref{tab:optical-data}.

In the direction of the optical counterpart, the Galactic foreground extinction, corresponding to a reddening of $E_{B-V}^\mathrm{Gal} = 0.04 \mathrm{\,mag}$, was retrieved from the Galactic Dust Reddening and Extinction Service\footnote{\url{https://irsa.ipac.caltech.edu/applications/DUST/}} \citep{2011ApJ...737..103S}. The Galactic total column density of hydrogen, accounting for contributions from both atomic and molecular hydrogen, was queried using the \texttt{NHtot} tool\footnote{\url{https://www.swift.ac.uk/analysis/nhtot/}}, yielding $N_\mathrm{H}^\mathrm{Gal} = 3.68\times10^{20} \mathrm{\,cm^{-2}}$ \citep{2013MNRAS.431..394W}.

\subsubsection{GRANDMA observations}
The Global Rapid Advanced Network Devoted to the Multi-messenger Addict \citep[GRANDMA;][]{2020MNRAS.497.5518A} started observing with the following telescopes: the TRT at the Spring Brook Observatory (SBO), the 1.88 m telescope at the Kottamia Astronomical Observatory (KAO), the AZT-8 telescope at the Lisnyky Observatory, the 70 cm telescope at the Abastumani Observatory, and the 60 cm telescope at the Pico dos Dias Observatory (OPD) approximately 11 hours after $T_0$. 
Detections of the optical afterglow and upper limits (ULs) are shown in Table~\ref{tab:optical-data} and are consistent with the results from other collaborations. We used the STDPipe pipeline and its web interface to perform forced photometry and template subtraction on a set of images obtained from different telescopes \citep{Karpov_2025}. For data processing of images in Bessel $R$ and Bessel $I$ filters, we used the Gaia DR3 synphot catalog~\citep{2023A&A...674A..33G}, and for data in SDSS $i$ filter, we used the Pan-STARRS DR2 catalog~\citep{2020ApJS..251....7F}. 

\subsubsection{Other optical observations}

The optical afterglow follow-up observations were also carried out with 
the 1-meter telescope at the Las Cumbres Observatory (LCO) node located at CTIO in Chile as part of the Global Supernova Project \citep[GSP;][]{2025GCN.39489....1L}, 
the Schmidt telescope at the Xinglong Observatory, 
the Multi-channel Photometric Survey Telescope \citep[Mephisto;][]{2025GCN.39511....1Z}, 
the Nordic Optical Telescope (NOT), 
the Lulin Observatory as part of the Kinder project \citep{2025ApJ...983...86C,2025GCN.39509....1A,2025ApJS..281...20A}, 
the Space-based Multi-band Astronomical Variable Objects Monitor/the Visible Telescope \citep[SVOM/VT;][]{10.1088/1674-4527/ae75cd}, 
the acquisition camera of the Very Large Telescope \citep[VLT;][]{2025GCN.39487....1Z}, 
and the 6-meter alt/az mounted telescope (Big Telescope Alt-azimuth; BTA) operated by the Special Astrophysical Observatory of the Russian Academy of Sciences (SAO RAS). 

All raw images were processed using telescope-specific preprocessing procedures, which included bias subtraction, flat-field correction, and cosmic-ray removal. Astrometric solutions were initially obtained for individual images through the \texttt{Astrometry.net} service\footnote{\url{https://nova.astrometry.net}} \citep{2010AJ....139.1782L}. Subsequently, the images were resampled and combined with \texttt{SWarp} \citep{2010ascl.soft10068B} to improve the signal-to-noise ratio of the images. Transient identification, astrometric refinement, and point-spread function (PSF) photometry at the transient location were then carried out using the Python package \texttt{STDPipe} \citep{2021ascl.soft12006K,Karpov_2025}.
Photometric calibration was performed using field stars selected from reference catalogs appropriate to the filter system of each observation. For images acquired with the Johnson-Cousins $VRI$ filters, we adopted the methodology of Gaia DR3 synthetic photometry \citep{2023A&A...674A..33G,2023ApJS..269...58X,2024ApJS..271...13H}, which provides synthetic photometric measurements derived from the low-resolution BP/RP (XP) spectra collected by the Gaia mission \citep{2023A&A...674A..33G}. For observations obtained with the Sloan $griz$ (as well as $r'i'$) filters, reference magnitudes were taken from the Pan-STARRS DR2 catalog \citep{2020ApJS..251....7F}. In particular, the photometric calibration of observations obtained with instrument-specific filter systems, including those of SVOM/VT and Mephisto, followed the corresponding instrument-specific data reduction procedures \citep{10.1088/1674-4527/ae75cd,2025ApJ...979...38C}. 

The refined optical photometric results of EP250226a, combined with the magnitudes reported in the General Coordinates Network (GCN) Circulars\footnote{\url{https://gcn.nasa.gov/circulars/events/ep250226a}}, are listed in Table~\ref{tab:optical-data}.

\subsection{Spectroscopic observations}\label{sec:Spectroscopic_Observations}

We also carried out spectroscopic observations of the optical counterpart of EP250226a/GRB 250226A using the X-shooter instrument \citep{2011A&A...536A.105V}, a multiwavelength medium-resolution spectrograph mounted on Unit Telescope 3 (Melipal) of the European Southern Observatory (ESO) Very Large Telescope (VLT), under program 114.27PZ (PIs: Tanvir, Malesani, Vergani).
X-shooter enables simultaneous spectral coverage across three arms: the ultraviolet-blue (UVB; 300--560 nm), visible (VIS; 550--1020 nm), and near-infrared (NIR; 1020--2100 nm) bands, with corresponding resolving powers of approximately R $\sim$ 5400, 8900, and 5600, respectively.
The full data set was reduced using the ESO Reflex environment, which provides an automated workflow for organizing the data and executing a sequence of data reduction steps \citep{2013A&A...559A..96F}.
The VLT/X-shooter spectra cover the wavelength range 3000--21000 \AA, and consist of $2\times600$ s exposures. The mid-time of the spectroscopic observation was 08:43:57 UTC on 26 February 2025 (2.15 hr after $T_0$).

In the VLT/X-shooter spectra, we clearly detected a continuum down to $4000\mathrm{\,\AA}$. We also identified a broad Lyman-$\alpha$ absorption feature at $\sim5230\mathrm{\,\AA}$, together with multiple narrow absorption features from O I, O I*, Si II, Si II*, C II, C II*, C IV, Si IV, Al II, Fe II, Fe II*, Al III, Ni II, Ni II*, Zn II, Cr II, Mg II, Mg I, and Ca II. We fitted these absorption lines with Voigt profiles using the Python package \texttt{VoigtFit} \citep{2018arXiv180301187K} and obtained a common redshift of $z = 3.3140 \pm 0.0001$, consistent with the preliminary value reported by \citet{2025GCN.39487....1Z}. The identified absorption lines are marked in Figure~\ref{fig:VLTspec}.

For the absorber at $z = 3.314$, we fitted the Lyman-$\alpha$ damping trough and obtained a neutral hydrogen column density of $\log(\NHI/\mathrm{cm^{-2}})=21.52^{+0.16}_{-0.17}$. 
This value is in good agreement with the median \HI~column density, $\log(\NHI/\mathrm{cm^{-2}})=21.59$, reported in statistical studies of GRB afterglows \citep[e.g.,][]{2019MNRAS.483.5380T,2009ApJS..185..526F,2006A&A...460L..13J}.
We also measured the equivalent widths (EWs) of absorption features using the Voigt profile fitting and compared them with those from a reference GRB sample \citep{2012A&A...548A..11D}. 
A line strength parameter (LSP) was introduced to represent the relative intensity of absorption lines in a GRB spectrum with respect to the reference sample using a single metric \citep{2012A&A...548A..11D}. We derived an LSP of $0.203 \pm 0.048$ for the afterglow spectrum of GRB 250226A (Figure~\ref{fig:VLTspec}), which is comparable to the average LSP of the reference sample \citep[a Gaussian distribution with a mean value of $\langle \mathrm{LSP} \rangle = -0.03$ and a standard deviation of $\sigma_\mathrm{LSP} = 0.72$;][]{2012A&A...548A..11D}.

\section{Data analysis}\label{sec:data_analysis}
\subsection{High-energy spectral analysis}\label{sec:HE_spectra}
We performed the time-averaged spectral analysis for EP data and GECAM-B using the Efficient Library for Spectral Analysis in High-Energy Astrophysics \citep[ELISA;][]{ELISA,2026arXiv261014587X}. 
The leave-one-out information criterion (LOOIC) and the widely applicable information criterion \citep[WAIC;][]{2010arXiv1004.2316W,2015arXiv150704544V} were used for spectral model diagnostics when employing the ELISA package.
The high-energy data obtained from EP and GECAM-B were divided into the following time intervals according to data availability, and spectral analysis was performed for each interval.

(I) The EP/WXT data extracted within $T_0 + [-0.8, +19.7]$ s were selected using the Bayesian blocks method \citep{2013ApJ...764..167S}. The EP/WXT spectrum can be described by an absorbed power-law model. We fixed the Galactic absorption column density at $N_\mathrm{H}^\mathrm{Gal} = 3.68\times10^{20}\mathrm{\,cm^{-2}}$. Because of the limited energy-band coverage of EP/WXT, the intrinsic column density ($N_\mathrm{H,X}$) and photon index ($\Gamma$) cannot be well constrained when both parameters are allowed to vary simultaneously. Therefore, we considered two cases: in the first case, $N_\mathrm{H,X}$ was left free to vary. In the second case, $N_\mathrm{H,X}$ was fixed at $0.91\times10^{22}\mathrm{\,cm^{-2}}$, evaluated at the burst redshift and derived from the first EP/FXT follow-up observation (see Table~\ref{tab:HE-data}). The intrinsic column density ($N_\mathrm{H,X}=0.91\times10^{22}\mathrm{\,cm^{-2}}$) lies within the distribution of the intrinsic column density of the long GRB sample characterized by a log-normal distribution, with $\log(N_{\mathrm{H,X}}/\mathrm{cm}^{-2})=21.7\pm0.5$ \citep{2012MNRAS.421.1697C}. We note that the $N_{\mathrm{H,X}}$ value fixed in the EP/WXT spectral fitting may be regarded as a lower limit. The spectral fitting results of WXT spectrum are presented in Figure~\ref{fig:WXT} and reported in Table~\ref{tab:HE-data}. 
The X-ray luminosity was estimated using Equation~1 from \citet{2025arXiv251010267L}.
The X-ray emission reached a peak flux of $9.8 \times 10^{-9}\,\mathrm{erg\,s^{-1}\,cm^{-2}}$ in 0.5--4\,keV band \citep{ 2025GCN.39513....1J}, corresponding to a peak X-ray luminosity of $2.24 \times 10^{50}\,\mathrm{erg\,s^{-1}}$.

(II) The EP/WXT data and GECAM-B data within $T_0 + [-0.8, +19.7]$ s were extracted for joint spectral analysis. The joint spectrum can be fitted by an absorbed power-law model with a high-energy exponential cutoff, which can be written as
\begin{equation}\label{eq:Compt}
    N(E) = K \left(\frac{E}{E_0}\right)^{-\Gamma}
        \exp \left[-\frac{(2-\Gamma)E}{E_\mathrm{p}}\right],
\end{equation} where $E_0$ is the reference energy fixed at 1 keV. The corresponding energy of the peak of the $\nu F_\nu$ spectrum is parameterized as the peak energy $E_\mathrm{p}$. The amplitude is described with the parameter $K$. The EP/WXT data in the soft X-ray band were included in the joint spectral fitting, and two cases were considered: with $N_{\mathrm{H,X}}$ fixed or left free to vary. The two cases yielded consistent photon indices of $0.94^{+0.05}_{-0.06}$ and $1.03 \pm 0.07$, respectively. In the case where the intrinsic column density was allowed to vary, the fitted intrinsic column density is $(8.81^{+6.21}_{-4.74})\mathrm{\times10^{22}\,cm^{-2}}$, which is higher than the typical values inferred for most GRBs. The distribution of the intrinsic column density of the long GRB sample was estimated to follow a log-normal distribution, with $\log(N_{\mathrm{H,X}}/\mathrm{cm}^{-2})=21.7\pm0.5$ \citep{2012MNRAS.421.1697C}. The fitted intrinsic column density is approximately 2.5 standard deviations above the mean of this distribution. Therefore, the photon index derived with the intrinsic column density fixed, $0.94^{+0.05}_{-0.06}$, is considered more reliable.

We also fitted the joint spectrum using an absorbed Band function \citep{1993ApJ...413..281B}. The low-energy photon index obtained from this fit is $0.95^{+0.06}_{-0.05}$, consistent with the photon index of $0.94^{+0.05}_{-0.06}$ derived from the absorbed cutoff power-law model. However, the high-energy photon index is not well constrained. The rest-frame spectral peak energy derived from the Band model is $(1.23^{+0.37}_{-0.25})\times10^3$ keV, consistent with the value of $(1.20^{+0.33}_{-0.23}) \times 10^{3}$ keV obtained from the absorbed cutoff power-law fit. The goodness of fit is comparable between the absorbed cutoff power-law model (stat/d.o.f. = 138.46/113) and the absorbed Band model (stat/d.o.f. = 138.69/112). Nevertheless, the absorbed cutoff power-law model provides a slightly better fit to the data.


We adopted the absorbed cutoff power-law model as our preferred model based on the above comparison. The photon index of the cutoff power-law model is $\Gamma = 0.94^{+0.05}_{-0.06}$, and the rest-frame spectral peak energy is $(1+z)E_\mathrm{p} = (1.20^{+0.33}_{-0.23}) \times 10^{3}$ keV. 
Following the standard analysis procedure developed for GECAM-B \citep{GHS_pipeline}, the isotropic-equivalent prompt gamma-ray energy was calculated to be $E_\mathrm{iso}=(7.32^{+3.75}_{-1.93})\times10^{53}$ erg, and the fluence was estimated to be $(3.17^{+1.19}_{-0.81})\times10^{-5}\mathrm{\,erg\,cm^{-2}}$. 
The spectral analysis results of joint EP/WXT and GECAM-B data are displayed in Figure~\ref{fig:WXT_GECAM} and listed in Table~\ref{tab:HE-data}. 
We note that $(1+z)E_\mathrm{p}$, $E_\mathrm{iso}$, and the fluence derived from the EP/WXT and GECAM-B joint spectrum are higher than those reported by Konus-Wind \citep{2025GCN.39542....1S}. This discrepancy may arise from differences in the adopted spectral models and in the energy ranges and time intervals covered by the joint spectral analysis.

(III) The EP/WXT light curve extends to $\sim250$ s (Figure~\ref{fig:HE-lc}). The EP/WXT data collected over the interval $T_0 + [0, +250]$ s were extracted for the time-averaged spectral analysis.
The time-averaged EP/WXT spectrum can be described by an absorbed power-law model. The spectral fitting procedure is the same as in part (I).
The spectral analysis results are displayed in Figure~\ref{fig:WXT-TimeAveraged} and reported in Table~\ref{tab:HE-data}. The unabsorbed X-ray flux was estimated to be $(9.26^{+1.76}_{-1.29})\times10^{-9}\,\mathrm{erg\,s^{-1}\,cm^{-2}}$ in 0.5--4\,keV band, corresponding to a time-averaged X-ray luminosity of $(3.12^{+0.59}_{-0.43})\times 10^{49}\,\mathrm{erg\,s^{-1}}$.

(IV) The X-ray afterglow observed by EP/FXT was independently fitted with an absorbed power-law model in the time intervals $T_{0} + [2725, 5440]$ s, 
$T_{0} + [43056, 46064]$ s, 
$T_{0} + [112193,115044]$ s,
and $T_{0} + [296553,299566]$ s. 
During the fitting process, the Galactic absorption column density was fixed at $N_\mathrm{H}^\mathrm{Gal}$ (see Section~\ref{sec:optical_obs}). The intrinsic absorption column density ($N_\mathrm{H,X}$) was evaluated at the burst redshift of $z=3.314$. 
The intrinsic column density, $N_\mathrm{H,X}=0.91\times10^{22}\mathrm{\,cm^{-2}}$, derived from the first EP/FXT spectrum was then fixed in the X-ray spectral analysis of the subsequent follow-up observations, as reported in Table~\ref{tab:HE-data}. 
The photon index ($\Gamma$) and the normalization were left free to vary.
The spectral analysis results of the EP/FXT data are shown in Figures~\ref{fig:FXT1}, and \ref{fig:FXT2}--\ref{fig:FXT34}, and are reported in Table~\ref{tab:HE-data}. We also fitted all EP/FXT data using a broken power-law model, allowing the break energy to lie within the X-ray band. The photon index below the break energy cannot be well constrained, while the photon index above the break energy is consistent with that obtained from the single power-law model. Moreover, the broken power-law model does not provide a statistically significant improvement in the goodness of fit compared with the single power-law model. Therefore, the results derived from the single power-law model are retained.




The spectral analysis of the EP/WXT data yields a photon index of $\Gamma = 1.81_{-0.46}^{+0.33}$ and an intrinsic column density of $N_{\mathrm{H,X}} = (15.05_{-6.53}^{+7.02})\times10^{22}\,\mathrm{cm^{-2}}$ (see Table~\ref{tab:HE-data}). 
However, when the intrinsic column density is fixed to the value obtained from the first EP/FXT observation ($0.91\times10^{22}\mathrm{\,cm^{-2}}$), the photon index becomes significantly harder, with $\Gamma = 0.95_{-0.32}^{+0.34}$. The posterior diagnostics for the above two cases are indistinguishable within the uncertainties (see Table~\ref{tab:HE-data}). Moreover, based on the joint spectral analysis of the EP/WXT and GECAM data (with $N_{\mathrm{H,X}}$ either fixed or left free to vary), as well as from an independent spectral analysis of the GECAM data, the photon index derived from a cutoff power-law model lies in a narrow range from $ 0.94$ to $1.03$ (see Table~\ref{tab:HE-data}). Therefore, these results suggest that the photon index is likely close to 0.94 rather than 1.81.

The X-ray flux densities at 3 keV were then derived for the afterglow modeling. The X-ray light curve is displayed in Figure~\ref{fig:EP_LC}.
The photon index derived from the EP/WXT detection is $0.94^{+0.05}_{-0.06}$, while those obtained from the four EP/FXT follow-up observations are $2.05 \pm 0.04$, $2.15\pm0.08$, $2.08_{-0.17}^{+0.18}$ and $1.63_{-0.45}^{+0.50}$, respectively (see Table~\ref{tab:HE-data}). Therefore, the X-ray spectra exhibit a clear softening trend over time, from the prompt phase to the afterglow phase (Figure~\ref{fig:EP_LC}).

\subsection{Empirical analysis of afterglow light curves}\label{sec:empirical_analysis}
The afterglow flux density can generally be characterized by a power-law dependence on both time ($t$) and frequency ($\nu$),
\begin{equation}\label{eq:flux_density}
F_{\nu} \propto t^{-\alpha} \cdot \nu^{-\beta},
\end{equation}
where $\alpha$ is the temporal decay index that describes the temporal decay of the afterglow flux density, and $\beta$ is the spectral index that describes the dependence of the afterglow flux density on frequency. The optical photometric measurements presented in Section~\ref{sec:optical_obs} were converted into flux densities at the effective wavelength of each filter by applying the standard magnitude-flux calibration. To account for Galactic foreground extinction, the optical data were corrected using the reddening value $E_{B-V}^\mathrm{Gal}$ (as described in Section~\ref{sec:optical_obs}) and the wavelength-dependent extinction coefficient, which was derived by interpolating the empirical extinction curves of \citet{2011ApJ...737..103S}. 
The optical and X-ray afterglows were fitted jointly using Equation~\ref{eq:flux_density}, yielding a temporal decay index of $\alpha_\mathrm{ox}=1.01 \pm 0.02$ and an optical-to-X-ray spectral index of $\beta_\mathrm{ox}=0.66 \pm 0.01$.


We further examined and compared the optical and X-ray afterglows separately by fitting each with a single power-law decay model, $F_{\nu} \propto t^{-\alpha}$, where $\alpha$ denotes the temporal decay index. For the optical afterglow, the observed temporal decay index is $\alpha_\mathrm{o,obs} = 1.02\pm0.02$. For the X-ray afterglow, the observed temporal decay index is $\alpha_\mathrm{x,obs} = 1.12\pm0.02$. 
The temporal decay indices derived from the independent fits to the optical and X-ray afterglows, as well as from the joint fit, consistently indicate a common decay behavior in the optical and X-ray bands, with the temporal decay indices close to $\alpha\sim1$ \citep[e.g.,][]{2022ApJ...940..169D,2025MNRAS.543.2404R}. The temporal decay index of the multiwavelength afterglow is $\alpha_\mathrm{ox}=1.01 \pm 0.02$, consistent with the predictions of the forward shock (FS) model for GRB afterglows \citep[e.g.,][]{1997ApJ...476..232M,1998ApJ...497L..17S} and with the temporal decay slopes observed in previous GRB samples \citep[e.g.,][]{2010ApJ...720.1513K,2011ApJ...734...96K}.
The afterglow light curves show nearly achromatic temporal decay, with no strong evidence of a rising phase, plateau, or jet break. 


Due to the limited wavelength coverage and temporal sampling of the optical photometric observations, the optical spectral index cannot be reliably constrained by fitting the optical photometric data alone. The temporal evolution of the optical spectral index therefore cannot be investigated.
The VLT/X-shooter spectrum, obtained at 2.15 hr after $T_0$, provides well-calibrated coverage from the optical to the near-infrared, allowing a more reliable calculation of the spectral index. We therefore fitted the VLT/X-shooter spectrum with a single power-law model, $F_{\nu} \propto \nu^{-\beta}$, where $\beta$ indicates the spectral index. The optical-to-infrared spectral index derived from the VLT/X-shooter spectrum is $\beta_\mathrm{o,obs} = 0.63\pm0.08$.
The X-ray photon index derived from the first EP/FXT follow-up observation at 1.13 hr after $T_0$ is $\Gamma= 2.05 \pm 0.04$ (Table~\ref{tab:HE-data}), corresponding to an X-ray spectral index of $\beta_\mathrm{x,obs} = \Gamma - 1 = 1.05 \pm 0.04$. The X-ray spectral index is also found to be consistent across the available EP/FXT observations (Figure~\ref{fig:EP_LC}), which is significantly larger than both the optical-to-X-ray spectral index, $\beta_\mathrm{ox}=0.66 \pm 0.01$, and the optical-to-infrared spectral index, $\beta_\mathrm{o,obs} = 0.63\pm0.08$. This difference suggests the presence of a spectral break between the optical and X-ray bands.

We adopted the temporal decay indices in the X-ray band ($\alpha_{\mathrm{x,obs}} \simeq 1.12$) and the optical band ($\alpha_{\mathrm{o,obs}} \simeq 1.02$), as well as the spectral indices in the X-ray band ($\beta_{\mathrm{x,obs}} \simeq 1.05$) and the optical band ($\beta_{\mathrm{o,obs}} \simeq 0.63$), to investigate the consistency of the observed indices with the closure relations predicted by the standard afterglow model \citep[e.g.,][]{1998ApJ...497L..17S,2000ApJ...543...66P,2013NewAR..57..141G,2018pgrb.book.....Z}. 

For an interstellar medium (ISM) slow cooling scenario ($\nu_\mathrm{m}<\nu_\mathrm{o}<\nu_\mathrm{c}<\nu_\mathrm{x}$), the electron energy-distribution indices derived from the temporal decay indices in optical and X-ray bands are $p(\alpha_\mathrm{o})=\frac{4\alpha_\mathrm{o}}{3}+1 \simeq 2.36$ and $p(\alpha_\mathrm{x})=\frac{4\alpha_\mathrm{x}+2}{3} \simeq 2.16$, respectively. 
For a wind-like circumburst medium in the slow-cooling regime ($\nu_\mathrm{m}<\nu_\mathrm{o}<\nu_\mathrm{c}<\nu_\mathrm{x}$), the inferred electron indices are $p(\alpha_\mathrm{o}) = \frac{4\alpha_\mathrm{o} + 1}{3} \simeq 1.69$ and $p(\alpha_\mathrm{x})=\frac{4\alpha_\mathrm{x}+2}{3} \simeq 2.16$, respectively. The optical estimate, $p(\alpha_\mathrm{o}) \simeq 1.69$, falls below the standard afterglow expectation of $p > 2$, indicating tension with a wind-like interpretation. 
However, for the above scenarios, the power-law index of the electron energy distribution derived from the X-ray spectral index is $p(\beta_\mathrm{x})=2\beta_\mathrm{x}\simeq2.1$. This value is broadly consistent with those inferred from the temporal decay indices in the optical and X-ray bands, $p(\alpha_\mathrm{o})\simeq2.36$ and $p(\alpha_\mathrm{x})\simeq2.16$, respectively, further supporting the ISM environment interpretation.

Although observations limited to the first few days after the burst do not strongly discriminate between ISM-like and wind-like environments, the normal decay of the optical light curves ($\alpha_{\mathrm{o,obs}} \simeq 1.02$) may point to an ISM-like medium, where the relativistic jet propagates through a constant-density ambient medium and continuously sweeps up the surrounding medium at a roughly constant rate \citep{2026ApJ..1000...58S}. We adopted the ISM-like case as a relatively reliable foundation for afterglow modeling (see details in Section~\ref{sec:afterglow_modeling}).



\section{Results and discussion}\label{sec:Results_and_Discussions}
\subsection{Classification of EP250226a/GRB 250226A}\label{sec:classification}
Extragalactic fast X-ray transients represent a newly emerging class of short-lived astrophysical events detected primarily in the soft X-ray band. 
Motivated by this growing population, we further investigated the nature and classification of the transient. 
For context, GRBs are commonly classified according to their observed gamma-ray emission durations, with events having $T_{90} < 2$ s classified as short-duration GRBs and those with $T_{90} > 2$ s classified as long-duration GRBs. These empirical categories broadly correspond to two progenitor scenarios: the merger of compact objects (Type I GRBs) and the collapse of massive stars (Type II GRBs), respectively. However, this correspondence is not one-to-one, and the observed duration alone cannot  establish the physical origin of an individual burst. Moreover, the Amati relation, which describes the correlation between the rest-frame spectral peak energy and the isotropic-equivalent prompt gamma-ray energy, provides an additional empirical diagnostic for investigating the classification of GRBs \citep{2006MNRAS.372..233A}.

The classification of the EP transient EP250226a as a long-duration GRB is supported by several lines of observational and population evidence, as summarized below. Observationally, the most immediate support comes from the multiwavelength coherence. The initial EP/WXT detection of a fast X-ray transient coincided temporally and spatially with a dedicated gamma-ray trigger, GRB 250226A, detected by Fermi/GBM, INTEGRAL, and GECAM-B (Figure~\ref{fig:HE-lc}). 
The rest-frame spectral peak energy $(1+z)E_\mathrm{p}$, the isotropic-equivalent energy $E_\mathrm{iso}$, the fluence and the duration $T_{90}$ were used to derive the Amati relation and the $T_{90}$-fluence diagram (see Figure~\ref{fig:diagram}). 
Within the broader context of GRB population studies, GRB 250226A lies in the Type II GRB region in both the Amati diagram and the $T_{90}$-fluence diagram (Figure~\ref{fig:diagram}). The statistical placement of EP250226a within the fast X-ray transient population also supports the long GRB classification. \citet{2025ApJ...993L..37O} demonstrated that the cumulative redshift distribution of EP extragalactic fast X-ray transients, including EP250226a, is statistically indistinguishable from that of canonical long GRBs. This suggests that they may share the same progenitor channel, namely the massive stellar core-collapse model \citep{2025ApJ...993L..37O}. 
Furthermore, previous studies reconstructed the luminosity functions and formation rates of EP extragalactic fast X-ray transients, finding that their formation rate evolves similarly to that of long GRBs \citep{2025arXiv251010267L,2025ApJ...995L..53G}. This similarity provides additional support for classifying EP250226a as a member of the long GRB population.



Previous studies suggested the presence of GRB subclasses, such as X-ray-rich GRBs (XRRs) and X-ray flashes (XRFs), which may constitute a substantial fraction of the fast X-ray transient population. XRFs are generally characterized by low peak energies \citep[$E_\mathrm{p} \lesssim 50\,\mathrm{keV}$;][]{2018ApJ...866...97B,2005ApJ...622..977L,2026A&A...714A..82S}. We therefore assessed whether EP250226a belongs to the XRF, XRR, or classical GRB (C-GRB) category. Given the high peak energy of EP250226a ($E_\mathrm{p} > 100\,\mathrm{keV}$, see Table~\ref{tab:HE-data}), an XRF origin can be excluded. We further performed a quantitative fluence-ratio analysis following the methodology adopted in previous studies \citep[e.g.,][]{2008ApJ...679..570S,2018ApJ...866...97B}. The empirical definitions of XRFs, XRRs, and C-GRBs are given by

{\small
\begin{align*}
           & \rm{S}(25-50\,\rm{keV})/\rm{S}(50-100\,\rm{keV})\leq 0.72\,&(\rm C-GRBs),   \\
    0.72 < & \rm{S}(25-50\,\rm{keV})/\rm{S}(50-100\,\rm{keV})\leq 1.32\,&(\rm XRRs),   \\
    1.32 < & \rm{S}(25-50\,\rm{keV})/\rm{S}(50-100\,\rm{keV})\,&(\rm XRFs).
\end{align*}}

We estimated the fluence ratio using the flux ratio measured over the same time interval, given the lack of significant spectral evolution. For EP250226a, the fluence ratio is $\rm{S}(25-50\,\rm{keV})/\rm{S}(50-100\,\rm{keV})=0.57\pm0.06$, consistent with the criterion for C-GRBs. Furthermore, the observer-frame spectral peak energy of EP250226a was estimated to be $E_\mathrm{p}\simeq278$ keV, which is consistent with the distribution of observer-frame spectral peak energies of C-GRBs, ranging from 64.6 keV to 2602.8 keV, with a mean value of $257.7\pm14.1$ keV \citep{2018ApJ...866...97B}. In addition, EP250226a lies in the region of the Amati diagram occupied by Type II GRBs, as discussed above. Its X-ray and optical afterglow properties are also consistent with the predictions of the FS model commonly used to describe classical GRB afterglows (Section~\ref{sec:empirical_analysis}). Taken together, these prompt emission and afterglow properties support the classification of EP250226a as a C-GRB.


EP observations revealed significant diversity in the early high-energy emission properties of extragalactic fast X-ray transients. 
Some events exhibit only soft X-ray emission without a reported gamma-ray counterpart, such as EP241113a \citep{2026arXiv260326213D}, EP250108a/SN 2025kg \citep[e.g.,][]{2025arXiv250417034L,2025ApJ...988L..13R}, EP250223a (H. Sun et al., in preparation), EP251023a \citep{2026arXiv260624329J}, and EP260321a/SN 2026gzf \citep[e.g.,][]{2026arXiv260610014Y}, whereas others are accompanied by detectable gamma-ray emission, such as EP250416a/GRB 250416C \citep{2026arXiv260421624Z}, EP250404a/GRB 250404A \citep{2025ApJ...989L..39Y}, EP240315a/GRB 240315C \citep{2025NatAs...9..564L,2025NatAs...9.1375L}, EP240801a/XRF 240801B \citep{2025ApJ...988L..34J} and EP240219a/GRB 240219A \citep{2024ApJ...975L..27Y}. 
Further examples of EP events are discussed by \citet{2026arXiv261028734E}.
Among those with gamma-ray counterparts, the relative timing between the gamma-ray and soft X-ray signals also varies substantially, with the gamma-ray emission occurring before, after, or contemporaneously with the soft X-ray emission. 
In addition, EP-discovered fast X-ray transients typically exhibit durations on the order of a few hundred seconds \citep{2026arXiv260627048V,2025Galax..13...62W}, which can be longer than those of long GRBs (typically tens of seconds).
In the case of EP250226a/GRB 250226A, the duration of the soft X-ray emission ($\sim250$ s in the EP/WXT light curve) is longer than that of the gamma-ray emission ($\sim61$ s in the Fermi/GBM light curve). Although weak soft X-ray emission below the detection threshold cannot be excluded, the first peak of the soft X-ray light curve is broadly coincident with the gamma-ray emission (Figure~\ref{fig:HE-lc}). The subsequent soft X-ray emission extends beyond the gamma-ray duration. No strong evidence is found for significant soft X-ray activity preceding the gamma-ray trigger. This behavior differs from that observed in some previously reported EP-detected transients, such as EP240315a/GRB 240315C, for which soft X-ray emission was detected prior to the onset of the gamma-ray emission \citep[e.g.,][]{2025NatAs...9..564L}. Taken altogether, the temporal behaviors mentioned above suggest that the prompt-emission properties of extragalactic fast X-ray transients are more diverse than expected from a simple extension of the classical GRB population. The growing sample of EP-detected transients points to a broader phenomenology of relativistic high-energy transients and may probe a wider range of physical conditions and emission mechanisms than those presented by classical GRBs alone \citep{2026arXiv260714317C}.

\subsection{Extinction of the host galaxy}\label{sec:extinction}
%

The VLT/X-shooter spectrum was obtained at a mid-time of 2.15 hr post-trigger. The VLT/X-shooter spectroscopic data were converted into an \texttt{XSPEC}-compatible format for the optical-to-X-ray spectral energy distribution (SED) analysis using the \texttt{ftflx2xsp} task in HEASoft (version 6.36), following the methodology of \citet{2018MNRAS.479.1542Z}. To enable a contemporaneous comparison, the X-ray spectrum at the same mid-time was simulated. In this simulation, we adopted a temporal decay index of 1.01 to determine the normalization of a power-law model (see Section~\ref{sec:empirical_analysis}) and fixed the remaining spectral parameters to match those of the first EP/FXT follow-up observation (see Table~\ref{tab:HE-data}). The X-ray spectrum was generated with the \texttt{fakeit} task in HEASoft (version 6.36) using the associated EP/FXT response matrix file and ancillary response file. The observed VLT/X-shooter spectrum and the simulated X-ray spectrum at 2.15 hr post-trigger were combined for the SED analysis (Figure~\ref{fig:SED}).

The SED was analyzed using \texttt{XSPEC} \citep[version 12.14.1;][]{1996ASPC..101...17A,1999ascl.soft10005A}, following the methodology described by \citet{2026ApJ...997..101L}.
The unattenuated broadband optical-to-X-ray afterglow spectrum was modeled using either a power-law or a broken power-law function.
To account for extinction and absorption in both the Galactic and the host galaxy, the model included two dust and gas components, described using the \texttt{XSPEC} models \texttt{TBabs}, \texttt{zTBabs}, and \texttt{zdust}. 
For the host galaxy, the extinction law of the Small Magellanic Cloud (SMC) was adopted to model the optical extinction of the host galaxy. 
We considered two cases for the SED modeling. In the first case, for either the absorbed power-law or the absorbed broken power-law model, all model parameters were allowed to vary freely. In the second case, the X-ray photon index in the absorbed broken power-law model was fixed at $\Gamma_\mathrm{x}=2.1$, as determined from the EP/FXT observations (Table~\ref{tab:HE-data}). This X-ray photon index corresponds to an X-ray spectral index of $\beta_\mathrm{x}=\Gamma_\mathrm{x}-1=1.1$. The spectral change between the optical and X-ray bands was fixed at 0.5, as predicted by the standard afterglow model, assuming that the cooling break lies between the optical and X-ray bands \citep[e.g.,][]{1998ApJ...497L..17S}. Therefore, the optical spectral index was set to $\beta_\mathrm{o}=\beta_\mathrm{x}-0.5=0.6$.
The fitting results show that the absorbed single power-law model is not sufficiently constrained by the data. The absorbed broken power-law model with all parameters free provides a statistically better fit to the broadband optical-to-X-ray afterglow SED than the model with the spectral-index difference fixed at 0.5. The model with free photon indices yields a fit statistic of 601.66 for 431 degrees of freedom, compared with 618.99 for 433 degrees of freedom for the model with fixed photon indices. We therefore adopt the results from the model with free photon indices, which yields an X-ray photon index of $2.24\pm0.07$ and an optical photon index of $1.33^{+0.09}_{-0.10}$. The inferred host galaxy color excess is $E_{B-V}=0.06\pm0.01\mathrm{\,mag}$, indicating negligible host galaxy extinction (Figure~\ref{fig:SED}).

\subsection{Multiwavelength afterglow modeling}\label{sec:afterglow_modeling}



Given the moderate redshift of EP250226a ($z=3.314$), the Lyman-$\alpha$ absorption from the host galaxy occurs at an observed wavelength of $\sim5230\,\mathrm{\AA}$ in the observer frame (Figure~\ref{fig:VLTspec}). Therefore, we used the flux densities from filters whose effective wavelengths exceed 5230 \AA, along with the flux density at 3 keV, to model the afterglow. 
The optical filters retained for the afterglow modeling are $r$, $r'$, Mephisto-r, $R$, $i$, $i'$, VT-R, $I$, and $z$ bands.
The optical flux densities were corrected for the Galactic extinction, while the extinction in the host galaxy was neglected (Section~\ref{sec:extinction}).
We modeled the multiwavelength afterglow as synchrotron emission from a FS assuming a top-hat jet propagating through a uniform ISM, using the Python package \texttt{VegasAfterglow} \citep{2025arXiv250710829W}.
The afterglow model parameters include the isotropic-equivalent kinetic energy of the jet ($E_\mathrm{k,iso}$), the initial bulk Lorentz factor ($\Gamma_0$), the half-opening angle of the jet core ($\theta_\mathrm{c}$), the viewing angle between the jet axis and the line of sight ($\theta_\mathrm{v}$), the number density of the uniform ISM ($n_\mathrm{ISM}$), the power-law index of the electron energy distribution ($p_\mathrm{FS}$), the fraction of shock energy carried by relativistic electrons ($\epsilon_\mathrm{e,FS}$), the fraction carried by magnetic fields ($\epsilon_\mathrm{B,FS}$), and the fraction of accelerated electrons ($\chi$).
The corresponding prior ranges of the model parameters are described in Table~\ref{tab:VegasAfterglow}. 
We fixed the fraction of accelerated electrons $\chi = 1$ in the afterglow modeling, implying efficient particle acceleration. Since no jet break is evident in the current dataset, the jet opening angle is weakly constrained. The model parameter space was explored via Bayesian inference, implemented with the Python packages \texttt{Bilby} \citep{2019ApJS..241...27A} and \texttt{PyMultiNest} \citep{2014A&A...564A.125B}. 

All parameters were initially allowed to vary freely in our afterglow modeling, resulting in substantial degeneracies among the inferred parameters. The derived posterior distributions favor a family of solutions characterized by relatively high isotropic-equivalent kinetic energies, bulk Lorentz factors, and ambient densities in a uniform ISM, extending beyond the commonly accepted parameter range for GRB afterglows (Figures~\ref{fig:afterglow_model1} and Posterior-1 in Table~\ref{tab:VegasAfterglow}). 
To reduce these parameter degeneracies, we attempted to fix either the initial bulk Lorentz factor or the number density of the uniform ISM, and to fix both parameters simultaneously. 
Firstly, we estimated the initial bulk Lorentz factor ($\Gamma_{0}$) based on the empirical relation between the isotropic-equivalent prompt gamma-ray energy ($E_\mathrm{iso}$) and the initial Lorentz factor, as reported by \citet{2010ApJ...725.2209L}, which can be written as
\begin{equation}\label{eq:Eiso-gamma}
    \Gamma_{0}\simeq182{\left(\frac{E_{\mathrm{iso}}}{10^{52}\mathrm{~erg}}\right)}^{0.25\pm0.03}.
\end{equation}
Using this relation and $E_\mathrm{iso} \simeq 7.32 \times 10^{53}$ erg (Section~\ref{sec:HE_spectra}), we obtained $\Gamma_0 \approx 500$, which was then adopted as a fixed parameter in the subsequent afterglow modeling. The inferred initial Lorentz factor is representative of the value expected from the empirical relation, suggesting that the jet is strongly relativistic and may be associated with an efficient central engine. The derived posterior distribution is shown in Figure~\ref{fig:afterglow_model2} and listed as Posterior-2 in Table~\ref{tab:VegasAfterglow}. 
Secondly, previous broadband afterglow studies inferred circumburst densities spanning approximately 0.1 $\mathrm{cm^{-3}}$ to 100 $\mathrm{cm^{-3}}$, with a typical value of around $10\mathrm{\,cm^{-3}}$ \citep[e.g.,][]{2002ApJ...577..155Y,2003ApJ...590..992F,2003ApJ...594..674B,2018ApJ...866..162G,2022MNRAS.511.2848A}. We therefore fixed the number density of the uniform ISM at $n_{\mathrm{ISM}} = 10\mathrm{\,cm^{-3}}$. The derived posterior distribution is depicted in Figure~\ref{fig:afterglow_model3} and reported as Posterior-3 in Table~\ref{tab:VegasAfterglow}. 
Thirdly, we also tried to fix both parameters simultaneously at the above values. The derived posterior distribution is displayed in Figure~\ref{fig:afterglow_model4} and presented as Posterior-4 in Table~\ref{tab:VegasAfterglow}. 

The derived Posterior-1, Posterior-2, and Posterior-3 yield relatively higher Bayesian evidence values and are therefore statistically favored (see $\ln\mathcal{Z}$ in Table~\ref{tab:VegasAfterglow}). Nevertheless, the inferred parameters in these solutions tend toward the extreme end of the parameter space for GRB afterglows. In the Posteriors 1--3, the inferred values of $\Gamma_0$ and/or $n_\mathrm{ISM}$ lie beyond the commonly accepted parameter ranges for GRB afterglows. Specifically, the derived Posterior-1 and Posterior-2 require much higher densities of $\sim10^{3.8}\mathrm{\,cm^{-3}}$ and $\sim10^{4.96}\mathrm{\,cm^{-3}}$, respectively. Likewise, the Lorentz factor in Posterior-3 ($\Gamma_0 \sim 10^{3.66}$) is an unusually high value compared with the $\Gamma_0 \sim 10^{2.70}$ inferred from Equation~\ref{eq:Eiso-gamma} and adopted in Posterior-4. Given the physically implausible posterior values inferred for some parameters when all parameters were allowed to vary freely (or when one parameter was fixed), we further adopted Gaussian priors for the initial bulk Lorentz factor and the number density of the uniform ISM. The Gaussian priors were centered on the estimated values of 500 and 10, respectively, with appropriate standard deviations of 150 and 5, yielding $\Gamma_0 \sim \mathcal{N}(500, 150^2)$ and $n_{\mathrm{ISM}} \sim \mathcal{N}(10, 5^2)$. The derived posterior distribution is shown in Figure~\ref{fig:afterglow_model5} and reported as Posterior-5 in Table~\ref{tab:VegasAfterglow1}. 
In Posterior-4, $\Gamma_0$ was fixed at 500, whereas in Posterior-5 it was allowed to vary freely, yielding $\Gamma_0=857.48_{-82.44}^{+74.24}$. Similarly, the ISM number density was fixed at $n_{\mathrm{ISM}}=10\mathrm{\,cm^{-3}}$ in Posterior-4, while Posterior-5 yielded $n_{\mathrm{ISM}}=13.93_{-3.46}^{+3.05}\mathrm{\,cm^{-3}}$. The remaining posterior parameters show no significant differences between the two posterior distributions. The Bayesian evidence values for Posterior-4 and Posterior-5 are $\ln\mathcal{Z} = -293.64\pm0.16$ and $\ln\mathcal{Z} = -279.01\pm0.16$, respectively, indicating a strong preference for Posterior-5 over Posterior-4.
For Posteriors 1--5, the power-law index of the electron energy distribution is well constrained to be $p_\mathrm{FS} \sim 2.2$, which is broadly consistent with the values commonly reported for most GRBs \citep{2010ApJ...716L.135C,2022MNRAS.511.2848A,2025ApJ...994L..46C}. The afterglow modeling supports a scenario in which the afterglow emission is dominated by the FS, in good agreement with the observed light curve and its smooth, steadily decaying pattern. Furthermore, the fraction of shock energy in magnetic fields is much smaller than that in relativistic electrons, with $\epsilon_\mathrm{B} \ll \epsilon_\mathrm{e}$ (Table~\ref{tab:VegasAfterglow} and Table~\ref{tab:VegasAfterglow1}). 

We also tested a wind-like environment in the afterglow modeling, considering different jet structures, including top-hat, Gaussian, and power-law profiles. All model parameters were allowed to vary freely, with uniform priors adopted for each parameter. The wind-like model provided a significantly poorer fit to the multiwavelength afterglow data compared with the ISM-like model. These additional tests allow us to assess alternative jet geometries and further support our preference for a uniform ISM environment as a more suitable description of the circumburst medium.


We adopt Posterior-5 as the fiducial model for the subsequent analysis based on the above discussion.
We emphasize that this inference depends on the adopted model and parameter choices. Therefore, the derived parameters should be interpreted within the framework of our modeling assumptions.
The multiwavelength afterglow light curves and the best-fitting model are shown in Figure~\ref{fig:afterglow_model4}.
The best-fitting isotropic-equivalent kinetic energy of the jet is $\log (E_\mathrm{k,iso}/\mathrm{erg}) = 55.61_{-0.25}^{+0.29}$, 
with a jet-core half-opening angle of $\log \theta_\mathrm{c} = -1.41\pm0.05$ (i.e., $\theta_\mathrm{c}\simeq1^\circ.99-2^\circ.50$). 
The viewing angle between the jet axis and the line of sight is $\log \theta_\mathrm{v}=-1.64\pm0.06$ (i.e., $\theta_\mathrm{v}\simeq1^\circ.14-1^\circ.51$). 
The initial bulk Lorentz factor is constrained to be $\Gamma_0=857.48_{-82.44}^{+74.24}$.
The uniform ISM number density is $n_\mathrm{ISM}=13.93_{-3.46}^{+3.05}\mathrm{\,cm^{-3}}$. 
The electron energy distribution is characterized by a power-law index of $p_\mathrm{FS} = 2.23\pm0.01$. 
The fractions of shock energy in relativistic electrons and magnetic field are $\log \epsilon_\mathrm{e,FS} = -1.96_{-0.23}^{+0.20}$ and $\log \epsilon_\mathrm{B,FS} = -6.52_{-0.27}^{+0.23}$, respectively.

The afterglow modeling suggests a highly collimated narrow jet observed nearly on-axis \citep{2010ApJ...711..641C,2020ApJ...900..112Z}, as indicated by the inferred values of $\theta_\mathrm{c}$ and $\theta_\mathrm{v}$.
We examined the energy release of the burst. A relation between the jet-corrected gamma-ray energy ($E_\gamma$) and the isotropic-equivalent gamma-ray energy ($E_\mathrm{iso}$) was reported for pre-Swift GRBs, i.e., $E_\gamma = (1-\cos{\theta_\mathrm{c}})E_\mathrm{iso}$ \citep{2001ApJ...562L..55F,2003ApJ...594..674B}.
In the small-angle approximation for $\theta_\mathrm{c}$, the jet-corrected gamma-ray energy is given by $E_\gamma \simeq E_\mathrm{iso}\theta_\mathrm{c}^2/2$.
Adopting the jet-core half-opening angle ($\theta_\mathrm{c}\simeq0.039$ rad) and the isotropic-equivalent gamma-ray energy ($E_\mathrm{iso}\simeq7.32\times10^{53}$ erg), the jet-corrected gamma-ray energy of GRB 250226A was estimated to be $E_\gamma \simeq5.54\times10^{50}\mathrm{\,erg}$, which is consistent with a roughly constant energy reservoir found in long GRBs \citep[$\sim10
^{50}-10^{52}$ erg;][]{2001ApJ...562L..55F,2003ApJ...594..674B,2010ApJ...711..641C,2015PhR...561....1K,2022MNRAS.511.2848A}.

We further investigated the gamma-ray radiative efficiency of the burst. The radiative efficiency is defined as $\eta_{\gamma}=\frac{E_\mathrm{iso}}{E_\mathrm{iso}+E_\mathrm{k,iso}}$ \citep{2004ApJ...613..477L,2018ApJS..236...26L}. Using the isotropic-equivalent gamma-ray energy ($E_\mathrm{iso}\simeq7.32\times10^{53}$ erg) and the isotropic-equivalent kinetic energy ($E_\mathrm{k,iso}\simeq4.07\times10^{55}$ erg), we calculated the gamma-ray radiative efficiency to be $\eta_{\gamma}\simeq 1.77\%$. 
We note that some GRBs have large values of $E_\mathrm{k,iso}$, and a statistical comparison is shown in Figure 9 of \citet{2026ApJ...997..246G}.
Although the radiative efficiency is sensitive to the assumed circumburst environment and the estimated microphysical parameters, the inferred gamma-ray efficiency of GRB 250226A remains low and is comparable to that of GRB 230204B \citep[$\eta_\gamma\sim4\%$;][]{2026ApJ...997..246G,2026ApJ..1000...58S}. 
Such low radiative efficiency indicates that the released energy of the jet is predominantly retained in kinetic form, with only a small fraction being converted into prompt gamma-ray emission.

We noted that the fraction of shock energy in magnetic field ($\log \epsilon_\mathrm{B,FS}$) typically ranges from $-6$ to $-2$ \citep{2015PhR...561....1K}.
In our afterglow modeling, the inferred magnetic energy fraction is also relatively low ($\log \epsilon_\mathrm{B,FS} \simeq -6.52$), placing GRB 250226A among this class of low-$\epsilon_\mathrm{B}$ bursts. Such a low value of $\epsilon_\mathrm{B}$ suggests that the post-shock magnetic field is likely dominated by shock compression of the pre-existing seed magnetic field in the ambient ISM \citep{2009MNRAS.400L..75K,2010MNRAS.409..226K,2014ApJ...785...29S,2015ApJS..219....9W}. Several GRBs were reported to exhibit particularly low magnetic energy fractions, including 
GRB 110731A \citep{2015ApJ...804..105F}, 
GRB 130427A \citep{2016ApJ...818..190F}, 
GRB 190530A \citep{2022MNRAS.511.1694G}, 
GRB 230204B \citep{2026ApJ...997..246G,2026ApJ..1000...58S},
and GRB 241030A \citep{2026arXiv260318956D}. 
In addition, synchrotron self-Compton (SSC) modeling may significantly improve the constraints on the model parameters. However, due to the lack of observational data in the GeV band, we do not attempt SSC modeling in this work.

\section{Summary} \label{sec:Summary}
Fast X-ray transients constitute an emerging population of short-lived astrophysical phenomena observed in the soft X-ray band. 
EP observations have revealed a diverse population of extragalactic fast X-ray transients with a wide range of high-energy emission properties. These events can exhibit soft X-ray emission either with or without gamma-ray counterparts, and the relative timing between X-ray and gamma-ray signals varies significantly.
The fast X-ray transient EP250226a was identified as gamma-ray burst GRB 250226A. EP250226a/GRB 250226A was observed by multiple facilities, obtaining a rich set of observational data (Section~\ref{sec:Observations_and_Data_Reduction}).

We performed a comprehensive multiwavelength study of the prompt and afterglow emission of EP250226a/GRB 250226A, integrating observations across the gamma-ray, X-ray, and optical bands. In addition, the spectroscopic redshift of the transient was determined to be $z = 3.314$ (Section~\ref{sec:Spectroscopic_Observations}). The transient is empirically classified as a C-GRB based on the fluence ratio in the high-energy spectrum (Section~\ref{sec:classification}). 
The isotropic-equivalent gamma-ray energy is measured to be $E_\mathrm{iso}\simeq7.32\times10^{53}$ erg.
The prompt emission observations indicate that the burst is a typical Type II GRB, as shown by the Amati relation (Figure~\ref{fig:diagram}).

Achromatic behavior is observed in both the X-ray and optical light curves. The light curves of the multiwavelength afterglow can be fitted with a temporal decay index of $\alpha_\mathrm{ox}=1.01 \pm 0.02$, an X-ray spectral index of $\beta_\mathrm{x,obs} = 1.05 \pm 0.04$, and an optical spectral index of $\beta_\mathrm{o,obs} = 0.63\pm0.08$. Hence, a cooling break frequency occurs between the optical and X-ray bands. The temporal decay index of the multiwavelength light curves is close to 1, indicating that the afterglow emission is dominated by a FS (Section~\ref{sec:empirical_analysis}). The optical-to-X-ray SED suggests that the host galaxy extinction is negligible (Section~\ref{sec:extinction}).

The multiwavelength afterglow can be described by a top-hat jet propagating through a homogeneous ISM. The power-law index of the electron energy distribution is constrained to be approximately 2.2. Afterglow modeling suggested a highly collimated, narrow jet viewed nearly on-axis. A large isotropic-equivalent kinetic energy was derived ($E_\mathrm{k,iso}\simeq4.07\times10^{55}$ erg), yielding a low gamma-ray radiative efficiency ($\eta_{\gamma}\simeq 1.77\%$). The fraction of shock energy in the magnetic field is low ($\log \epsilon_\mathrm{B,FS} \simeq -6.52$), suggesting that the post-shock magnetic field is primarily governed by shock-compression of the pre-existing seed field in the surrounding ISM (Section~\ref{sec:afterglow_modeling}).

\begin{figure*}[htbp]
  \centering
  \includegraphics[width=\linewidth]{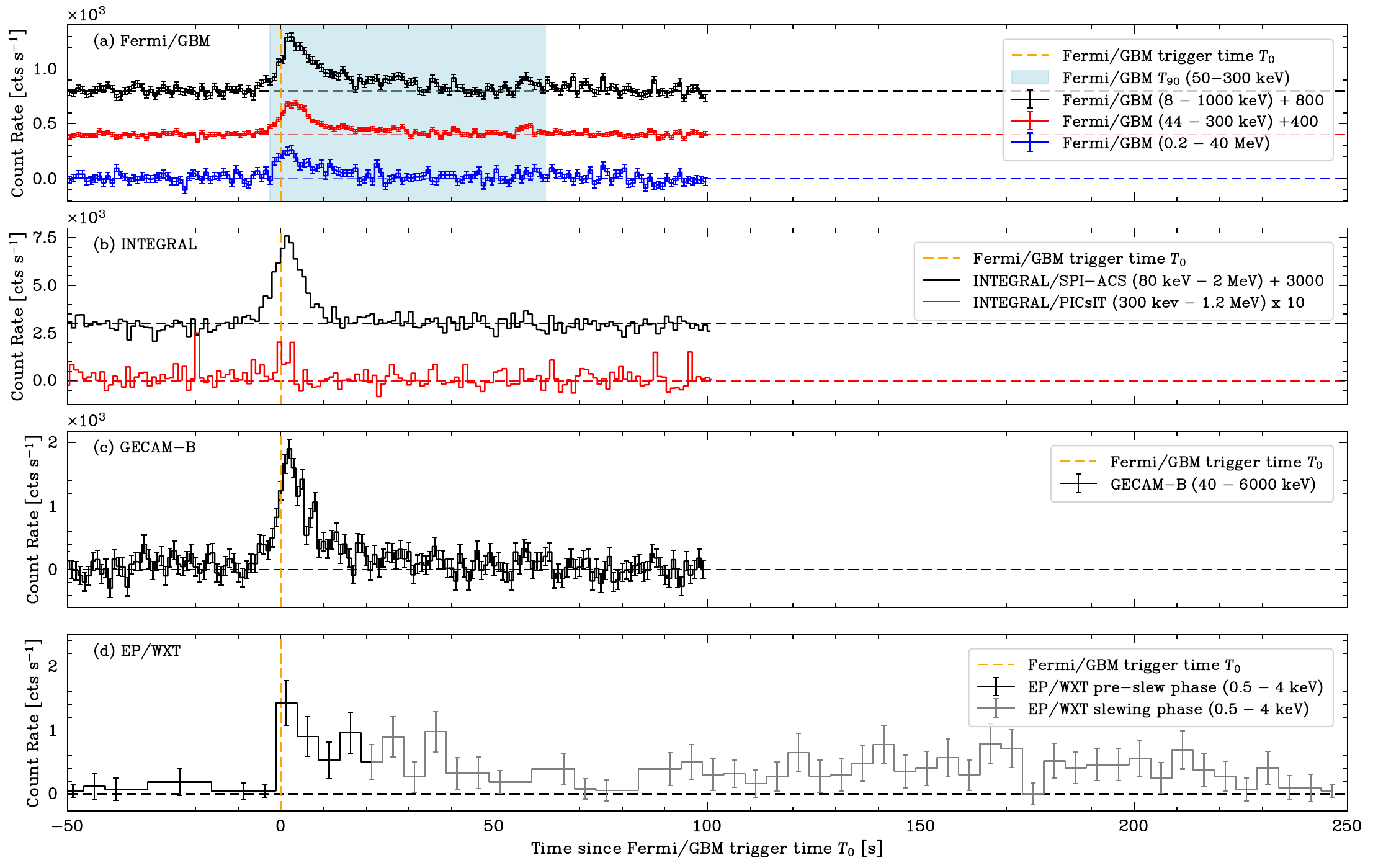}
  \caption{
  High-energy light curves of the transient observed by several satellites. Panel (a) presents the Fermi/GBM light curve. Panel (b) displays the INTEGRAL light curve. Panel (c) depicts the GECAM-B light curve. Panel (d) illustrates the EP/WXT light curve. The vertical orange dashed lines mark $T_0$, while the horizontal dashed lines indicate the background levels of the color-coded light curves. The light curves are binned at a time resolution of 1 s, except for the EP/WXT light curve, which is binned at a time resolution of 5 s.}
  \label{fig:HE-lc}
\end{figure*}

\begin{figure*}[ht!]
    \centering

    \begin{subfigure}{0.3\textwidth}
        \centering
        \includegraphics[width=\linewidth]{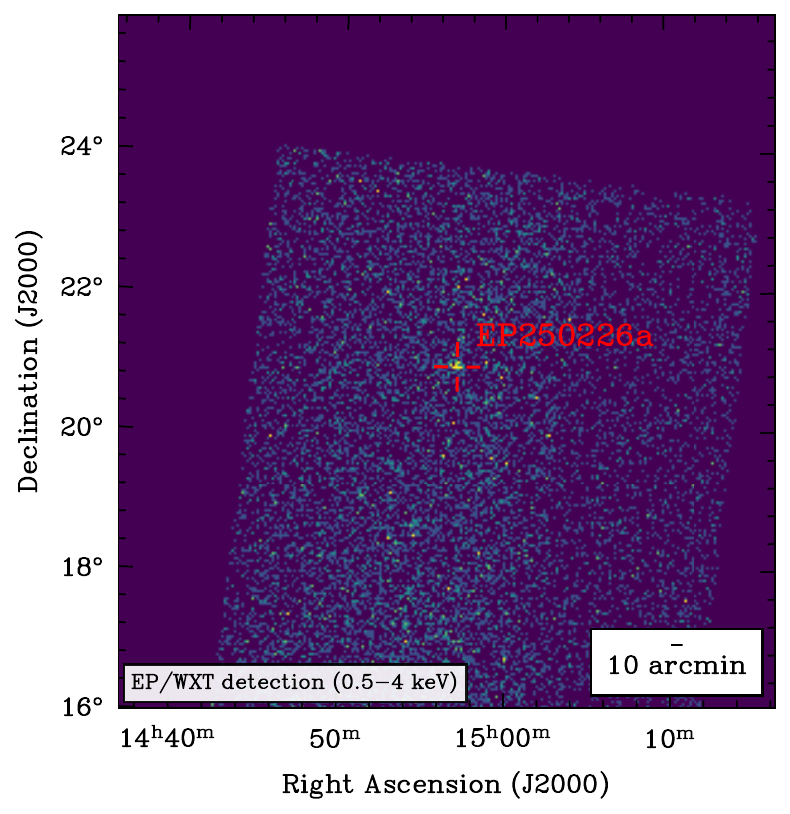}
        \caption{EP/WXT X-ray detection}
    \end{subfigure}
    \hfill
    \begin{subfigure}{0.343\textwidth}
        \centering
        \includegraphics[width=\linewidth]{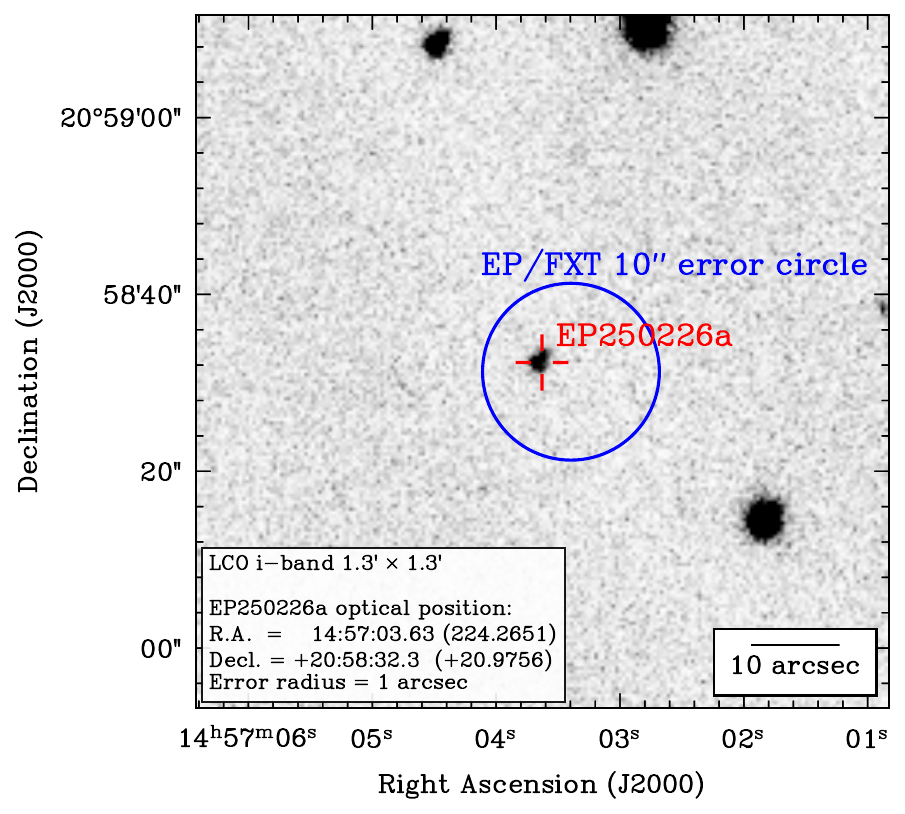}
        \caption{LCO optical afterglow detection}
    \end{subfigure}
    \hfill
    \begin{subfigure}{0.343\textwidth}
        \centering
        \includegraphics[width=\linewidth]{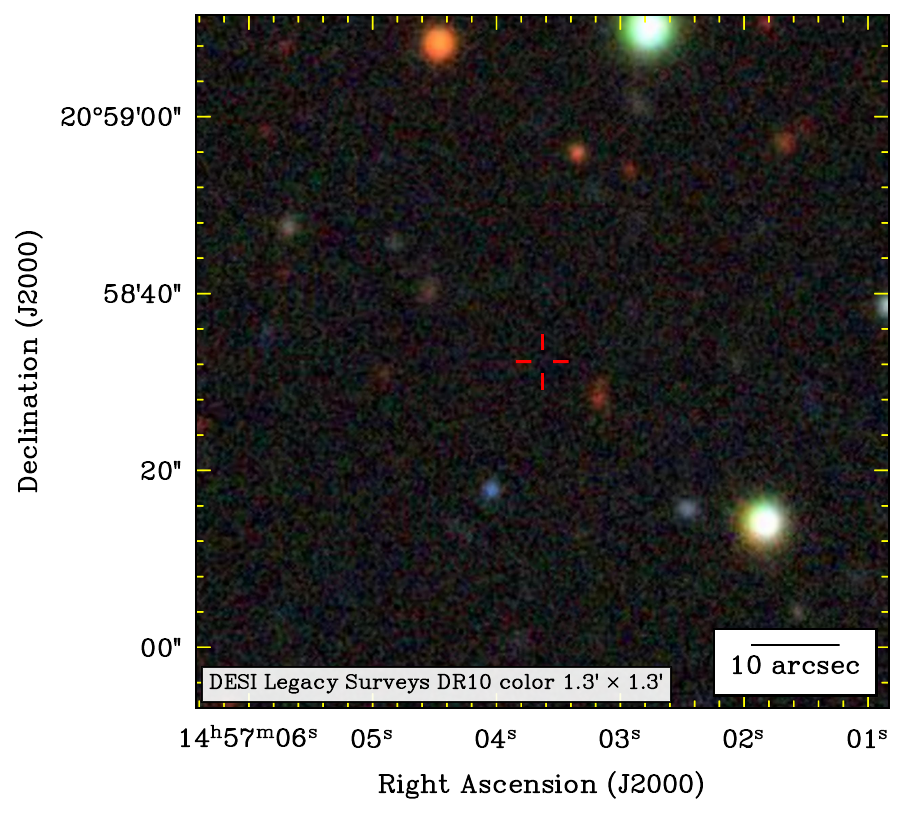}
        \caption{Archival image}
    \end{subfigure}

    \caption{Sky field of EP250226a. 
    Panel (a): EP/WXT detection image in the 0.5--4 keV band. 
    Panel (b): Photometric image observed by LCO with a mid-time of $\sim2.85$ hr post-trigger. 
    The source magnitude was measured to be $i = 19.33 \pm 0.04$ mag in the AB system. 
    The blue circle indicates the EP250226a localization calculated from the ground-analyzed EP/FXT data. 
    Panel (c): Archival image combined from the DESI Legacy Surveys DR10 $grz$ bands 
    \citep{2019AJ....157..168D,2022ASPC..532....7B}. 
    The localization of EP250226a is marked by four red lines forming a cross.}
    \label{fig:fc}
\end{figure*}

\begin{figure*}[htbp]
    \centering
    
    \begin{subfigure}{0.49\linewidth}
        \centering
        \includegraphics[width=\linewidth]{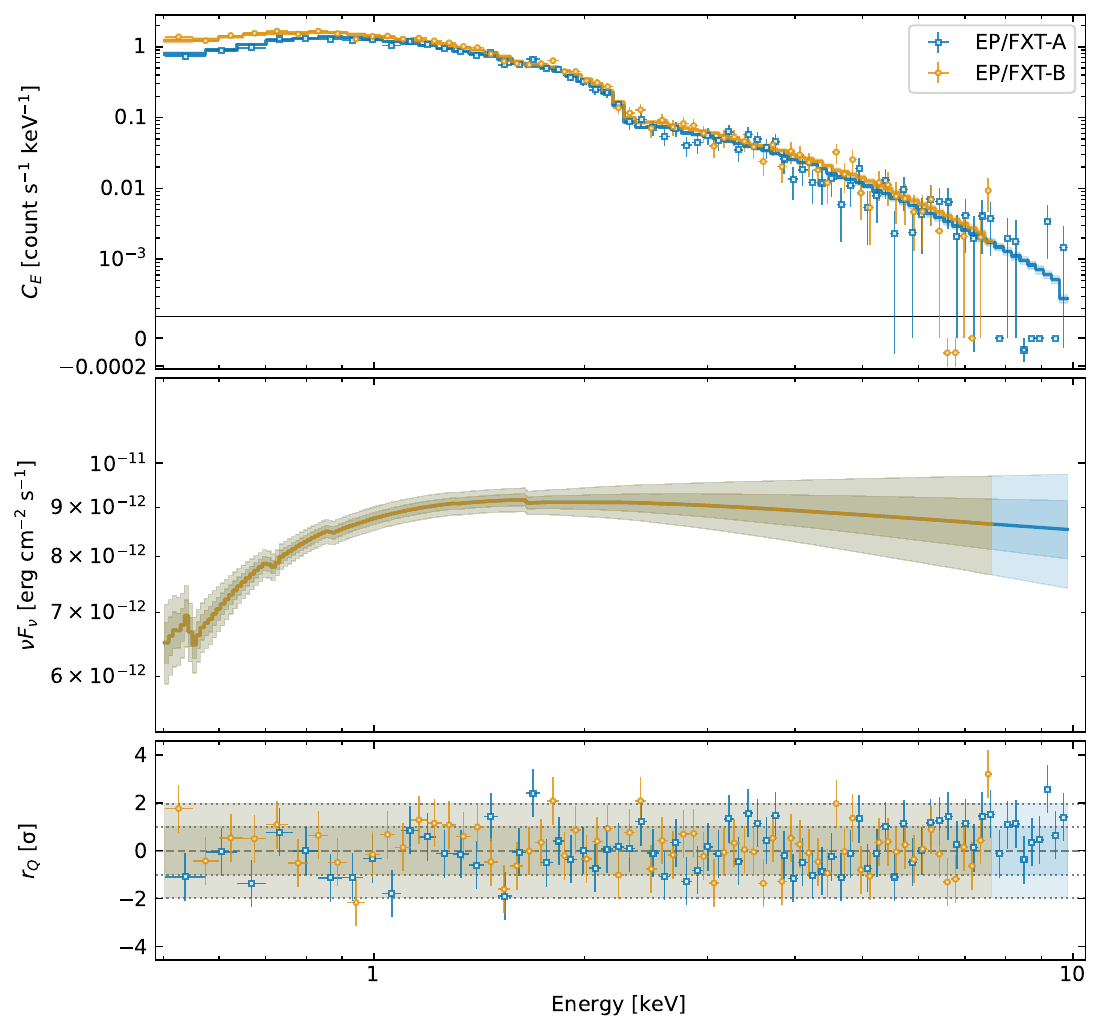}
        \caption{EP/FXT spectrum in $T_{0} + [2725,5440]$ s, power-law model, and residuals.}
    \end{subfigure}
    \hfill
    \begin{subfigure}{0.49\linewidth}
        \centering
        \includegraphics[width=\linewidth]{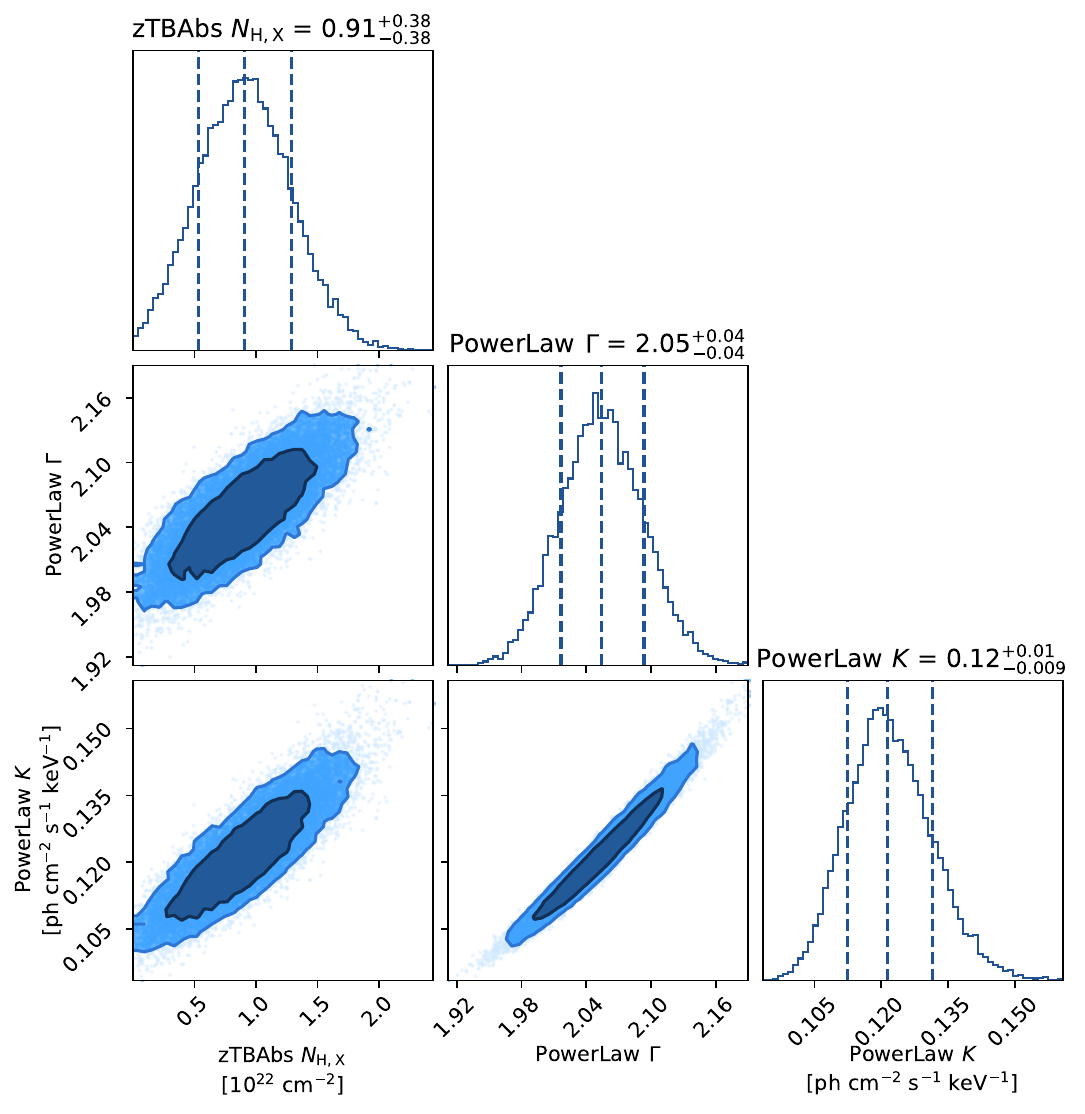}
        \caption{Posterior distributions of model parameters.}
    \end{subfigure}
    \caption{Time-integrated spectral analysis of the X-ray afterglow. Panel (a): The EP/FXT spectrum observed during $T_{0} + [2725,5440]$ s. The X-ray spectrum is fitted with an absorbed power-law model. Panel (b): Posterior distributions of the parameters of the absorbed power-law model.}
    \label{fig:FXT1}
\end{figure*}

\begin{figure}[htbp]
  \centering
  \includegraphics[width=\linewidth]{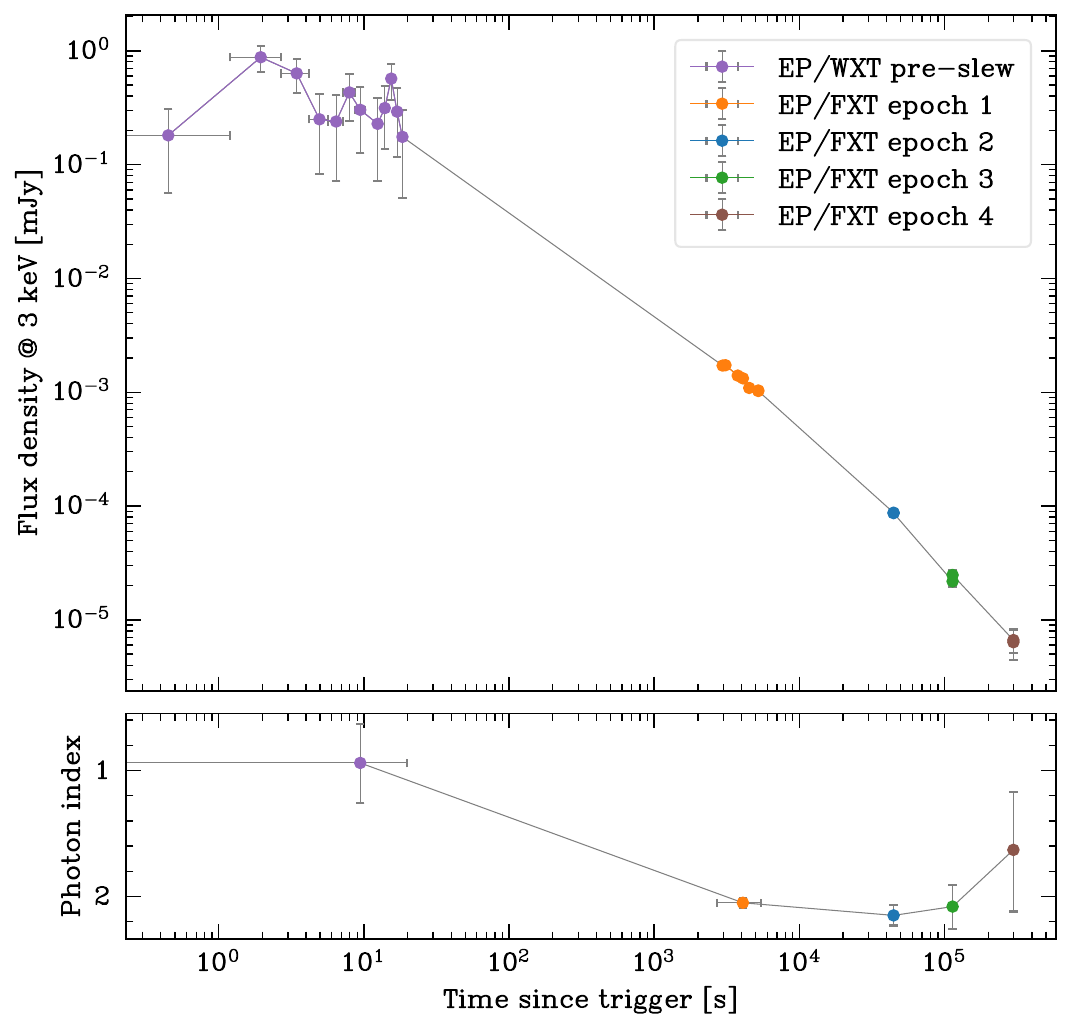}
  \caption{Upper panel: X-ray flux density light curve at 3 keV derived from EP/WXT and EP/FXT observations. Different colors denote the EP/WXT detection and the four EP/FXT follow-up epochs. Lower panel: Temporal evolution of photon index of X-ray afterglow.}
  \label{fig:EP_LC}
\end{figure}

\begin{figure*}[htbp]
    \begin{subfigure}{0.49\linewidth}
        \centering
        \includegraphics[width=\linewidth]{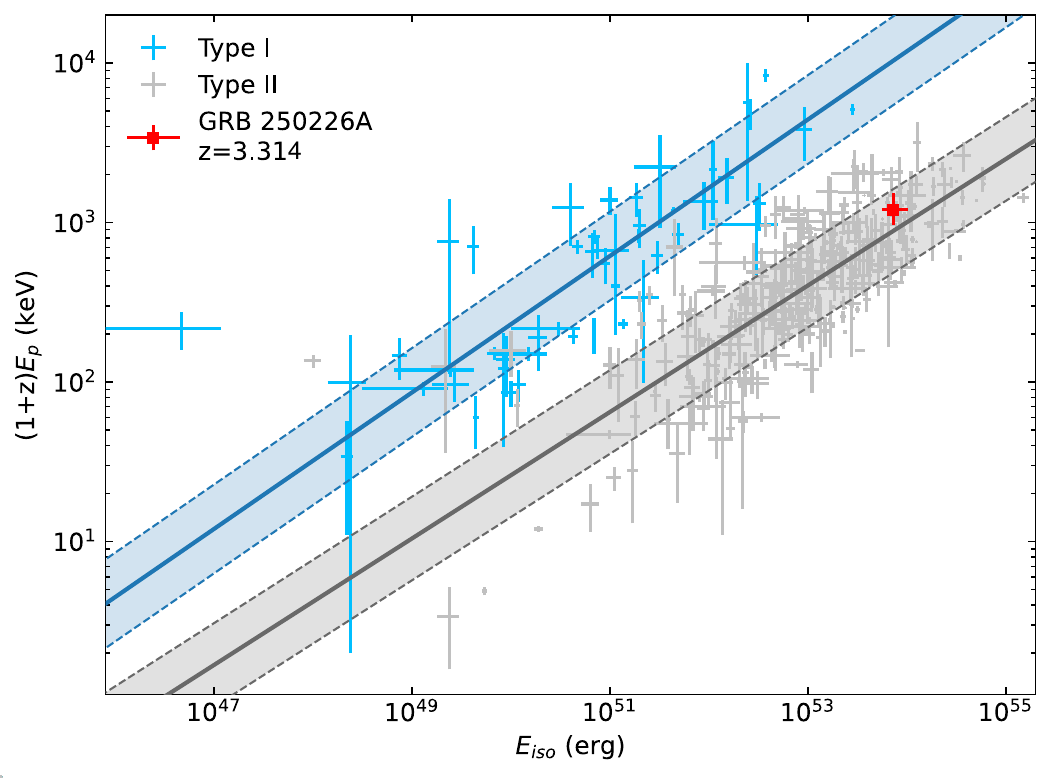}
        \caption{Amati diagram of GRB 250226A.}
    \end{subfigure}
    \hfill
    \begin{subfigure}{0.49\linewidth}
        \centering
        \includegraphics[width=\linewidth]{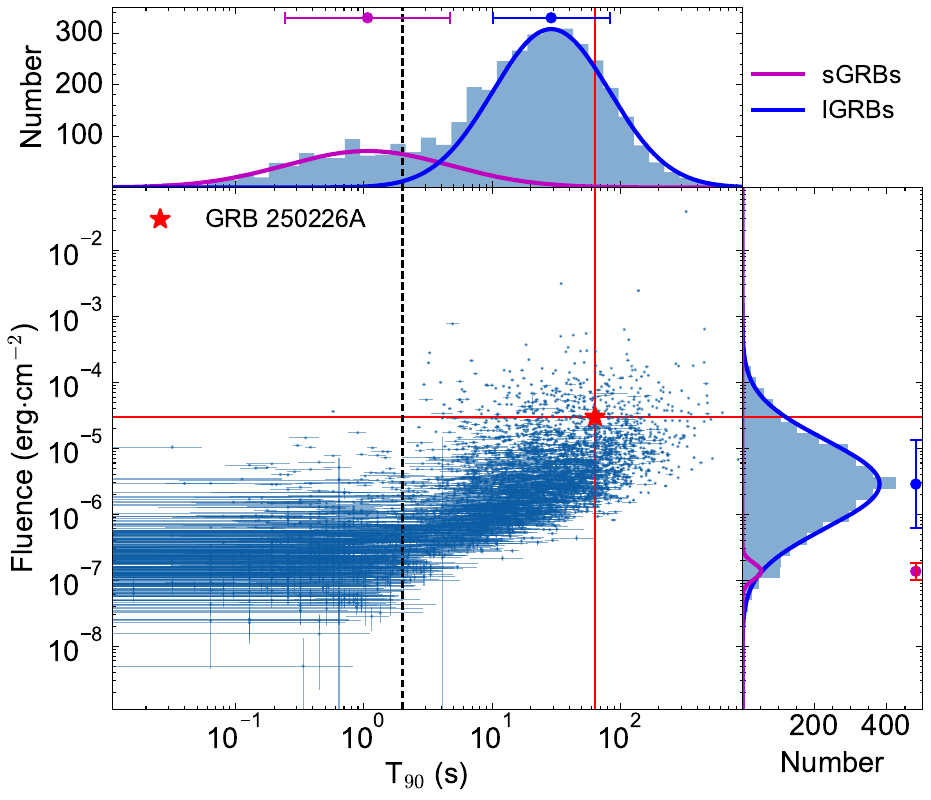}
        \caption{Position of GRB 250226A in the $T_{90}$-fluence diagram.}
    \end{subfigure}

    \caption{Amati and $T_{90}$-fluence diagrams for GRB 250226A. Panel (a): Amati diagram of GRB 250226A. Panel (b): Position of GRB 250226A in the $T_{90}$-fluence diagram. The vertical dashed line indicates the division between short and long GRBs at $T_{90}=2$ s. The red star marks the position of GRB 250226A in the diagram.}
    \label{fig:diagram}
\end{figure*}

\begin{figure*}[htbp]
    \centering

    \begin{subfigure}{0.49\linewidth}
        \centering
        \includegraphics[width=\linewidth]{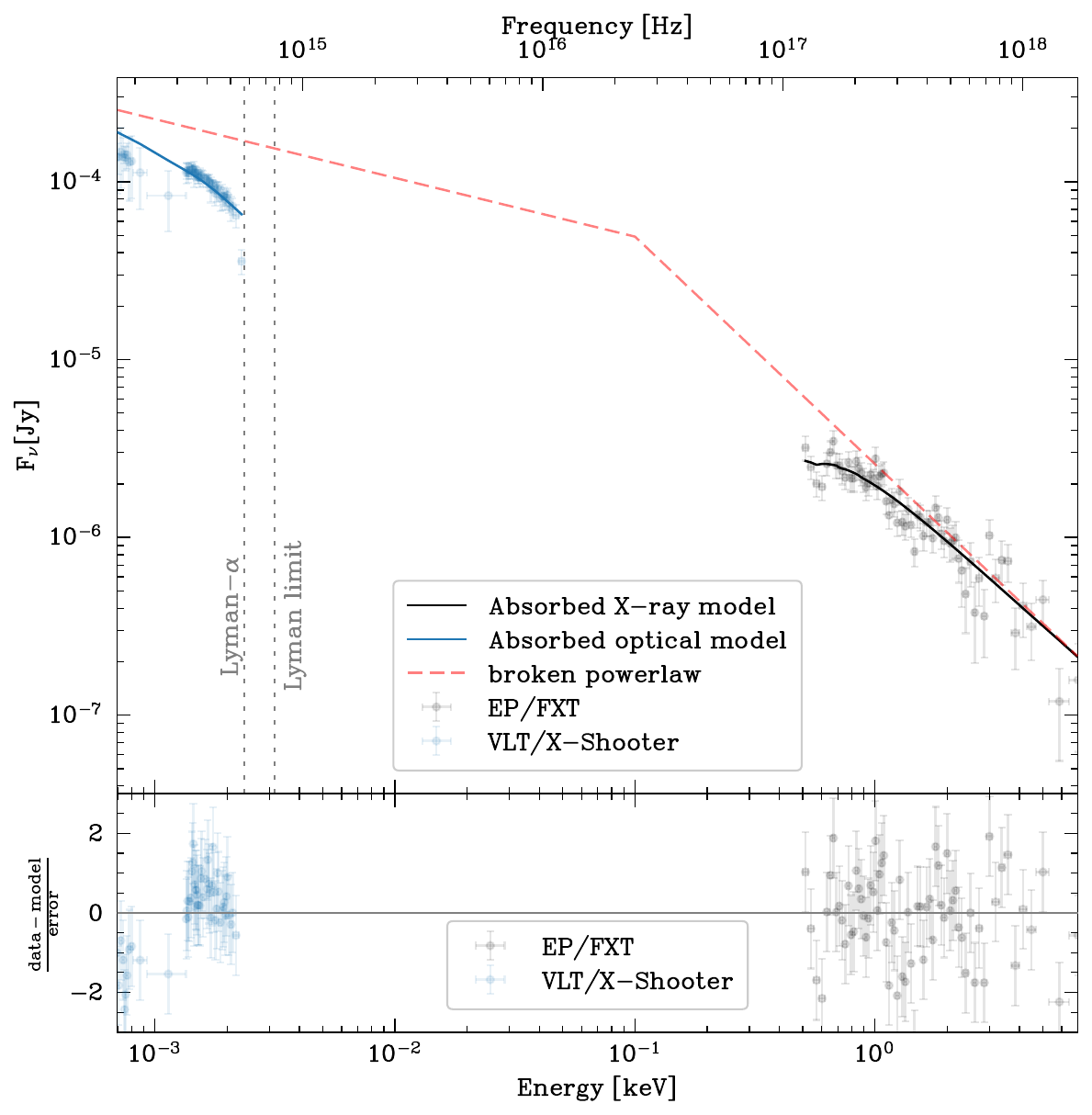}
        \caption{Joint spectrum combined from EP/FXT and VLT/X-shooter.}
    \end{subfigure}
    \hfill
    \begin{subfigure}{0.49\linewidth}
        \centering
        \includegraphics[width=\linewidth]{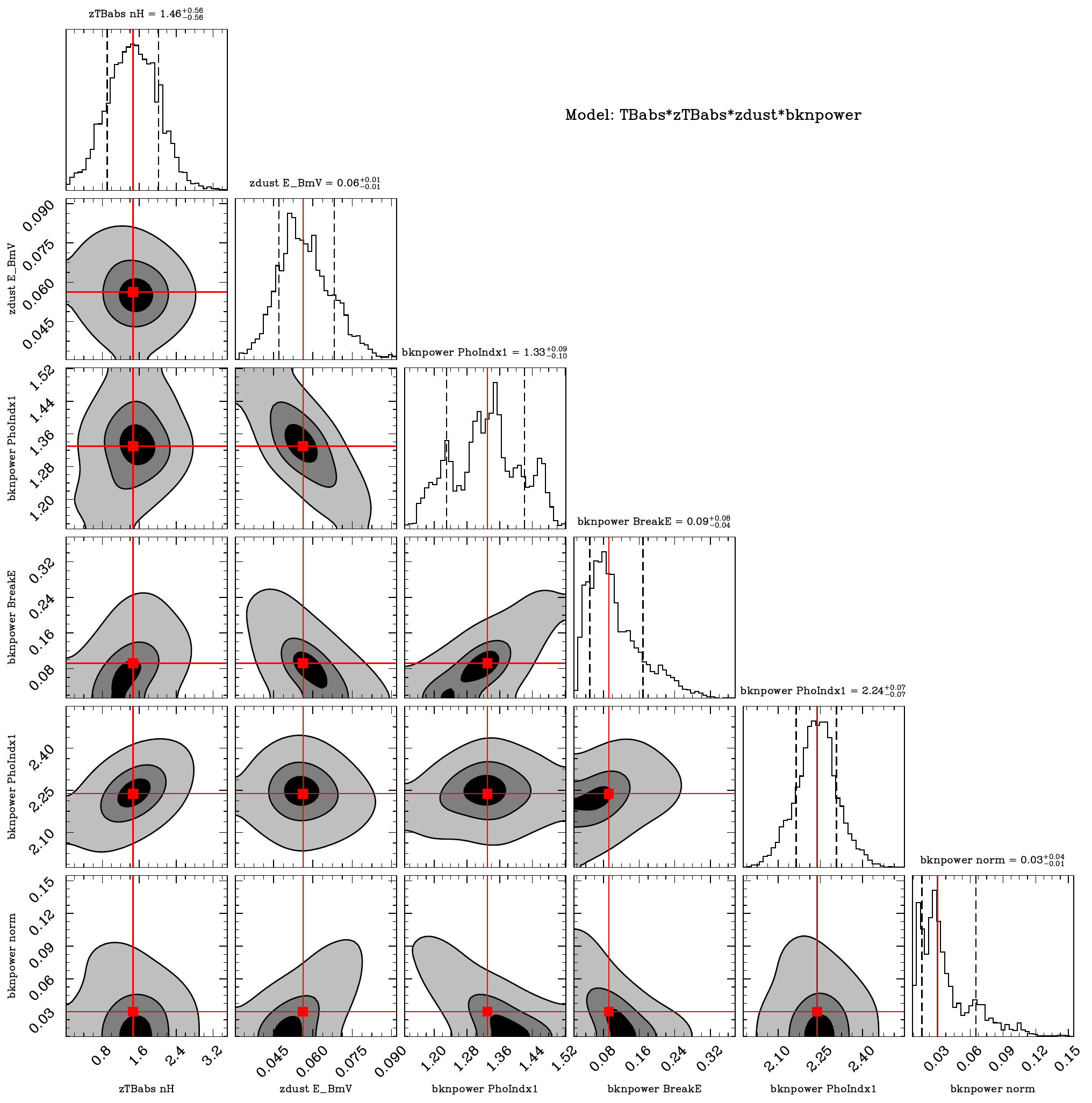}
        \caption{Posterior distributions of model parameters.}
    \end{subfigure}

    \caption{Spectral energy distribution of the afterglow and the corresponding posterior distributions of the model parameters. Panel (a): The observed VLT/X-shooter spectrum (light-blue points) and the simulated X-ray spectrum (gray points) at 2.15 hr post-trigger are combined for the SED analysis. The unabsorbed broken power-law model is displayed by a light-red dashed line, while the absorbed optical and X-ray models are shown by blue and black solid lines, respectively. Panel (b): Posterior distributions of the parameters of the absorbed broken power-law model.}
    \label{fig:SED}
\end{figure*}

\begin{figure*}[htbp]
    \centering

    \begin{subfigure}{0.7\linewidth}
        \centering
        \includegraphics[width=\linewidth]{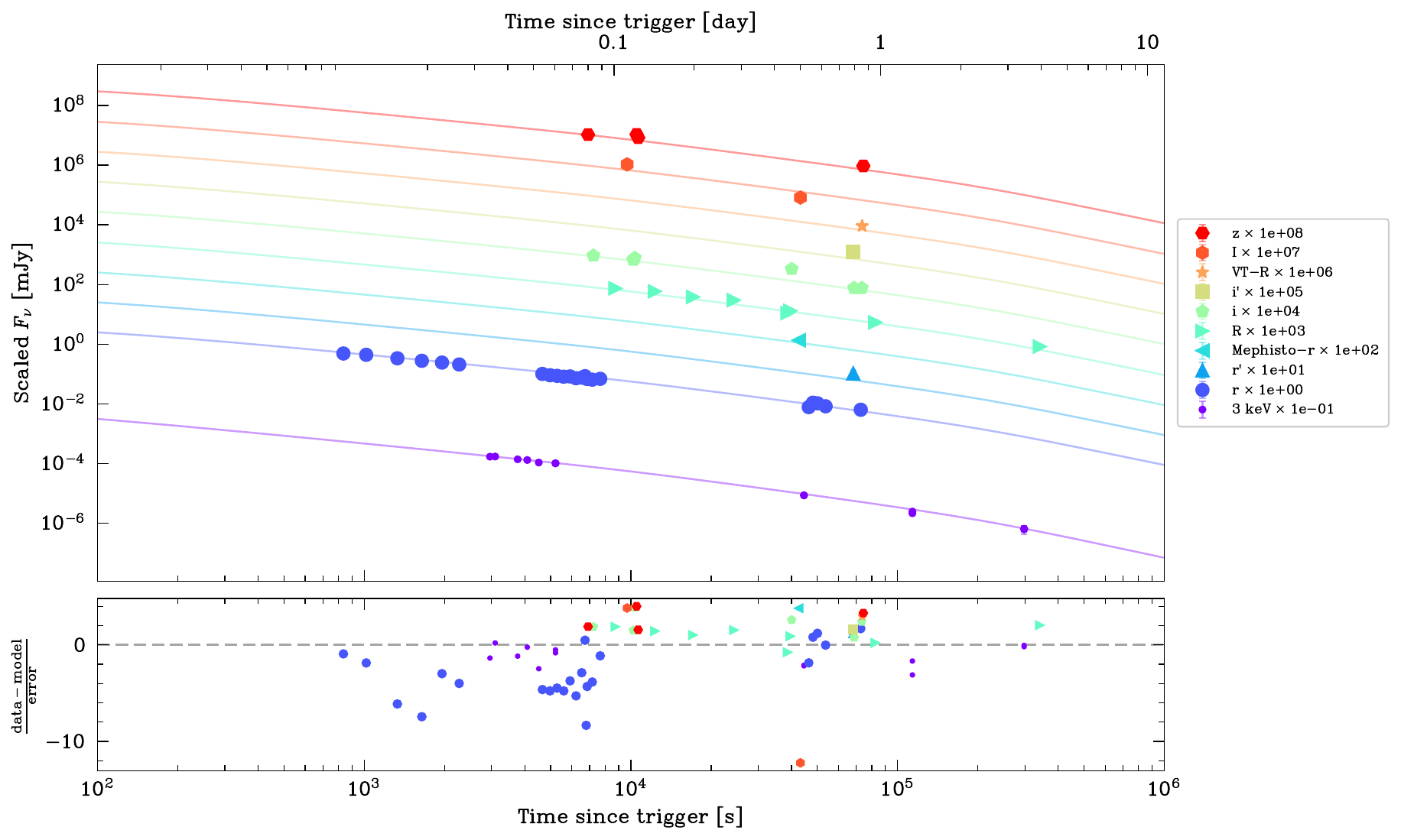}
    \end{subfigure}

    \vspace{0.5cm}

    \begin{subfigure}{0.75\linewidth}
        \centering
        \includegraphics[width=\linewidth]{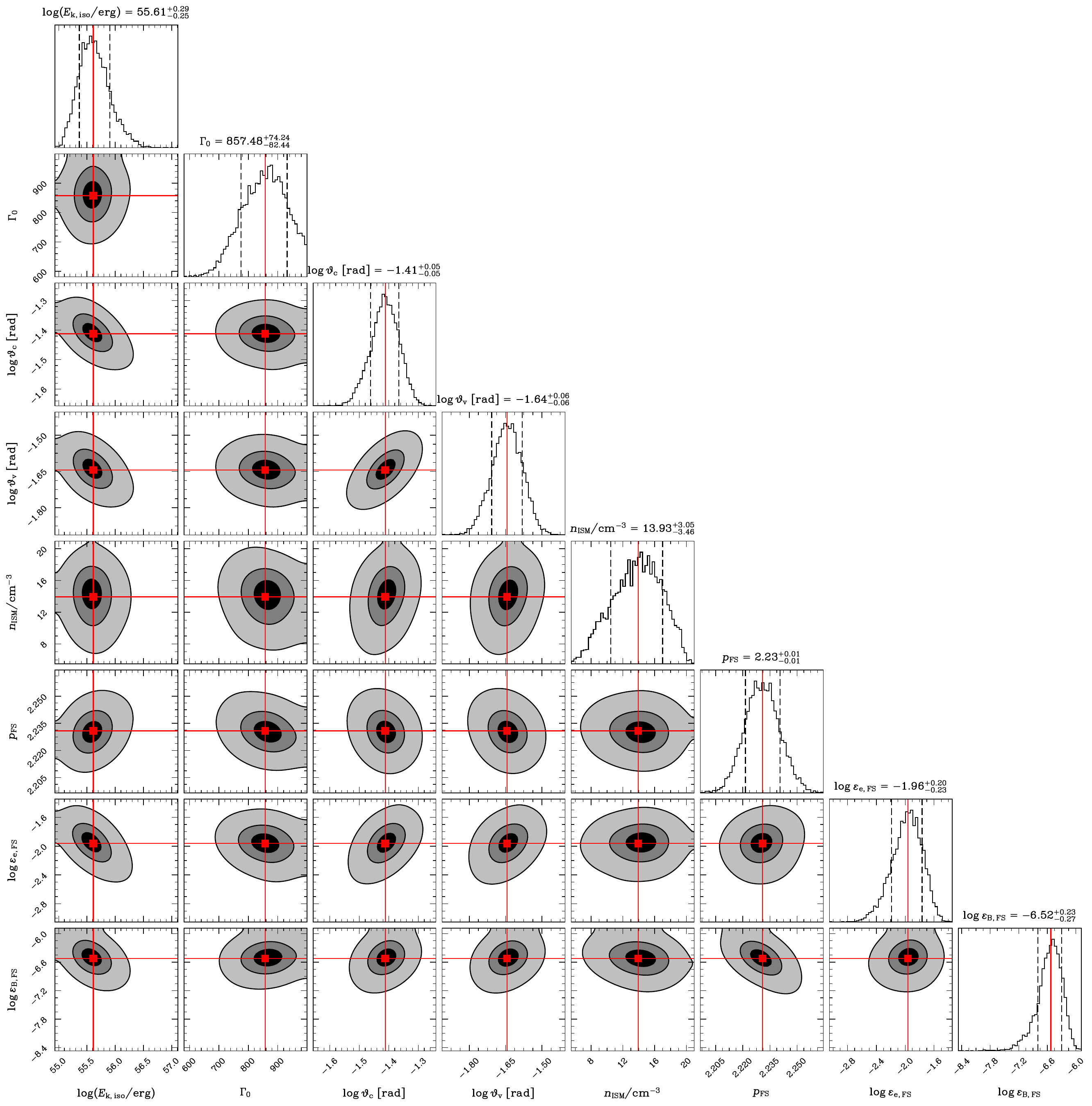}
    \end{subfigure}

    \caption{Multiwavelength light curves of the afterglow (top panel) and the corresponding marginalized posterior distribution (bottom panel). The afterglow is modeled as a top-hat jet propagating through a uniform ISM, with the initial bulk Lorentz factor and the ISM number density assigned Gaussian priors of $\Gamma_0 \sim \mathcal{N}(500, 150^2)$ and $n_{\mathrm{ISM}} \sim \mathcal{N}(10, 5^2)$, respectively. The light curves and posterior distributions are obtained using Posterior-5 from Table~\ref{tab:VegasAfterglow1}.}
    \label{fig:afterglow_model5}
\end{figure*}

The soft X-ray sensitivity of EP opens a new observational window for investigating the prompt soft X-ray emission from high-energy transients.
For EP250226a/GRB 250226A, the soft X-ray emission lasts longer than the gamma-ray emission, while no significant soft X-ray emission is detected before the gamma-ray trigger, distinguishing it from some previously reported EP transients with earlier soft X-ray activity (Figure~\ref{fig:HE-lc}). 
Beyond individual events, the growing sample of EP-detected events suggests that fast X-ray transients represent a broader phenomenological class than a simple extension of the classical GRB population. This growing sample will enable systematic population studies, providing opportunities to characterize their observational properties, identify potential subclasses, and constrain their progenitors and emission mechanisms. 
EP observations, combined with multiwavelength follow-up observations, will also facilitate investigations into the possible connections between fast X-ray transients and other classes of high-energy phenomena, 
including off-axis GRBs \citep[e.g.,][]{2024A&A...690A.101W,2025ApJ...985...21Z,2025ApJ...986..106G}, 
dirty fireballs \citep[e.g.,][]{2026arXiv260326213D}, 
low-luminosity GRBs \citep[e.g.,][]{2025ApJ...988L..71W,2026arXiv260606213C}, XRFs \citep[e.g.,][]{2025ApJ...988L..34J}, 
and supernova-associated fast X-ray transients \citep[e.g.,][]{2025arXiv250417034L,2025ApJ...988L..13R,2026arXiv260610014Y,2026arXiv260606213C}.
Furthermore, the detection of high-redshift fast X-ray transients demonstrates the potential of EP to probe relativistic explosions in the early Universe \citep[e.g.,][]{2025NatAs...9..564L,2025ApJ...988L..71W,2026arXiv260927127A}.

\clearpage

\clearpage
\begin{table*}[ht!]
\caption{Spectral analysis results for prompt emission and afterglow emission of EP250226a.}
\label{tab:HE-data}
\centering
\scriptsize
\begin{tabular}{ccc|ccc|c|cc|cc|c}
\hline\hline
Data &
Time Slice &
$T_\mathrm{mid}$ &
$N_\mathrm{H,X}$ &
$\Gamma$ &
$(1+z)E_\mathrm{p}$ &
stat/(d.o.f.) &
\multicolumn{2}{c|}{LOOIC} &
\multicolumn{2}{c|}{WAIC} &
Figure
\\
&
[$\mathrm{s}$] &
[$\mathrm{s}$] &
[$10^{22}\mathrm{\,cm^{-2}}$] &
&
[$10^3$\,keV] &
&
Deviance &
$p$ &
Deviance &
$p$ &
\\
\hline
EP/WXT & [-0.8,19.7] & 9.5 &
$15.05_{-6.53}^{+7.02}$ &
$1.81_{-0.46}^{+0.33}$ &
--- &
21.82/21 &
$23.94 \pm 6.11$ & 1.54 &
$23.91 \pm 6.11$ & 1.52 &
\ref{fig:WXT}
\\

EP/WXT & [-0.8,19.7] & 9.5 &
0.91 (fixed) &
$0.95_{-0.32}^{+0.34}$ &
--- &
21.64/22 &
$24.03 \pm 5.93$ & 1.54 &
$23.99 \pm 5.93$ & 1.52 &
\ref{fig:WXT}
\\

EP/WXT+GECAM & [-0.8,19.7] & 9.5 &
$8.81_{-4.74}^{+6.21}$ &
$1.03 \pm 0.07$ &
$1.34_{-0.29}^{+0.48}$ &
137.59/112 &
$141.99 \pm 16.78$ & 3.52 &
$141.95 \pm 16.78$ & 3.50 &
\ref{fig:WXT_GECAM}
\\

EP/WXT+GECAM & [-0.8,19.7] & 9.5 &
0.91 (fixed) &
$0.94_{-0.06}^{+0.05}$ &
$1.20_{-0.23}^{+0.33}$ &
138.46/113 &
$142.45 \pm 17.15$ & 3.22 &
$142.44 \pm 17.15$ & 3.22 &
\ref{fig:WXT_GECAM}
\\

\hline

EP/WXT & [0,250] & 125.0 &
$3.96_{-2.32}^{+2.88}$ &
$1.48_{-0.23}^{+0.26}$ &
--- &
31.84/23 &
$35.83 \pm 6.57$ & 2.89 &
$35.71 \pm 6.55$ & 2.83 &
\ref{fig:WXT-TimeAveraged}
\\

EP/WXT & [0,250] & 125.0 &
0.91 (fixed) &
$1.24\pm0.15$ &
--- &
30.61/24 &
$33.60 \pm 6.24$ & 2.17 &
$33.55 \pm 6.23$ & 2.15 &
\ref{fig:WXT-TimeAveraged}
\\

\hline

EP/FXT & [2725,5440] & 4082.5 &
$0.91 \pm 0.38$ &
$2.05 \pm 0.04$ &
--- &
132.14/139 &
$135.92 \pm 14.37$ & 2.90 &
$135.82 \pm 14.35$ & 2.84 &
\ref{fig:FXT1}
\\

EP/FXT & [43056,46064] & 44560.0 &
0.91 (fixed) &
$2.15\pm0.08$ &
--- &
100.69/109 &
$103.21 \pm 10.86$ & 1.84 &
$103.20 \pm 10.86$ & 1.84 &
\ref{fig:FXT2}
\\

EP/FXT & [112193,115044] & 113618.5 &
0.91 (fixed) &
$2.08_{-0.17}^{+0.18}$ &
--- &
72.93/123 &
$75.10 \pm 12.87$ & 1.50 &
$75.09 \pm 12.87$ & 1.50 &
\ref{fig:FXT34}
\\

EP/FXT & [296553,299566] & 298059.5 &
0.91 (fixed) &
$1.63_{-0.45}^{+0.50}$ &
--- &
75.08/121 &
$78.43 \pm 14.41$ & 2.63 &
$78.41 \pm 14.40$ & 2.62 &
\ref{fig:FXT34}
\\

\hline
\end{tabular}

\tablefoot{
The uncertainties of the spectral parameters are reported as the $1\sigma$ credible intervals.}
\end{table*}

\begin{table*}[ht!]
\caption{Bayesian inference results of afterglow model parameters.}
\label{tab:VegasAfterglow1}
\centering
\begin{tabular}{cccccc}
\hline\hline
Parameter 
& Prior 
& Posterior-5 \\
\hline
$\log (E_{\mathrm{k,iso}}/\mathrm{erg})$ 
& Uniform prior $\mathcal{U}(50, 60)$
& $55.61_{-0.25}^{+0.29}$\\

$\Gamma_0$
& Gaussian prior $\mathcal{N}(500, 150^2)$
& $857.48_{-82.44}^{+74.24}$ \\

$\log \theta_{\mathrm{c}}$ [rad]
& Uniform prior $\mathcal{N}(-3, 0.2)$
& $-1.41\pm0.05$ \\

$\log \theta_{\mathrm{v}}$ [rad]
& Uniform prior $\mathcal{N}(-3, 0.2)$
& $-1.64\pm0.06$ \\

$n_{\mathrm{ISM}}/\mathrm{cm^{-3}}$
& Gaussian prior $\mathcal{N}(10, 5^2)$
& $13.93_{-3.46}^{+3.05}$ \\

$p_{\mathrm{FS}}$
& Uniform prior $\mathcal{N}(2.01, 3)$
& $2.23\pm0.01$ \\

$\log \epsilon_{\mathrm{e,FS}}$
& Uniform prior $\mathcal{N}(-10, 0)$
& $-1.96_{-0.23}^{+0.20}$ \\

$\log \epsilon_{\mathrm{B,FS}}$
& Uniform prior $\mathcal{N}(-10, 0)$
& $-6.52_{-0.27}^{+0.23}$ \\

$\chi$
& 1 (fixed) \\

\hline
$\ln \mathcal{Z}$
& ---
& $-279.01 \pm 0.16$ \\

Figure
& ---
& \ref{fig:afterglow_model5} \\

\hline
\end{tabular}

\tablefoot{The uncertainties of the above parameters are reported as the $1\sigma$ credible intervals. The additional subscript $\mathrm{FS}$ represents the contribution from the forward-shock component.}
\end{table*}

\clearpage
\begin{acknowledgements}
    We thank the anonymous referee for constructive and insightful comments that improved the current work significantly.
    This work is supported by Natural Science Foundation of China 12393813, National Key R\&D Program of China (2023YFE0101200), 
    and the Yunnan Revitalization Talent Support Program (YunLing Scholar Project).
    
    This work is based on the data obtained with Einstein Probe, a space mission supported by the Strategic Priority Program on Space Science of Chinese Academy of Sciences, in collaboration with the European Space Agency, the Max-Planck-Institute for extraterrestrial Physics (Germany), and the Centre National d'\'tudes Spatiales (France). W. Li, H. Zou, Y. Xiao, W. Guo, H. Sun and N. Li are supported by NSFC (12120101003, 12373010 and 12573049), the National Key R\&D Program (2022YFA1602902 and 2023YFA1607804), and Strategic Priority Research Program of CAS (XDB0550100 and XDB0550000).
    
    Y. Rajabov and N. Sasaki thank the GRANDMA collaboration, especially, Nu Khantanakorn, Aom Manasan for TRT, W. Corradi, T. Almeida, L. Fraga, F. Navarete, A. Takey, Y. Hendy, M. Abdelkareem for KAO, M. Pillas, S. Antier for coordinations, L. Nyambane, H. Muenter, N. Kochiashvili, O. Pyshna for the  GRANDMA operations, M. Coughlin, N. Guessoum, P. Hello, P-A. Duverne, S. Karpov, T. Pradier for the GRANDMA core team.
    EA's work was funded by the Science Committee of the Ministry of Science and Higher Education of the Republic of Kazakhstan (Grant No. AP26103591) and by the Nazarbayev University Faculty Development Competitive Research Grant Program (No. 040225FD4713).
    The NRIAG team acknowledges financial support from the Egyptian Science, Technology \& Innovation Funding Authority (STDF) under grant number 45779.
    
    A.~L.~Thakur, P. Barria,  J. Rodi, L. Piro, L. Natalucci and G. Gianfagna acknowledge support from the European Union's Horizon 2020 Programme under the AHEAD2020 project (grant agreement n. 871158).  A.~L.~Thakur, L. Piro and G. Gianfagna also acknowledge support by ASI (Italian Space Agency) through the Contract no. 2019-27-HH.0. P. Barria, L. Piro and L. Natalucci acknowledge support by Italian Research Center on High Performance Computing Big Data and Quantum Computing (ICSC), project funded by European Union - NextGenerationEU - and National Recovery and Resilience Plan (NRRP) - Mission 4 Component 2 within the activities of Spoke 2 (Fundamental Research and Space Economy), (CN 00000013 - CUP C53C22000350006).
    
    The GECAM (Huairou-1) mission is supported by the Strategic Priority Research Program on Space Science (Grant No. XDA15360000) of the Chinese Academy of Sciences. 
    
    This publication has made use of data collected at Lulin Observatory, partly supported by the TAOvA with the NSTC grant 114-2740-M-008-002.
    A.A. and T.-W.C. acknowledge financial support from the Yushan Fellow Program MOE-111-YSFMS-0008-001-P1, and from the NSTC 114-2112-M-008-021-MY3. 
    
    This work makes use of observations from the Las Cumbres Observatory network. The LCO team is supported by NSF grants AST-2308113 and AST-1911151. 
    
    The 60/90-cm Schmidt telescope at the Xinglong station is operated by WSGP group at NAOC and supported by the CAS Special Funds for Observatory and Facility Development, and the Zhengjia Enterprise Group. W. Li, H. Zou, Y. Xiao, W. Guo, and N. Li are supported by NSFC (12120101003 and 12373010), the National Key R\&D Program (2022YFA1602902 and 2023YFA1607804), and Strategic Priority Research Program of CAS (XDB0550100 and XDB0550000).
    
    Mephisto is developed at and operated by the South-Western Institute for Astronomy Research of Yunnan University (SWIFAR-YNU), funded by the ``Yunnan University Development Plan for World-Class University" and ``Yunnan University Development Plan for World-Class Astronomy Discipline". The Mephisto team acknowledges support from the Key Laboratory of Survey Science of Yunnan Province (202449CE340002), the ``Science \& Technology Champion Project'' (202005AB160002), and from two ``Team Projects" -- the ``Top Team'' (202305AT350002) and the ``Innovation Team'' (202105AE160021), funded by the ``Yunnan Revitalization Talent Support Program". B.K. acknowledges the support from the ``Special Project for High-End Foreign Experts", Xingdian Funding from Yunnan Province, and two National Key Research and Development Programs of China (2024YFA1611603, 2024YFA1611604).
    
    DBM is funded by the European Union (ERC, HEAVYMETAL, 101071865). Views and opinions expressed are, however, those of the authors only and do not necessarily reflect those of the European Union or the European Research Council. Neither the European Union nor the granting authority can be held responsible for them. The Cosmic Dawn Center (DAWN) is funded by the Danish National Research Foundation under grant DNRF140.
    A.S. acknowledges financial support from the Centre national d'\'etudes spatiales (CNES), France (ROR: \url{https://ror.org/04h1h0y33}) within the framework of the SVOM mission.
    N.R.T. is supported by STFC grant UKRI1200. 
    BS acknowledges the support of the French Agence Nationale de la Recherche (ANR), under grant ANR-23-CE31-0011 (project PEGaSUS).

    Observations with the SAO RAS telescopes are supported by the Ministry of Science and Higher Education of the Russian Federation. The renovation of telescope equipment is currently provided within the national project ``Science and universities.''
    
    We acknowledge the valuable discussions with Drs. Thomas Hussenot-Desenonges, Nino Kochiashvili and Damien Turpin.
    
    This research has made use of data obtained through the High Energy Astrophysics Science Archive Research Center online service, provided by the NASA/Goddard Space Flight Center. This research has made use of ``Aladin sky atlas'' developed at CDS, Strasbourg Observatory, France.
\end{acknowledgements}

%
  \bibliographystyle{aa} 
  \bibliography{main} 
\begin{appendix}




\onecolumn

\section{High-energy spectral analysis}
\begin{figure*}[htbp]
    \centering
    \begin{subfigure}{0.49\linewidth}
        \centering
        \includegraphics[width=\linewidth]{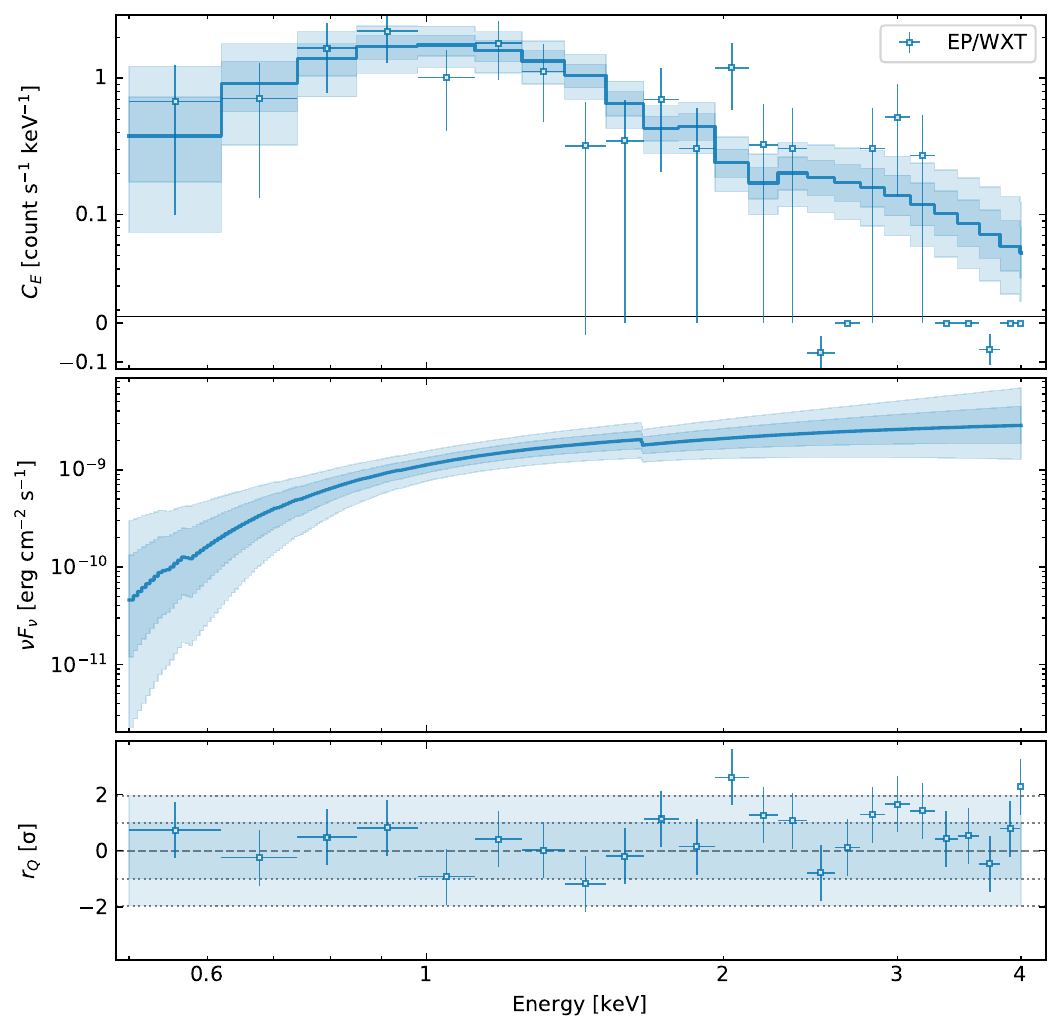}
        \caption{EP/WXT spectrum, power-law model, and residuals.}
    \end{subfigure}
    \hfill
    \begin{subfigure}{0.49\linewidth}
        \centering
        \includegraphics[width=\linewidth]{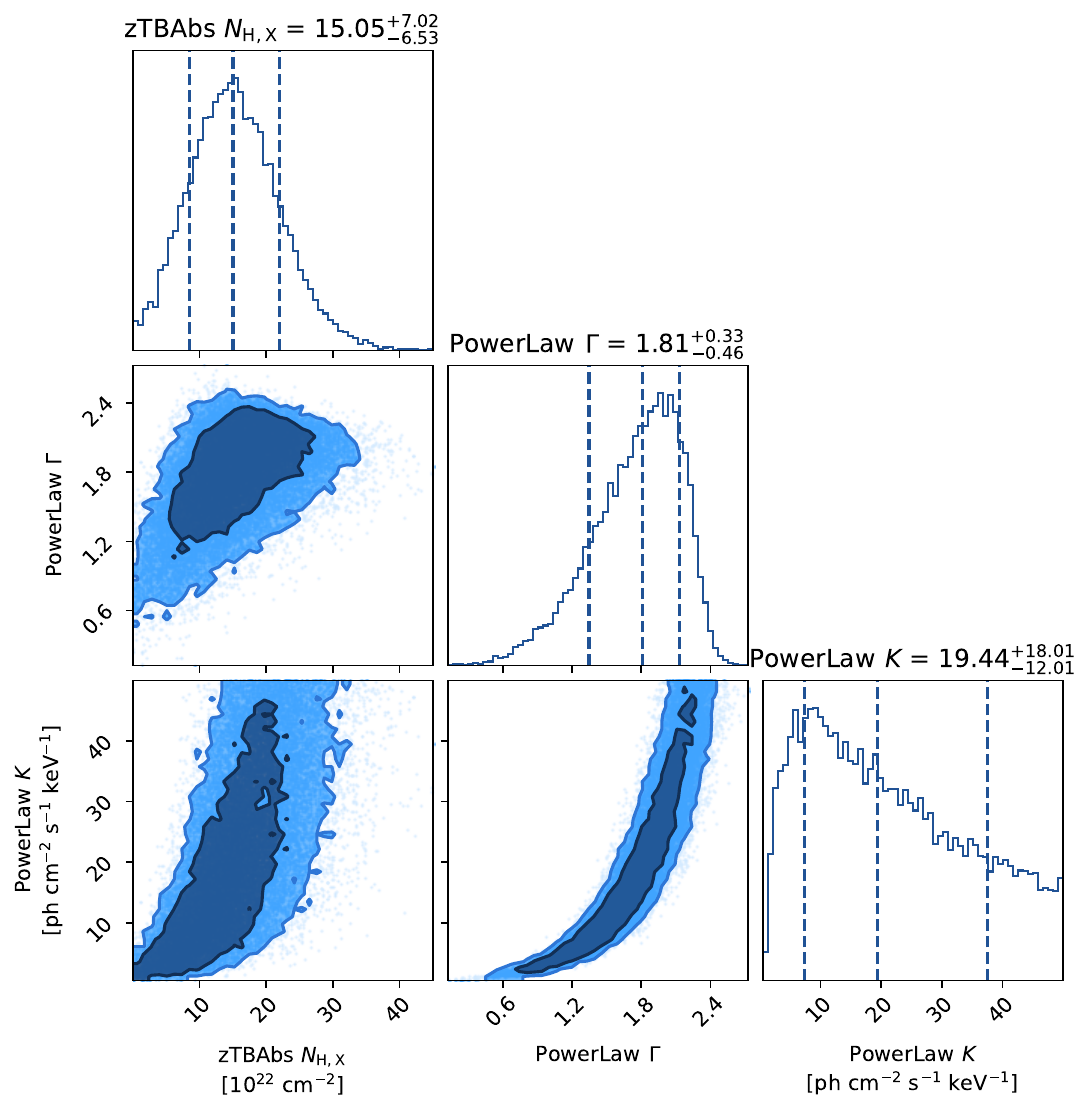}
        \caption{Posterior distributions of model parameters.}
    \end{subfigure}
    
    \vspace{0.5cm}
    
    \begin{subfigure}{0.49\linewidth}
        \centering
        \includegraphics[width=\linewidth]{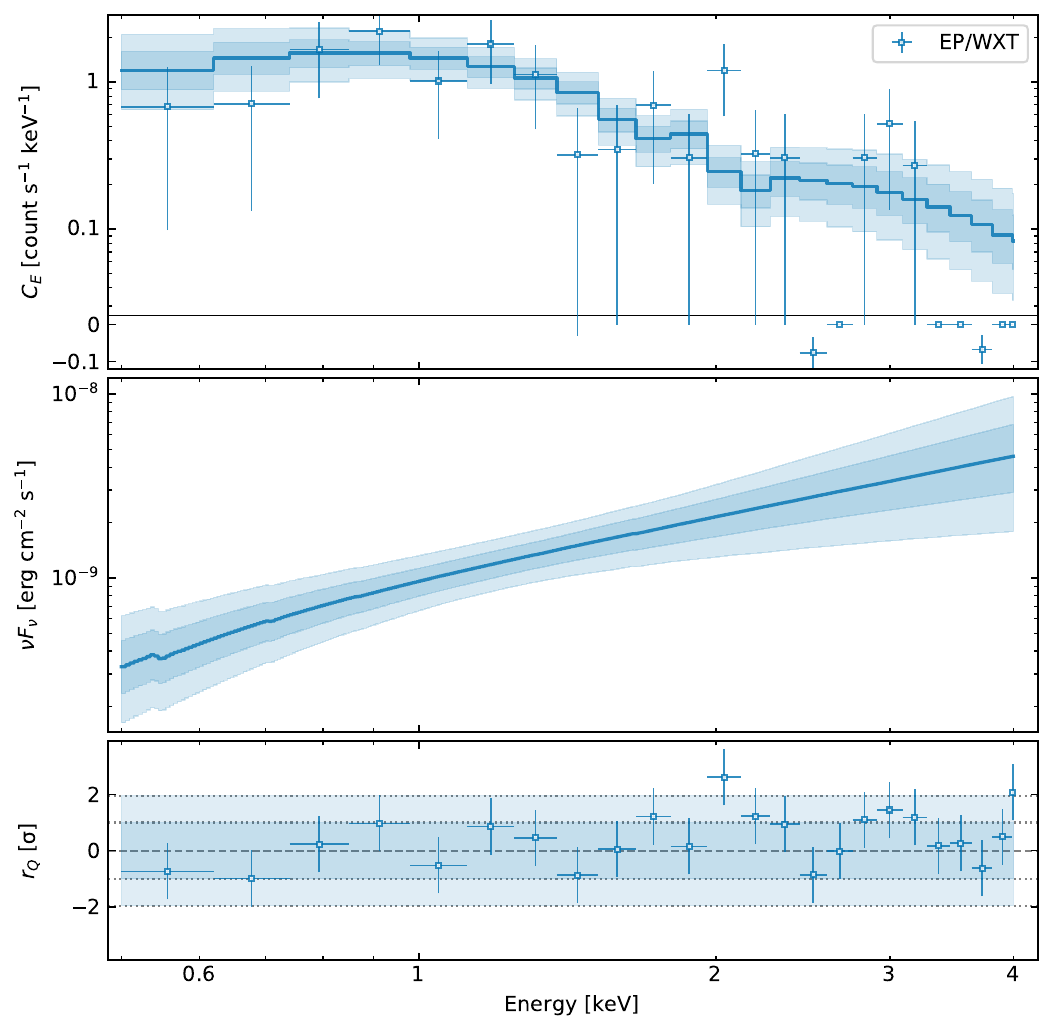}
        \caption{EP/WXT spectrum, power-law model, and residuals.}
    \end{subfigure}
    \hfill
    \begin{subfigure}{0.49\linewidth}
        \centering
        \includegraphics[width=\linewidth]{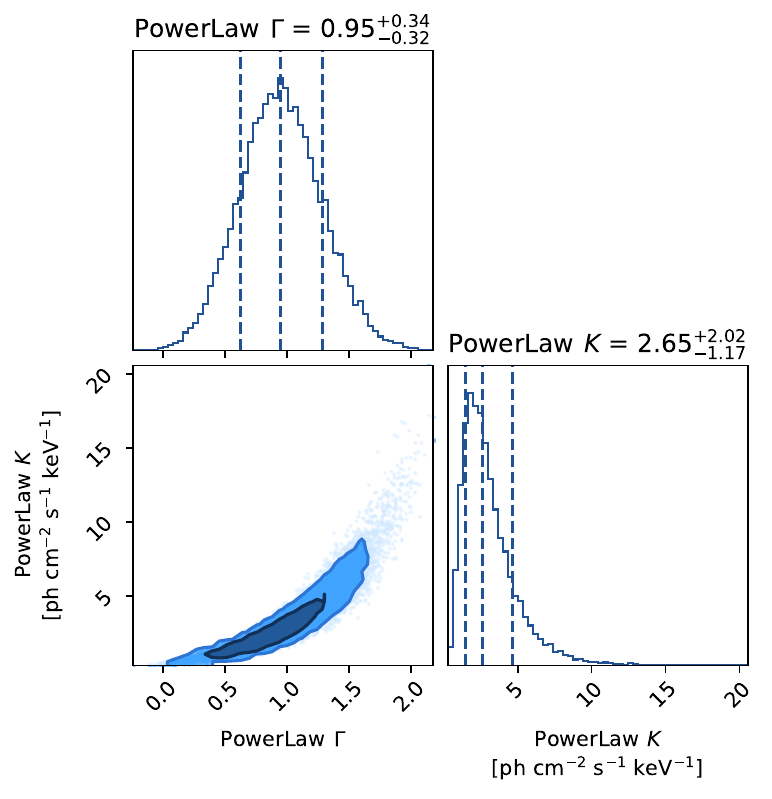}
        \caption{Posterior distributions of model parameters.}
    \end{subfigure}

    \caption{Time-integrated spectral analysis of EP/WXT pre-slew phase over $T_0 + [-0.8,19.7]$ s. Upper panel: The EP/WXT spectrum is fitted with an absorbed power-law model, with all parameters left free to vary. Lower panel: The spectrum is fitted with the same model, with the photon index free to vary while the intrinsic column density is fixed at $N_\mathrm{H,X}=0.91\times10^{22}\mathrm{\,cm^{-2}}$ (see details in Section~\ref{sec:HE_spectra}).}
    \label{fig:WXT}
\end{figure*}

\begin{figure*}[htbp]
    \centering
    \begin{subfigure}{0.49\linewidth}
        \centering
        \includegraphics[width=\linewidth]{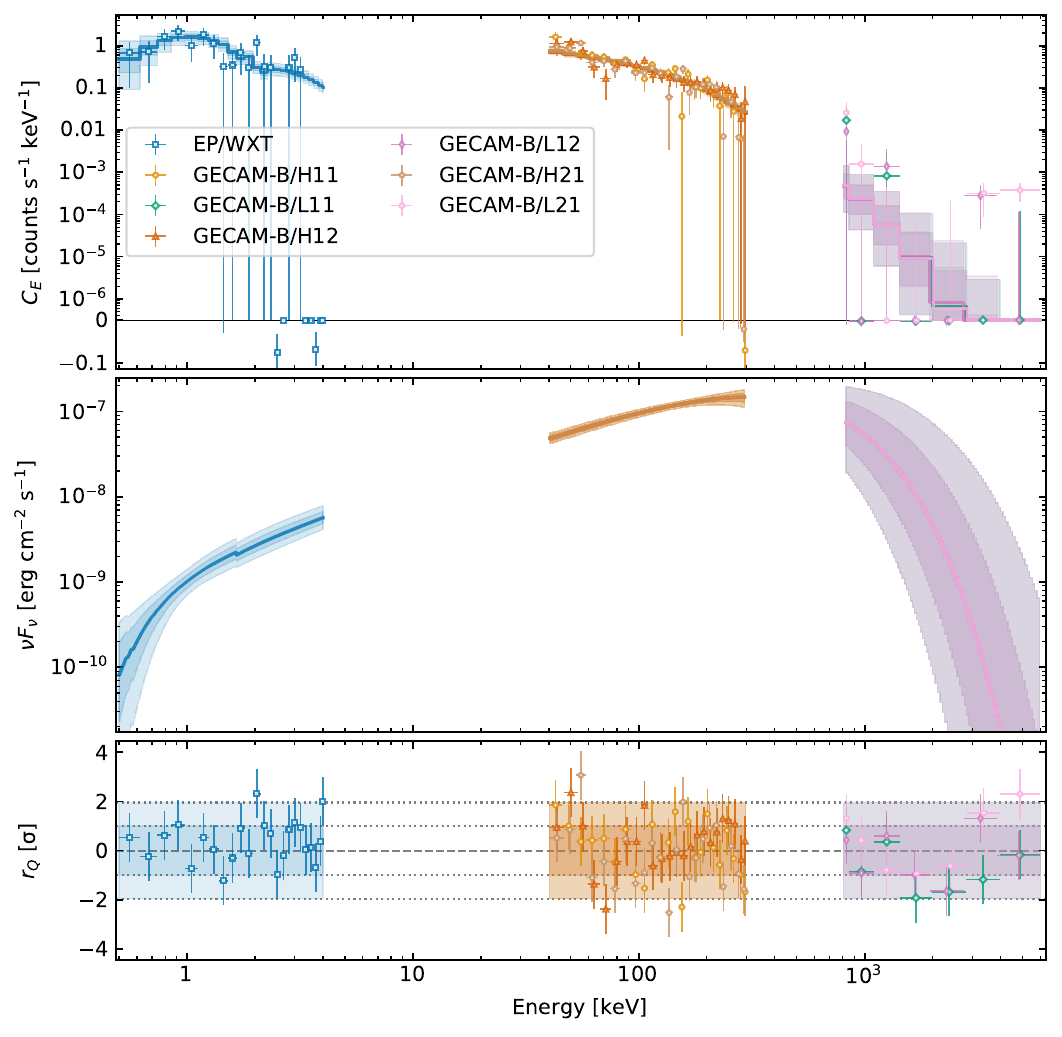}
        \caption{Joint spectrum from EP/WXT and GECAM-B, cutoff power-law model, and residuals.}
    \end{subfigure}
    \hfill
    \begin{subfigure}{0.49\linewidth}
        \centering
        \includegraphics[width=\linewidth]{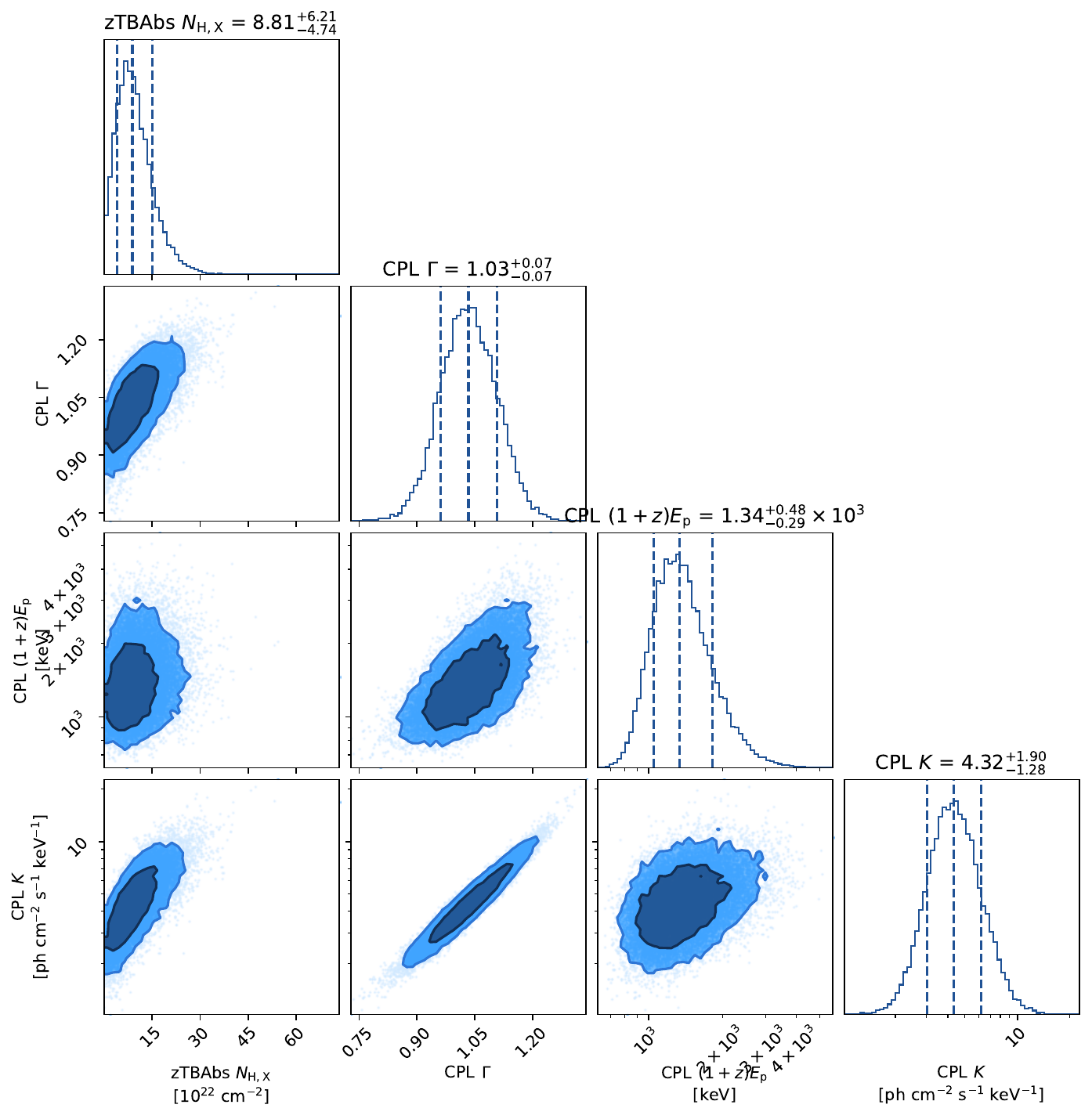}
        \caption{Posterior distributions of model parameters.}
    \end{subfigure}
    
    \vspace{2cm}
    
    \begin{subfigure}{0.49\linewidth}
        \centering
        \includegraphics[width=\linewidth]{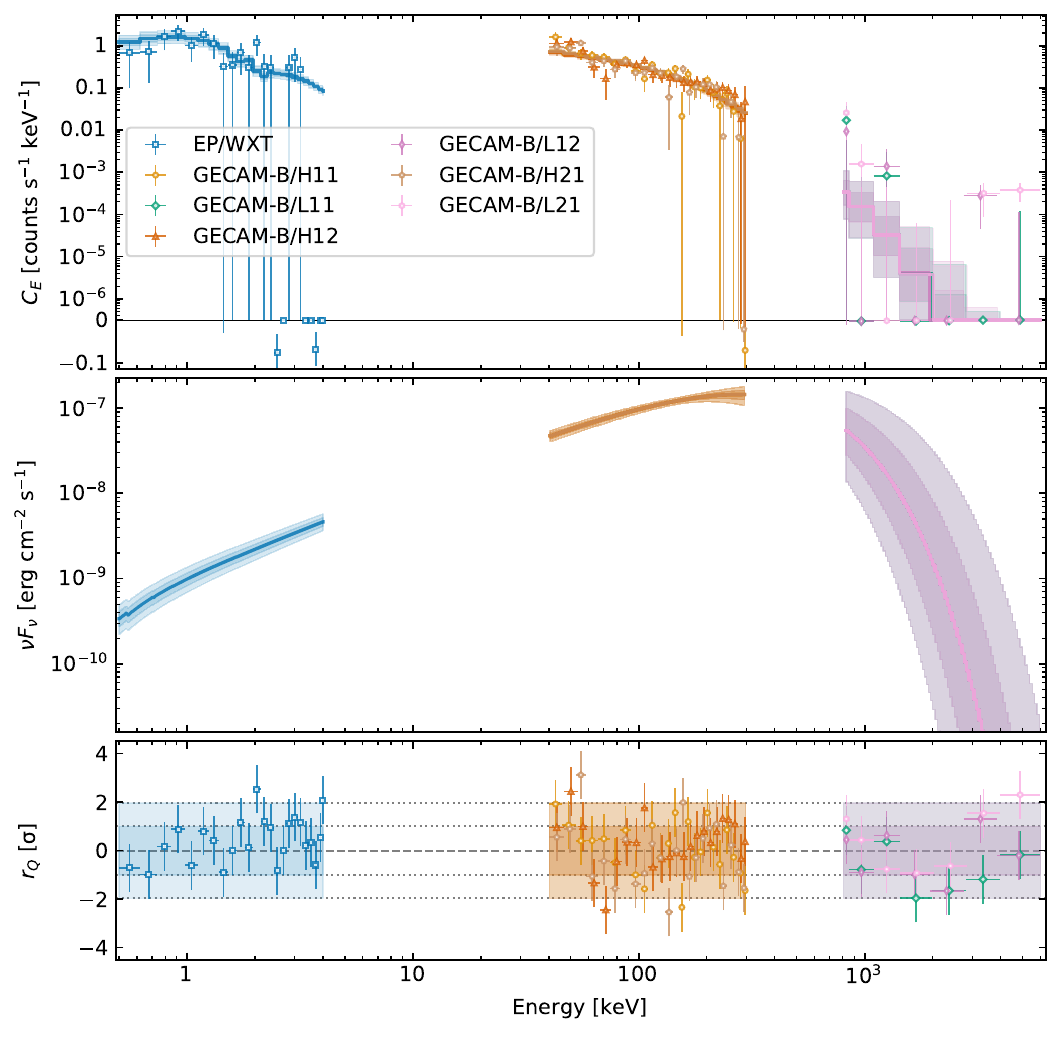}
        \caption{Joint spectrum from EP/WXT and GECAM-B, cutoff power-law model, and residuals.}
    \end{subfigure}
    \hfill
    \begin{subfigure}{0.49\linewidth}
        \centering
        \includegraphics[width=\linewidth]{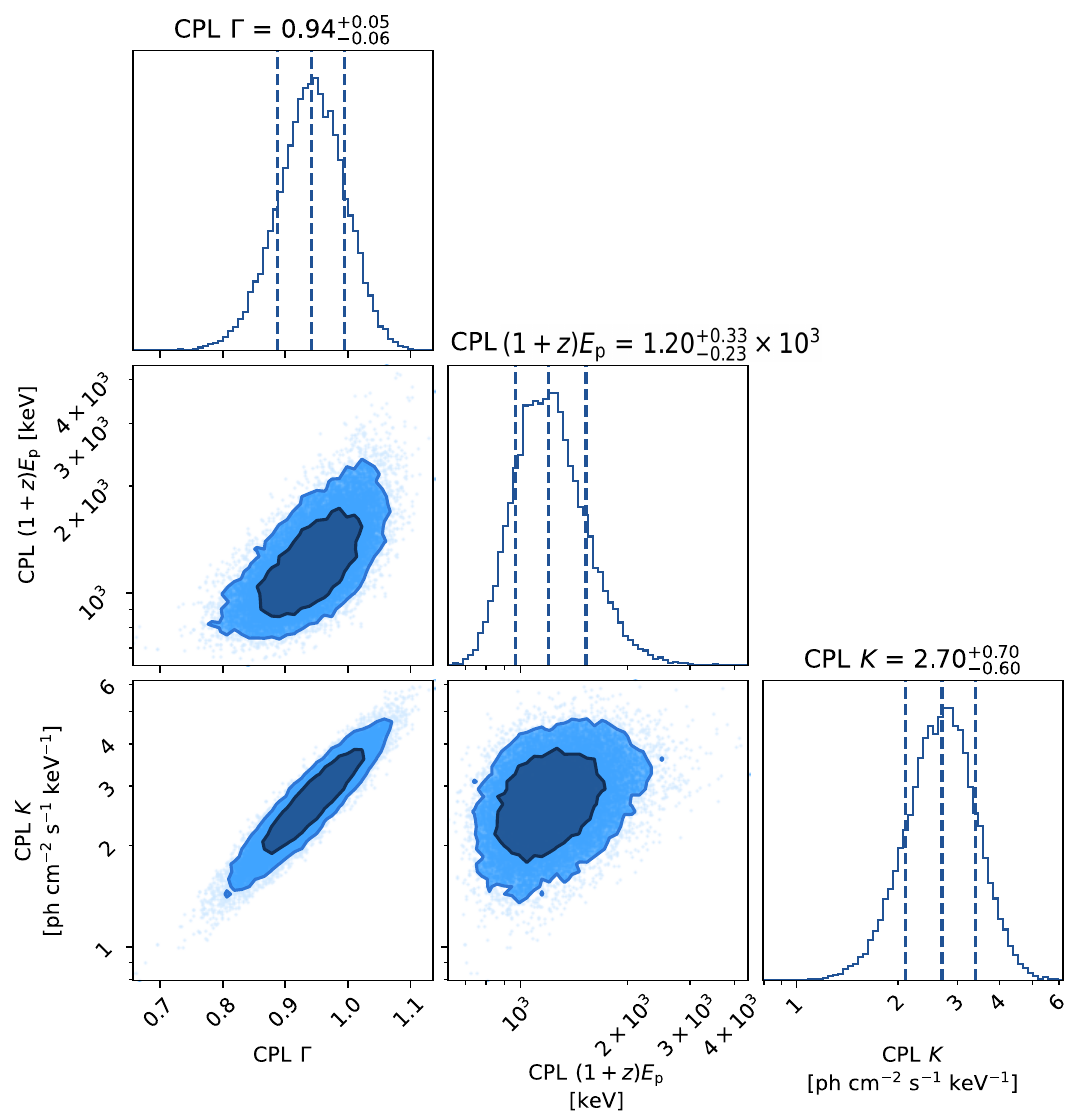}
        \caption{Posterior distributions of model parameters.}
    \end{subfigure}

    \caption{Time-integrated spectral analysis of EP/WXT and GECAM-B over $T_0 + [-0.8,19.7]$ s. Upper panel: The joint spectrum is fitted with an absorbed cutoff power-law model, with all parameters left free to vary. Lower panel: The spectrum is fitted with the same model, but the intrinsic column density is fixed at $N_\mathrm{H,X}=0.91\times10^{22}\mathrm{\,cm^{-2}}$ (see details in Section~\ref{sec:HE_spectra}).}
    \label{fig:WXT_GECAM}
\end{figure*}

\begin{figure*}[htbp]
    \centering
    \begin{subfigure}{0.49\linewidth}
        \centering
        \includegraphics[width=\linewidth]{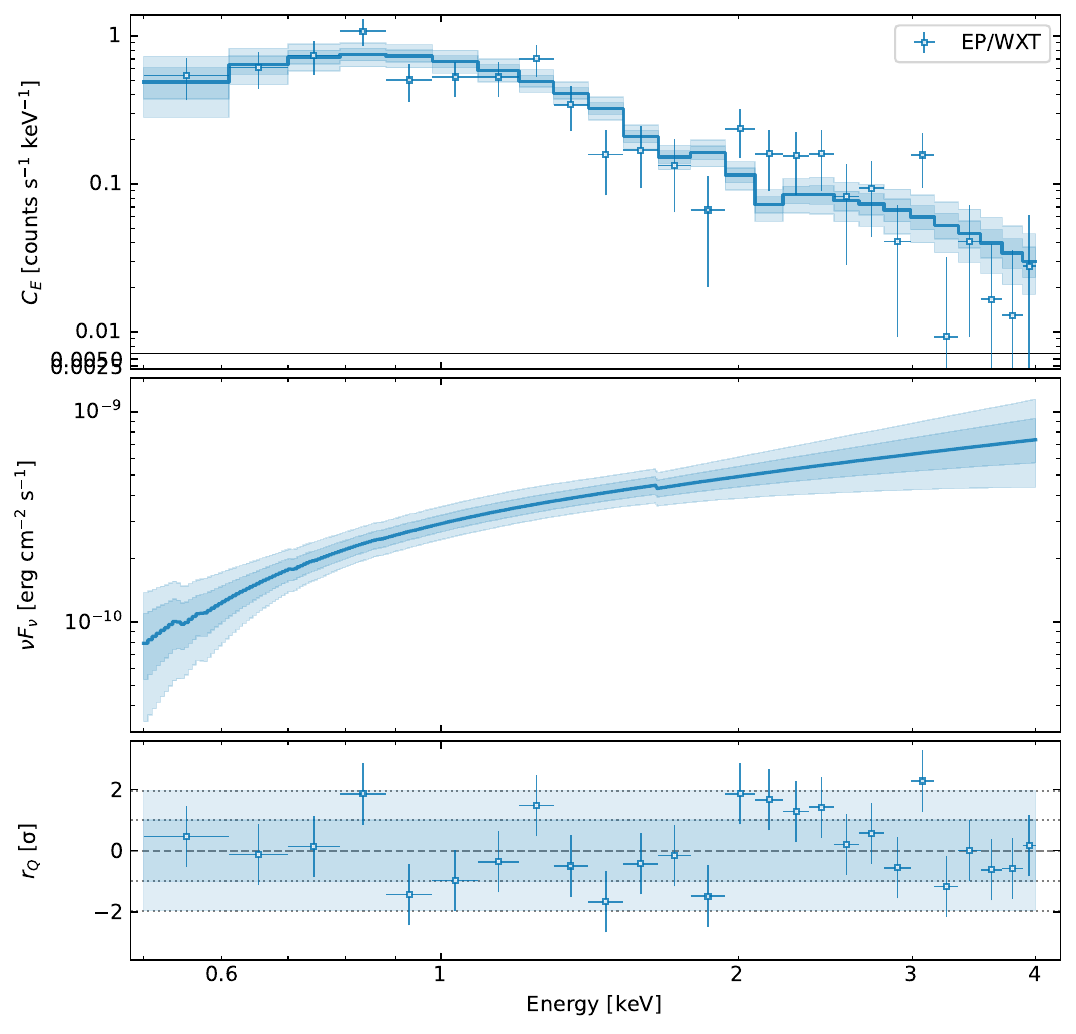}
        \caption{EP/WXT spectrum, power-law model, and residuals.}
    \end{subfigure}
    \hfill
    \begin{subfigure}{0.49\linewidth}
        \centering
        \includegraphics[width=\linewidth]{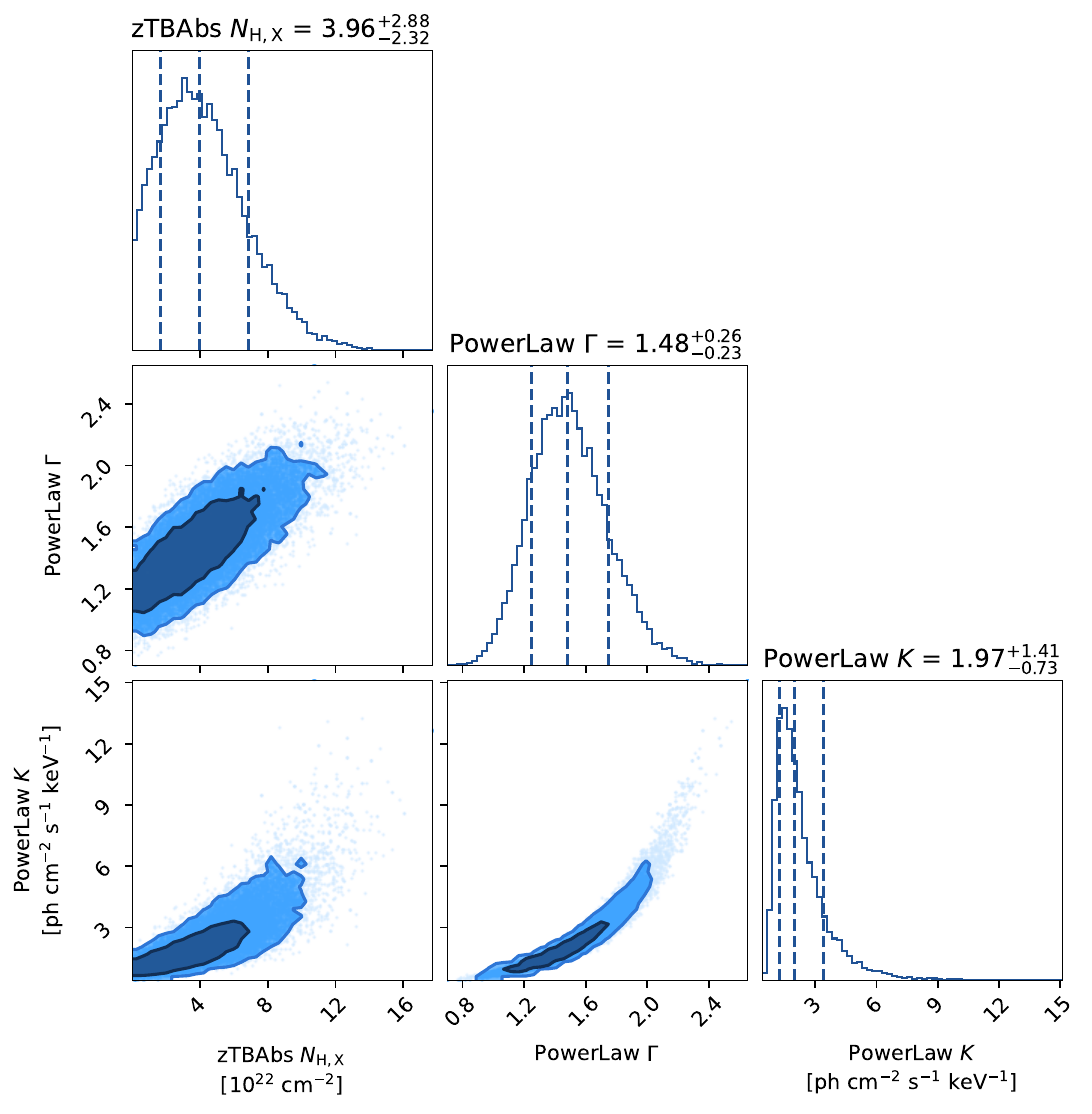}
        \caption{Posterior distributions of model parameters.}
    \end{subfigure}
    
    \vspace{2cm}
    
    \begin{subfigure}{0.49\linewidth}
        \centering
        \includegraphics[width=\linewidth]{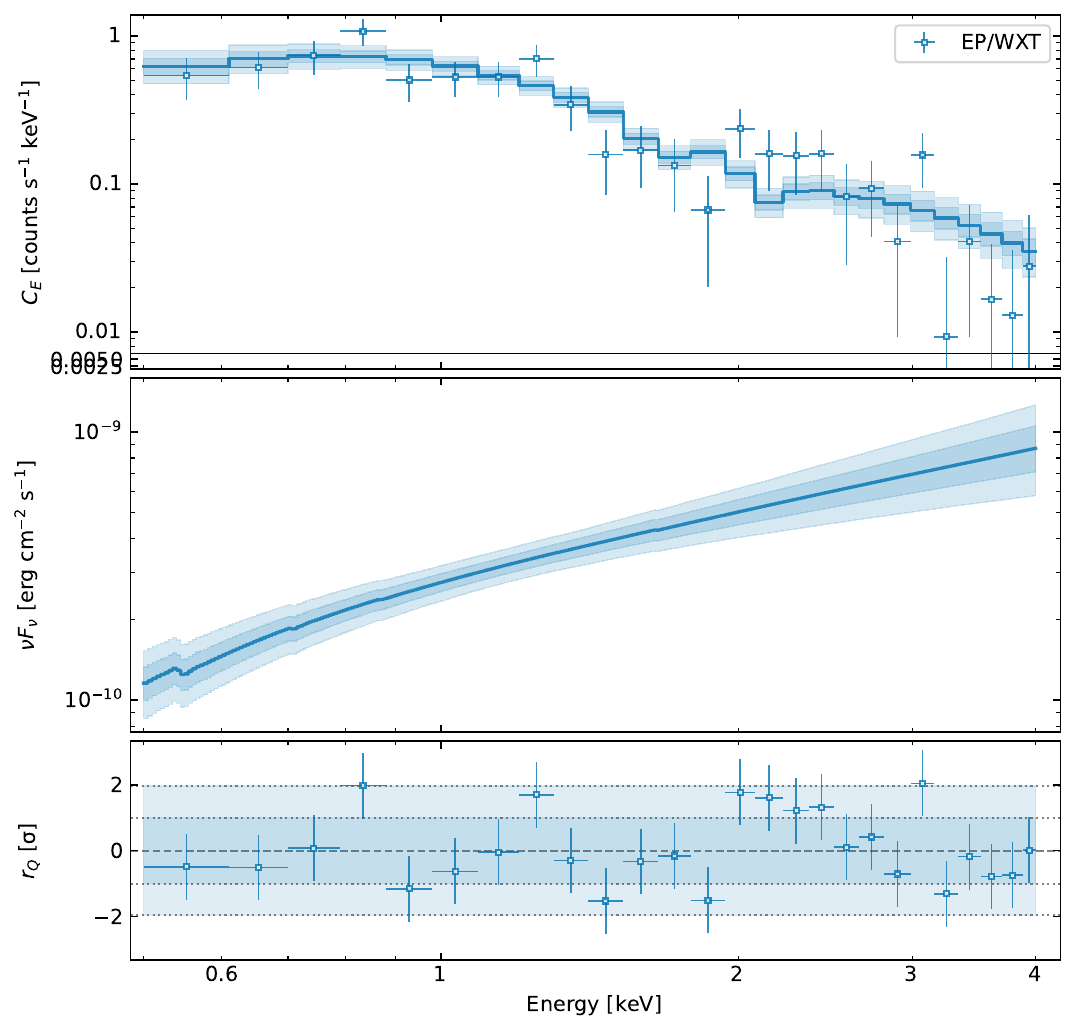}
        \caption{EP/WXT spectrum, power-law model, and residuals.}
    \end{subfigure}
    \hfill
    \begin{subfigure}{0.49\linewidth}
        \centering
        \includegraphics[width=\linewidth]{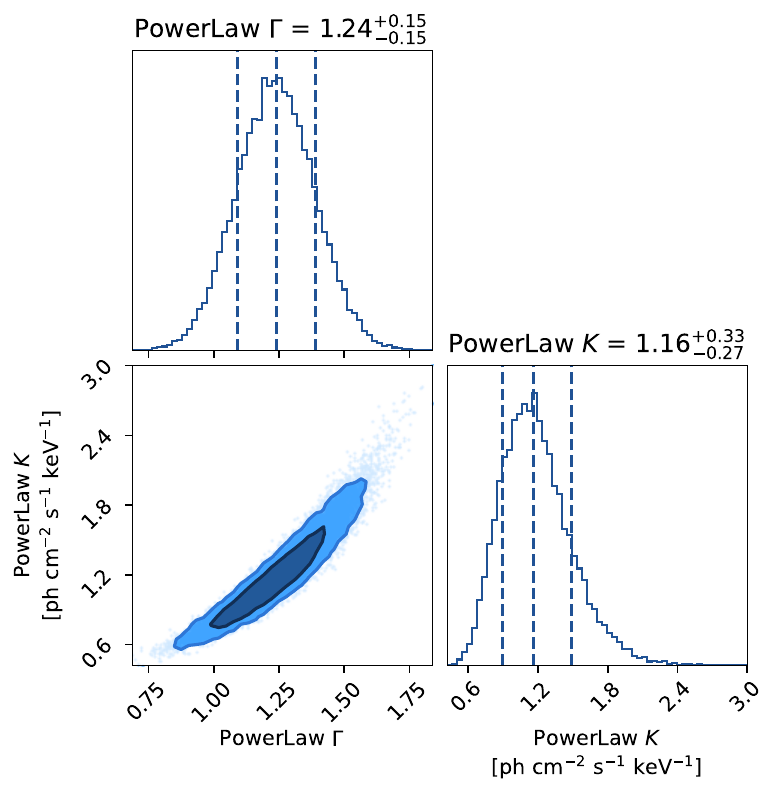}
        \caption{Posterior distributions of model parameters.}
    \end{subfigure}

    \caption{Time-integrated spectral analysis of EP/WXT detection over $T_0 + [0,250]$ s. Upper panel: The EP/WXT spectrum is fitted with an absorbed power-law model, with all parameters left free to vary. Lower panel: The spectrum is fitted with the same model, with the photon index free to vary while the intrinsic column density is fixed at $N_\mathrm{H,X}=0.91\times10^{22}\mathrm{\,cm^{-2}}$ (see details in Section~\ref{sec:HE_spectra}).}
    \label{fig:WXT-TimeAveraged}
\end{figure*}

\begin{figure*}[htbp]
    \centering
    
    
    \begin{subfigure}{0.49\linewidth}
        \centering
        \includegraphics[width=\linewidth]{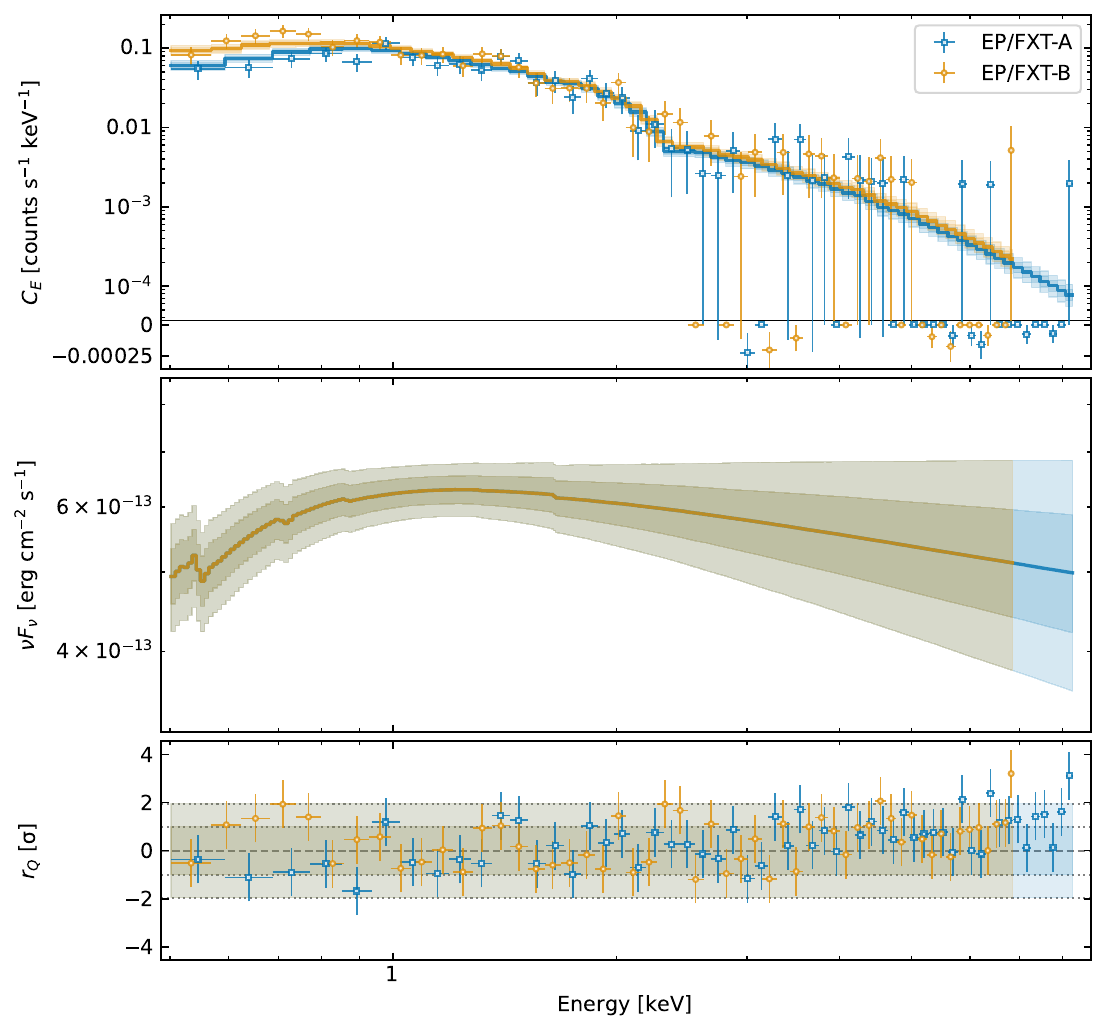}
        \caption{EP/FXT spectrum in $T_{0} + [43056,46064]$ s, power-law model, and residuals.}
    \end{subfigure}
    \hfill
    \begin{subfigure}{0.49\linewidth}
        \centering
        \includegraphics[width=\linewidth]{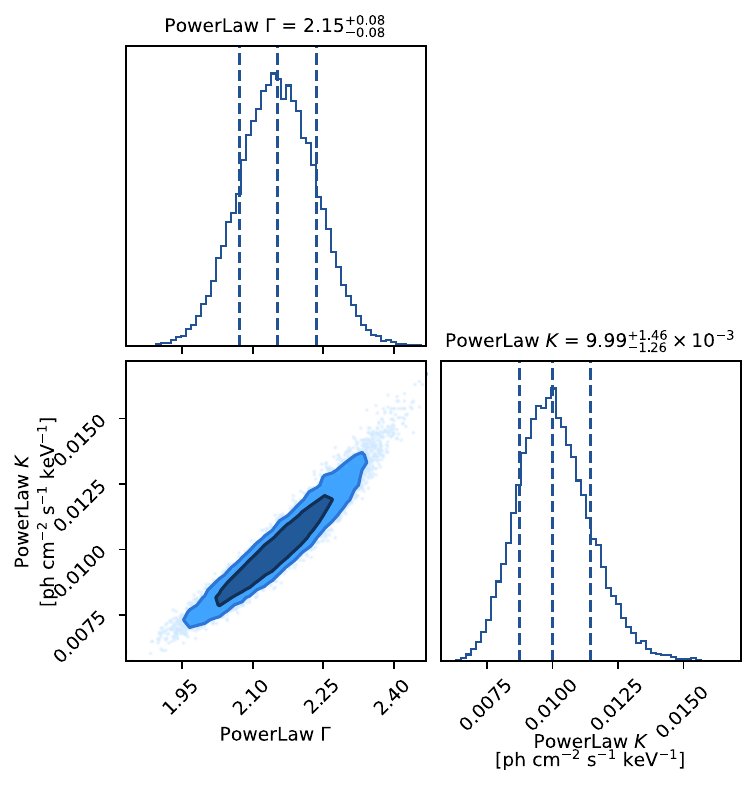}
        \caption{Posterior distributions of model parameters.}
    \end{subfigure}

    \caption{Time-averaged spectral analysis of the X-ray afterglow. Panel (a): The EP/FXT spectrum observed during $T_{0} + [43056,46064]$ s. The X-ray spectrum is fitted with an absorbed power-law model, with the intrinsic column density fixed at $N_\mathrm{H,X}=0.91\times10^{22}\mathrm{\,cm^{-2}}$ (see Section~\ref{sec:HE_spectra} for details). Panel (b): Posterior distributions of the parameters of the absorbed power-law model.}
    \label{fig:FXT2}
\end{figure*}

\begin{figure*}[htbp]
    \centering

    \begin{subfigure}{0.49\linewidth}
        \centering
        \includegraphics[width=\linewidth]{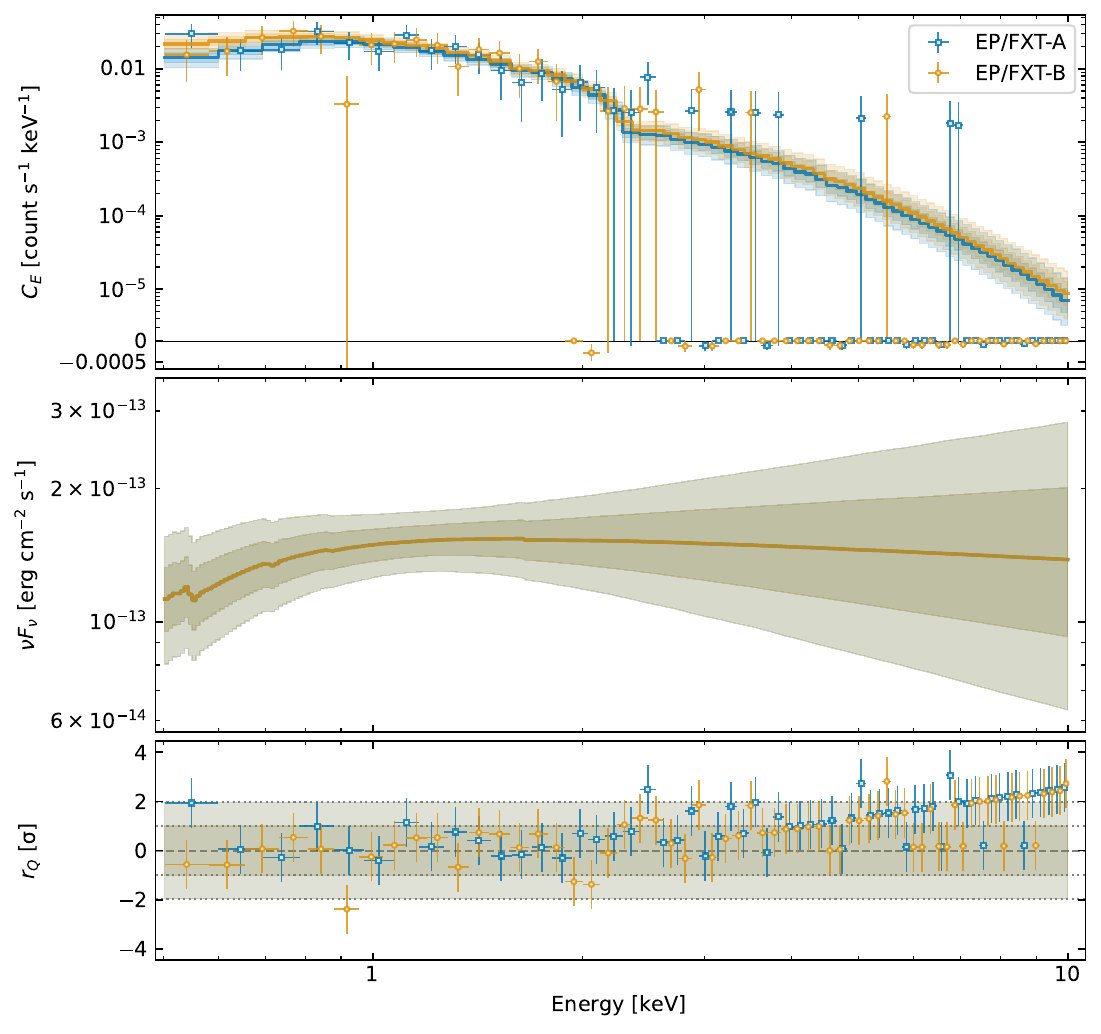}
        \caption{EP/FXT spectrum in $T_{0} + [112193,115044]$ s, power-law model, and residuals.}
    \end{subfigure}
    \hfill
    \begin{subfigure}{0.49\linewidth}
        \centering
        \includegraphics[width=\linewidth]{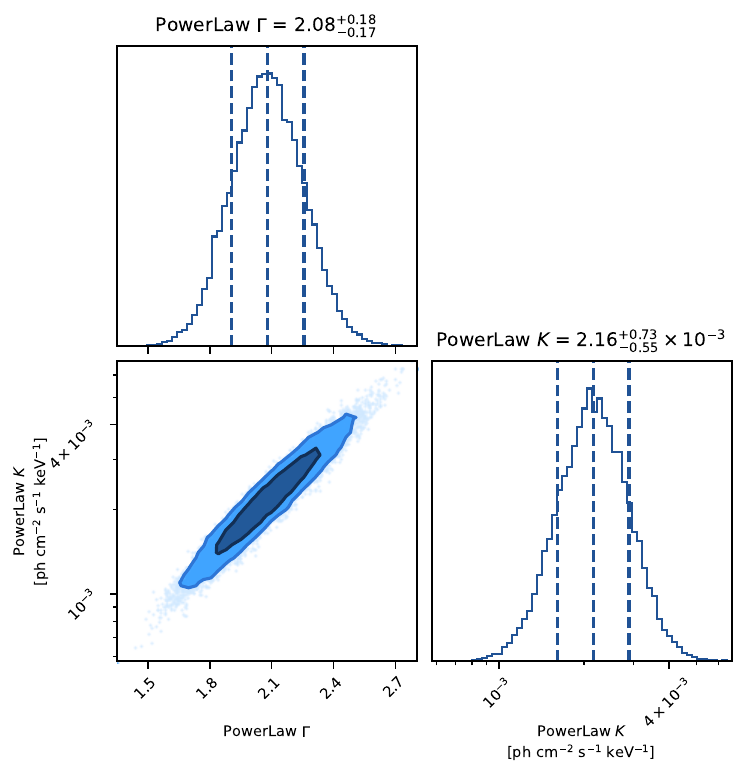}
        \caption{Posterior distributions of model parameters.}
    \end{subfigure}
    
    \vspace{2cm}
    
    \begin{subfigure}{0.49\linewidth}
        \centering
        \includegraphics[width=\linewidth]{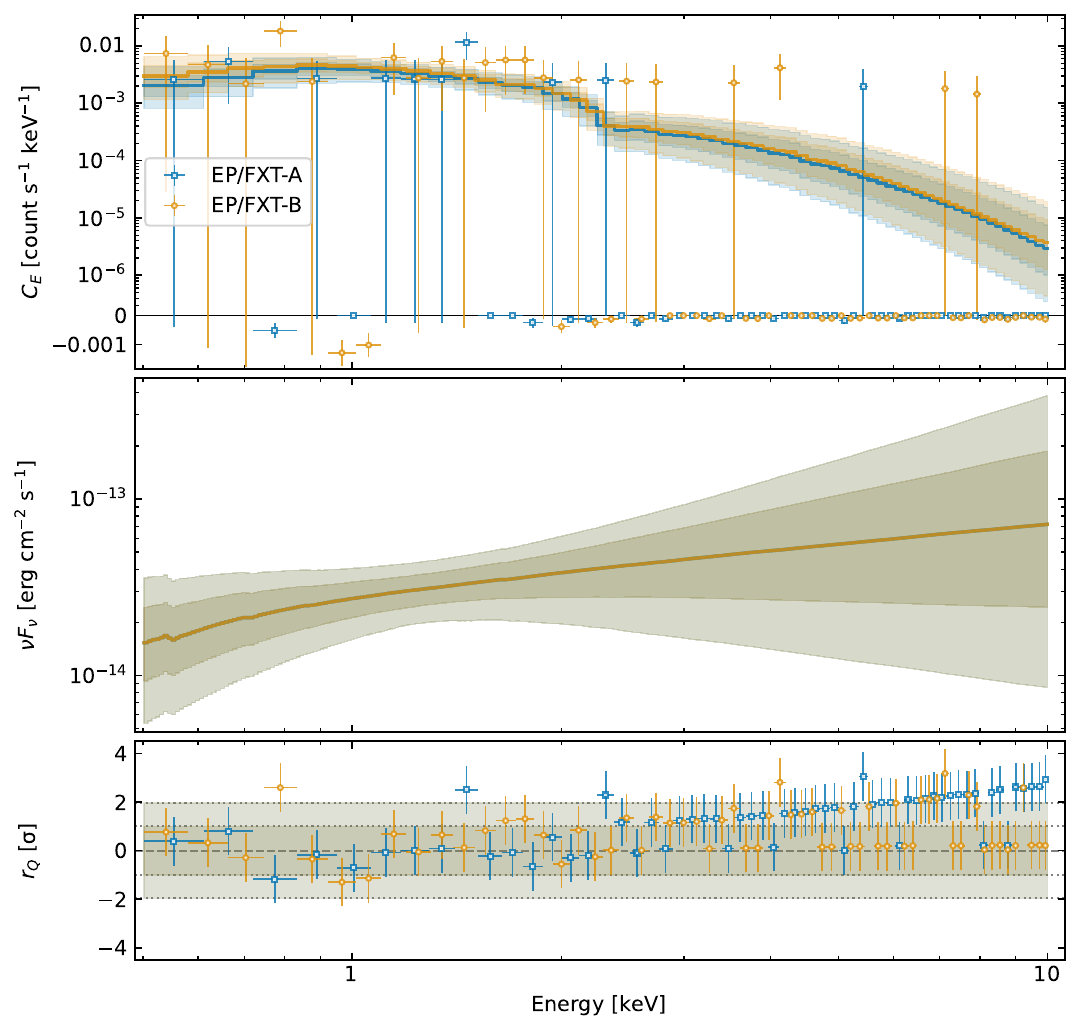}
        \caption{EP/FXT spectrum in $T_{0} + [296553,299566]$ s, power-law model, and residuals.}
    \end{subfigure}
    \hfill
    \begin{subfigure}{0.49\linewidth}
        \centering
        \includegraphics[width=\linewidth]{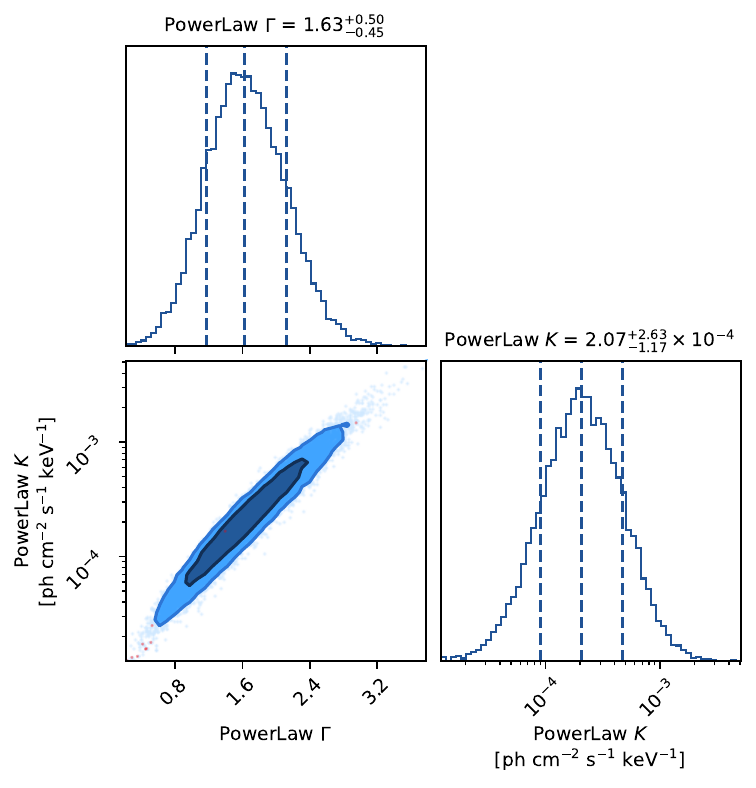}
        \caption{Posterior distributions of model parameters.}
    \end{subfigure}

    \caption{Time-integrated spectral analysis of the X-ray afterglow. The EP/FXT spectra observed during $T_{0} + [112193,115044]$ s (upper panel) and $T_{0} + [296553,299566]$ s (lower panel) are shown. The X-ray spectrum in each time interval is fitted with an absorbed power-law model. The intrinsic column density is fixed at $N_\mathrm{H,X}=0.91\times10^{22}\mathrm{\,cm^{-2}}$ throughout the X-ray spectral analysis. For details, see Section~\ref{sec:HE_spectra}.}
    \label{fig:FXT34}
\end{figure*}

We present the results of the high-energy spectral analysis in this section, covering both the prompt emission and afterglow phase. For the prompt emission, the EP/WXT spectral analysis results are shown in Figure~\ref{fig:WXT}, while the results of the joint spectral analysis of EP/WXT and GECAM-B are presented in Figure~\ref{fig:WXT_GECAM}. The time-averaged EP/WXT spectrum is also shown in Figure~\ref{fig:WXT-TimeAveraged}. For the X-ray afterglow emission, the EP/FXT spectra are presented in Figures~\ref{fig:FXT2}--~\ref{fig:FXT34}.

\clearpage

\section{Photometric data on the optical afterglow of EP250226a}
The refined optical photometry of EP250226a and the measurements reported in the GCN Circulars are summarized in Table~\ref{tab:optical-data}.
\begin{longtable}{cccccc}
\caption{Refined optical photometric results of EP250226a, combined with the results from GCN Circulars.}
\label{tab:optical-data}\\
\hline
\hline
$T_\mathrm{mid}$ &
Exposure &
Filter &
Magnitude &
Facility &
Reference \\
(s) &
(s) &
 &
(mag) &
 &
 \\
\hline
\endfirsthead

\caption[]{Refined optical photometric results of EP250226a, combined with the results from GCN Circulars.}\\
\hline
$T_\mathrm{mid}$ &
Exposure &
Filter &
Magnitude &
Facility &
Reference \\
(s) &
(s) &
 &
(mag) &
 &
 \\
\hline
\endhead

\hline
\endfoot

836 & 15 & $r$ & $17.27 \pm 0.10$ & CTIO/TRT & This work \\
1018 & 300 & $r$ & $17.38 \pm 0.01$ & CTIO/TRT & This work \\
1331 & 300 & $r$ & $17.67 \pm 0.01$ & CTIO/TRT & This work \\
1645 & 300 & $r$ & $17.88 \pm 0.01$ & CTIO/TRT & This work \\
1957 & 300 & $r$ & $18.03 \pm 0.02$ & CTIO/TRT & This work \\
2271 & 300 & $r$ & $18.19 \pm 0.02$ & CTIO/TRT & This work \\
4657 & 300 & $r$ & $18.98 \pm 0.04$ & CTIO/TRT & This work \\
4973 & 300 & $r$ & $19.09 \pm 0.05$ & CTIO/TRT & This work \\
5286 & 300 & $r$ & $19.14 \pm 0.05$ & CTIO/TRT & This work \\
5601 & 300 & $r$ & $19.21 \pm 0.05$ & CTIO/TRT & This work \\
5913 & 300 & $r$ & $19.19 \pm 0.04$ & CTIO/TRT & This work \\
6227 & 300 & $r$ & $19.34 \pm 0.05$ & CTIO/TRT & This work \\
6540 & 300 & $r$ & $19.29 \pm 0.05$ & CTIO/TRT & This work \\
6797 & 45 & $r$ & $19.35 \pm 0.02$ & ESO/VLT & This work \\
6852 & 300 & $r$ & $19.40 \pm 0.05$ & CTIO/TRT & This work \\
6907 & 60 & $z$ & $18.90 \pm 0.01$ & ESO/VLT & This work \\
7017 & 45 & $g$ & $20.58 \pm 0.01$ & ESO/VLT & This work \\
7165 & 300 & $r$ & $19.46 \pm 0.06$ & CTIO/TRT & This work \\
7673 & 300 & $r$ & $19.39 \pm 0.06$ & CTIO/TRT & This work \\
7811 & 900 & $V$ & $20.22 \pm 0.16$ & CTIO/TRT & This work \\
8744 & 900 & $R$ & $19.31 \pm 0.06$ & CTIO/TRT & This work \\
9678 & 900 & $I$ & $18.90 \pm 0.10$ & CTIO/TRT & This work \\
10200 & 120 & $i$ & $19.39 \pm 0.06$ & CTIO/LCO & This work \\
10345 & 120 & $i$ & $19.24 \pm 0.06$ & CTIO/LCO & This work \\
10505 & 120 & $z$ & $18.88 \pm 0.10$ & CTIO/LCO & This work \\
10650 & 120 & $z$ & $19.15 \pm 0.14$ & CTIO/LCO & This work \\
12306 & 180 & $R$ & $19.56 \pm 0.16$ & CTIO/TRT & This work \\
17092 & 720 & $R$ & $20.03 \pm 0.16$ & CTIO/TRT & This work \\
24304 & 600 & $R$ & $20.30 \pm 0.19$ & CTIO/TRT & This work \\
38665 & $6\times300$ & $R$ & $21.38 \pm 0.22$ & SBO/TRT & This work \\
40067 & $15\times300$ & $i$ & $20.17 \pm 0.25$ & Kinder/SLT & This work \\
40209 & $6\times300$ & $I$ & $19.91$ UL & SBO/TRT & This work \\
41572 & $3\times300$ & Mephisto-u & $22.46 \pm 0.36$ & Lijiang/Mephisto & This work \\
41572 & $3\times300$ & Mephisto-g & $22.24 \pm 0.18$ & Lijiang/Mephisto & This work \\
42596 & $3\times300$ & Mephisto-r & $21.16 \pm 0.05$ & Lijiang/Mephisto & This work \\
46436 & $52\times60$ & $r$ & $21.76 \pm 0.16$ & Xinglong/Schmidt & This work \\
48152 & $3\times300$ & $r$ & $21.39 \pm 0.18$ & Kinder/LOT & This work \\
50060 & $52\times60$ & $r$ & $21.44 \pm 0.12$ & Xinglong/Schmidt & This work \\
53697 & $52\times60$ & $r$ & $21.69 \pm 0.22$ & Xinglong/Schmidt & This work \\
68689 & $20\times150$ & $i$ & $21.76 \pm 0.12$ & KAO/1.88m & This work \\
72757 & 540 & $r$ & $21.98 \pm 0.06$ & ORM/NOT & This work \\
73394 & 540 & $i$ & $21.76 \pm 0.07$ & ORM/NOT & This work \\
74386 & 1200 & $z$ & $21.52 \pm 0.08$ & ORM/NOT & This work \\
82433 & $17\times300$ & $R$ & $22.16 \pm 0.26$ & OPD/60cm & This work \\
165689 & $26\times300$ & $R$ & $22.30$ UL & OPD/60cm & This work \\
341194 & 1680 & $R$ & $24.18 \pm 0.12$ & SAO RAS/BTA & This work \\
7235 & 3000 & $i$ & $19.05 \pm 0.03$ & COLIBR\'I/DDRAGO & \citet{2025GCN.39488....1M} \\
39598 & 3600 & $R$ & $21.19 \pm 0.07$ & Xinglong/2.16m & \citet{2025GCN.39515....1J} \\
43224 & 3600 & $I$ & $21.68 \pm 0.05$ & Xinglong/2.16m & \citet{2025GCN.39515....1J} \\
68184 & 300 & $i'$ & $21.22 \pm 0.29$ & SAAO/LCO & \citet{2025GCN.39524....1P} \\
68184 & 300 & $r'$ & $21.44 \pm 0.36$ & SAAO/LCO & \citet{2025GCN.39524....1P} \\
73620 & $18\times100$ & VT-B & $22.58\pm0.14$ & SVOM/VT & \citet{2025GCN.39514....1L} \\
73620 & $18\times100$ & VT-R & $21.56\pm0.10$ & SVOM/VT & \citet{10.1088/1674-4527/ae75cd} \\
\hline
\end{longtable}

\tablefoot{
$T_\mathrm{mid}$ represents the delay, measured in seconds, between the midtime of the photometric observations and the trigger time ($T_0$). 
The magnitudes are reported in the AB system and are not corrected for Galactic foreground extinction.
}

\clearpage
\section{Bayesian inference of the afterglow}
\begin{figure*}[ht!]
    \centering
    \includegraphics[width=0.7\linewidth]{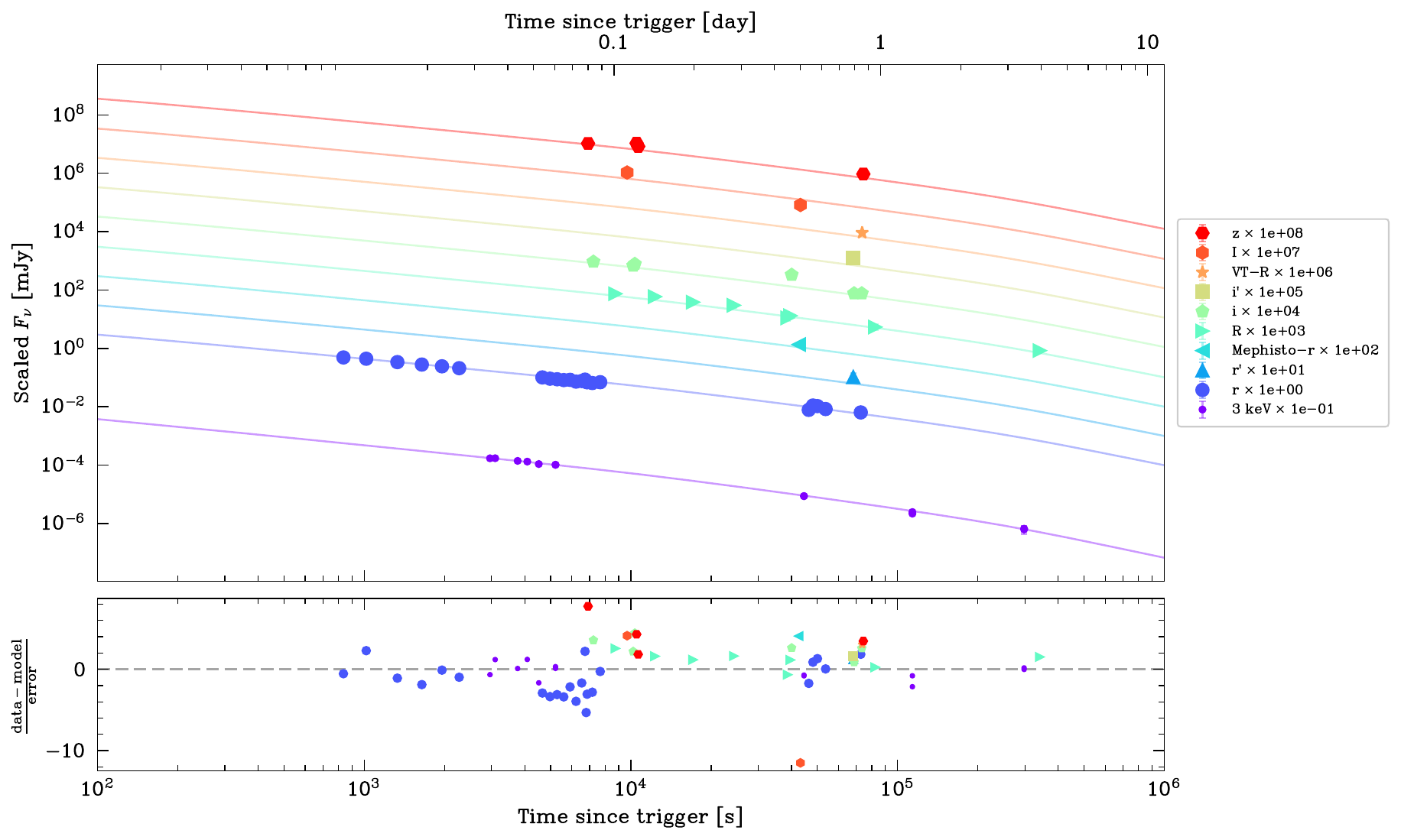}
    \includegraphics[width=0.7\linewidth]{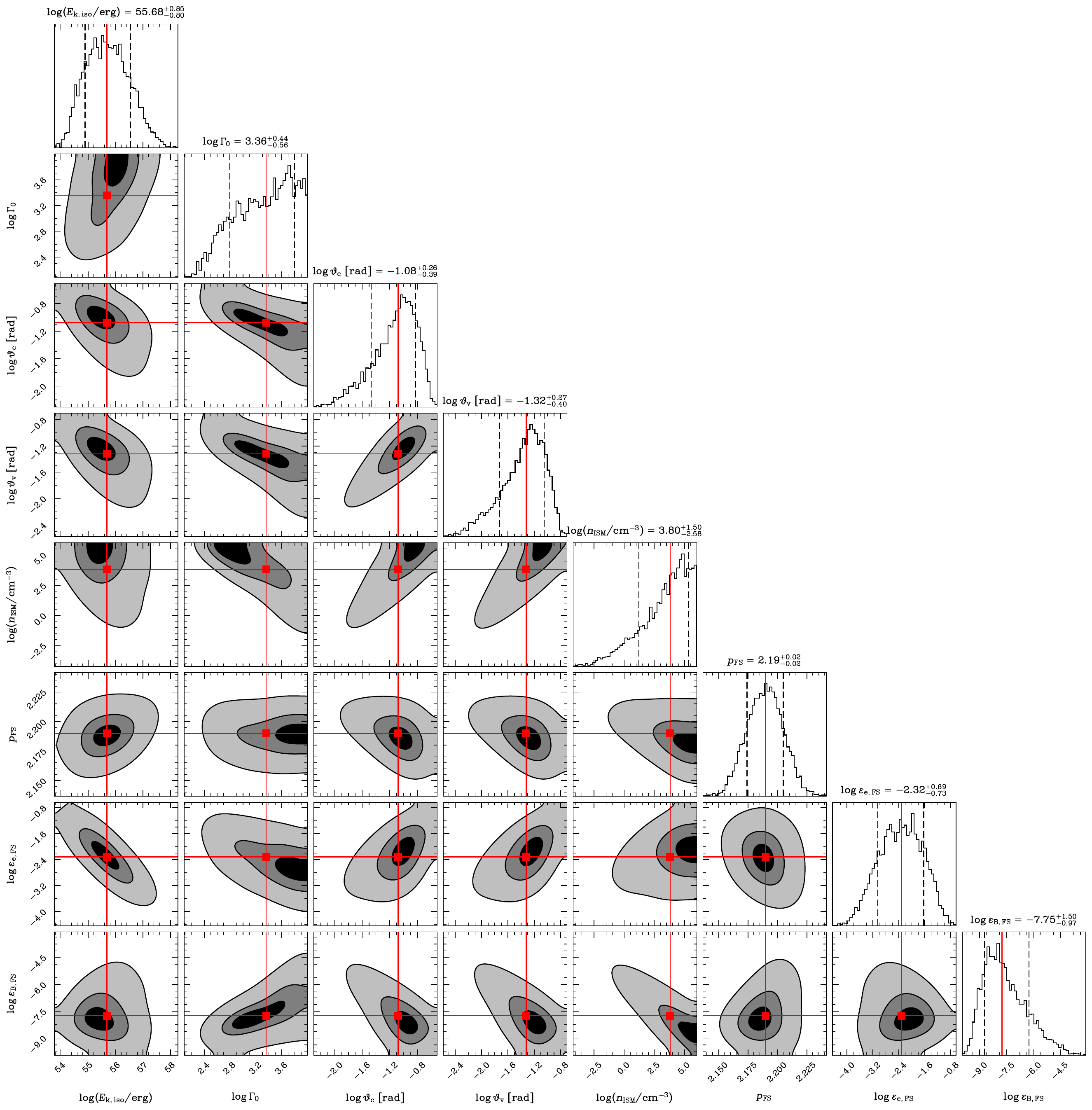}
    \caption{Multiwavelength afterglow light curves (top panel) along with the corresponding marginalized posterior distribution (bottom panel). The afterglow is described by a top-hat jet propagating through a uniform ISM, with all model parameters allowed to vary. The light curves and posterior distributions are obtained using Posterior-1 from Table~\ref{tab:VegasAfterglow}.}
    \label{fig:afterglow_model1}
\end{figure*}

\begin{figure*}[ht!]
    \centering
    \includegraphics[width=0.7\linewidth]{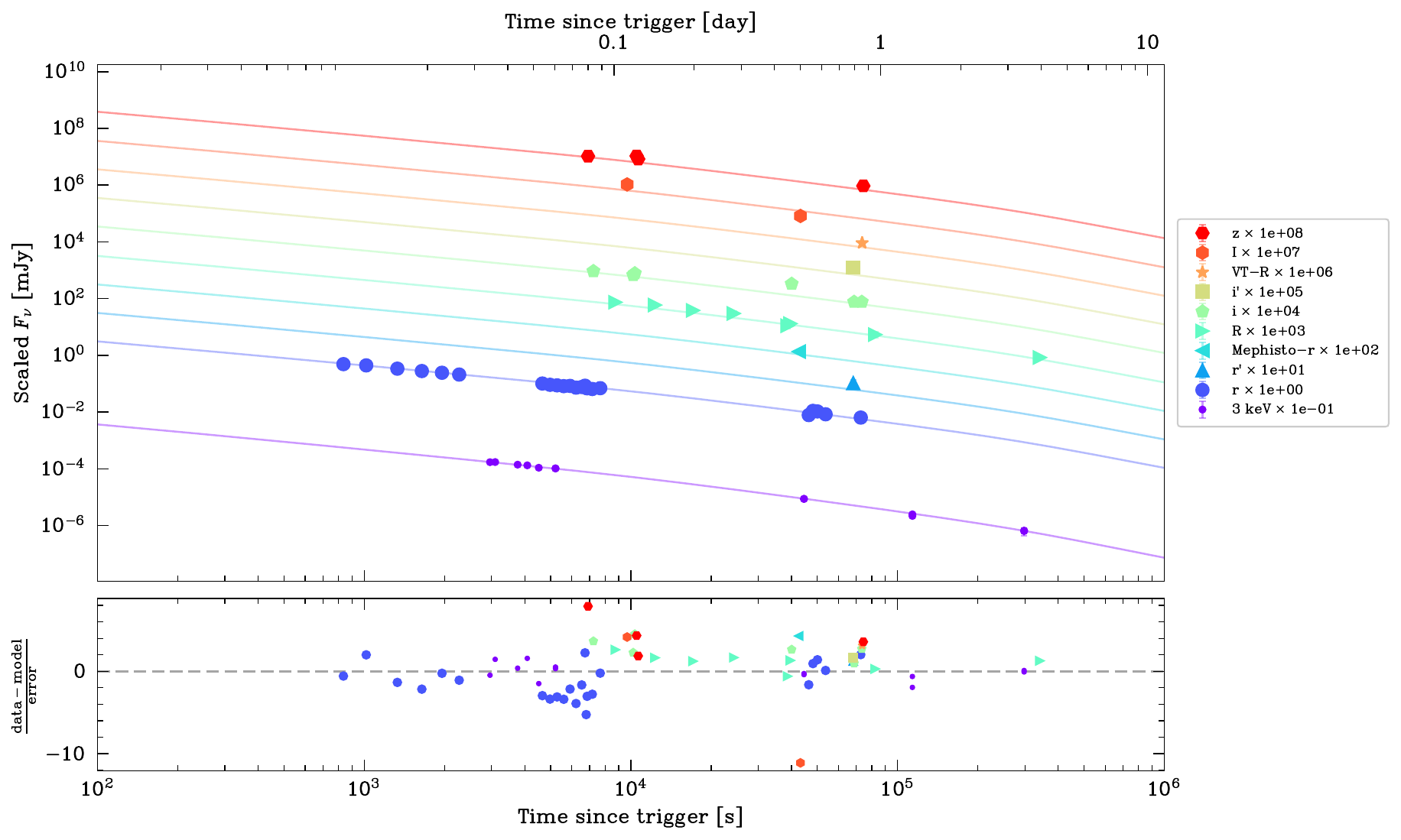}
    \includegraphics[width=0.7\linewidth]{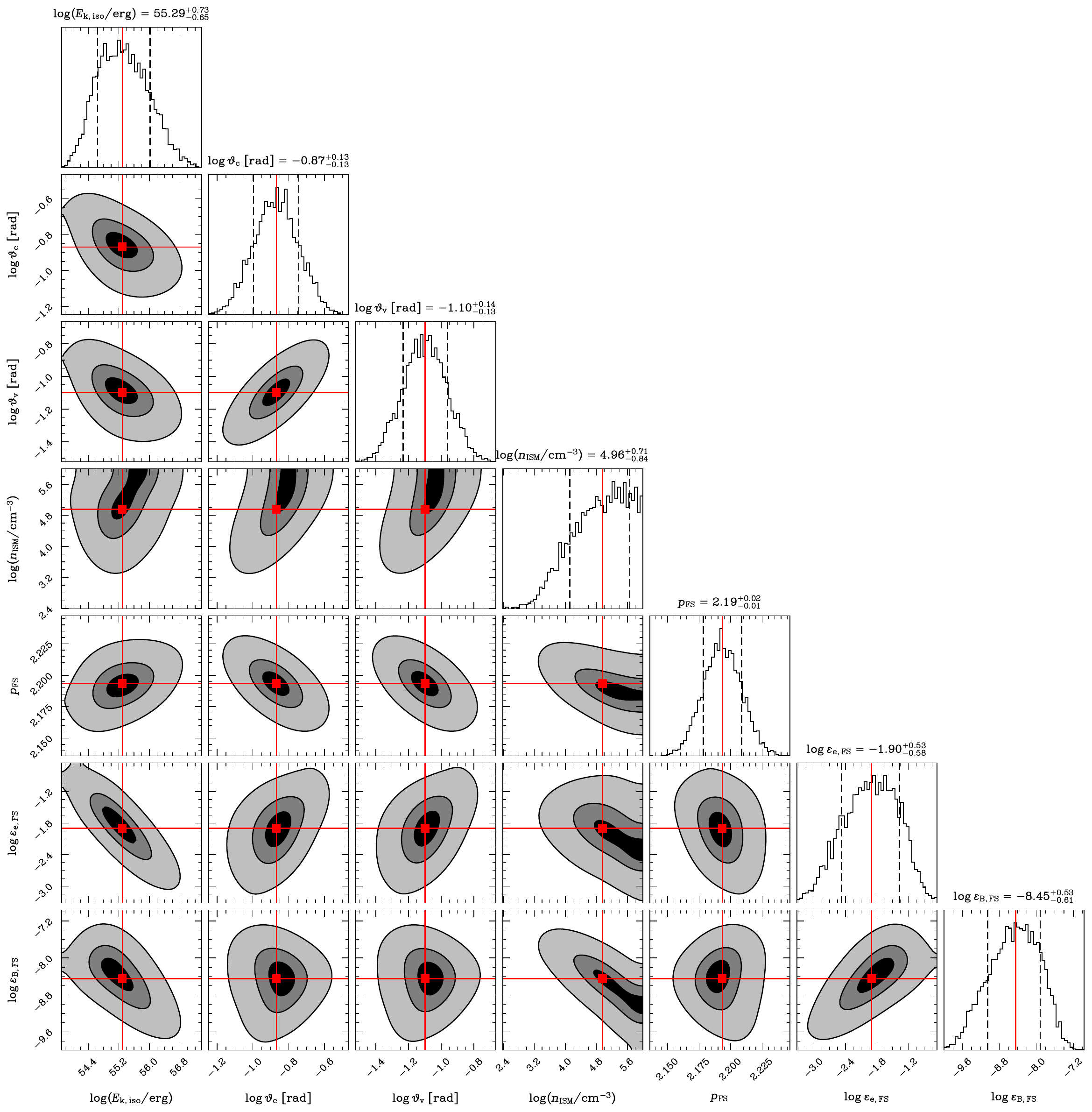}
    \caption{Multiwavelength afterglow light curves (top panel) along with the corresponding marginalized posterior distribution (bottom panel). The afterglow is modeled by a top-hat jet propagating through a uniform ISM, with the initial bulk Lorentz factor fixed at $\Gamma_0=500$. The light curves and posterior distributions are obtained using Posterior-2 from Table~\ref{tab:VegasAfterglow}.}
    \label{fig:afterglow_model2}
\end{figure*}

\begin{figure*}[ht!]
    \centering
    \includegraphics[width=0.7\linewidth]{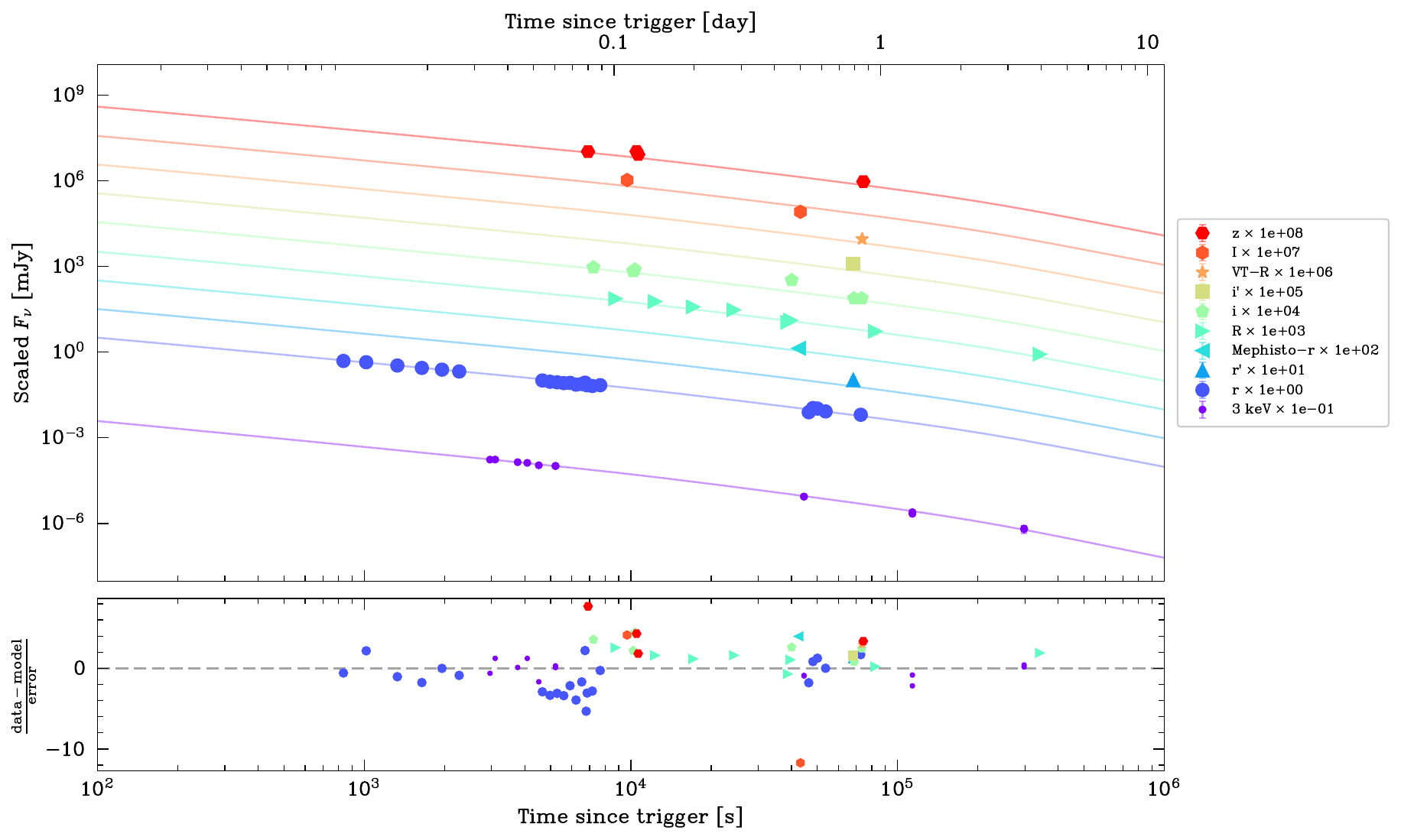}
    \includegraphics[width=0.7\linewidth]{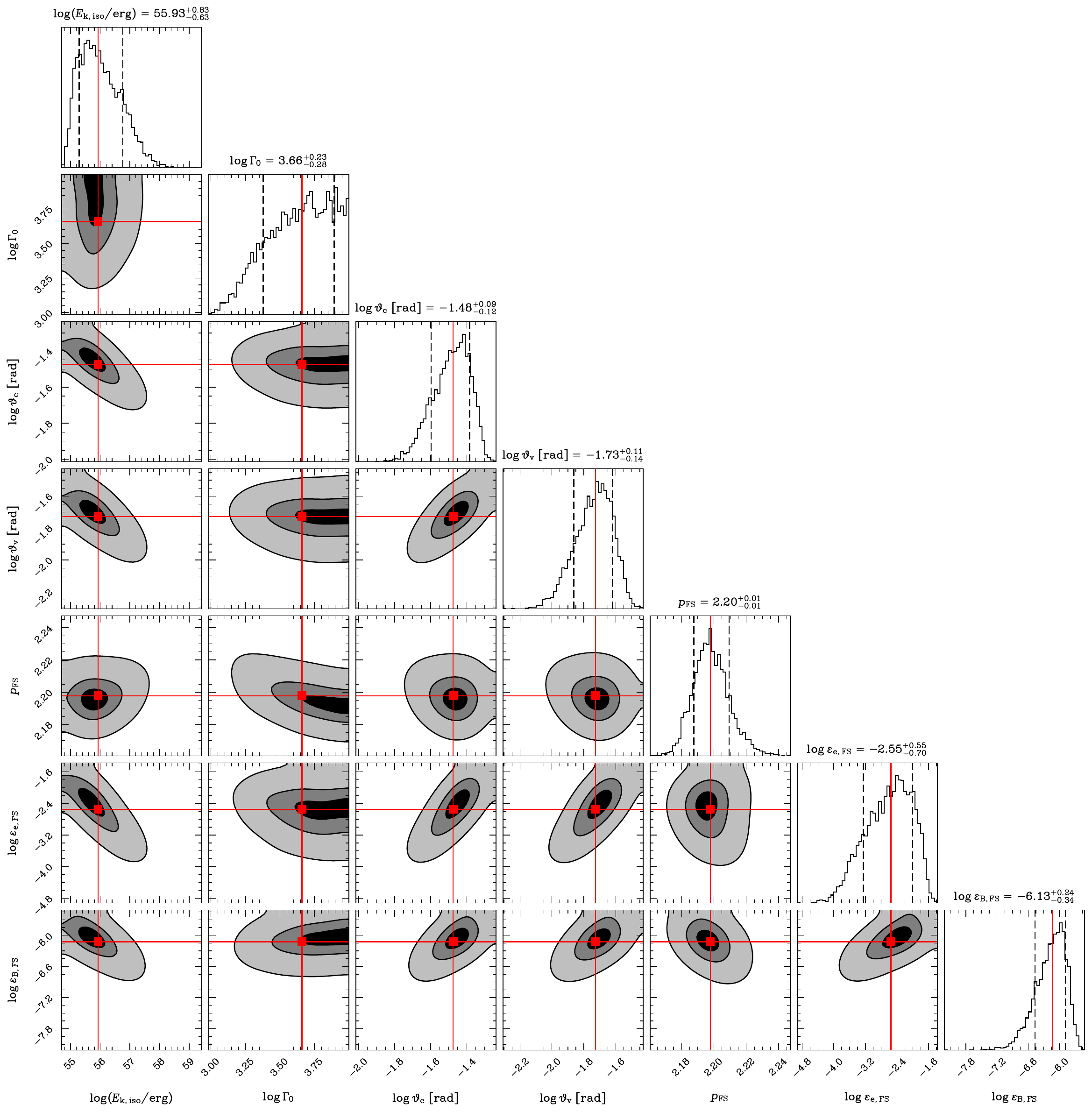}
    \caption{Multiwavelength afterglow light curves (top panel) along with the corresponding marginalized posterior distribution (bottom panel). The afterglow is characterized as a top-hat jet propagating through a uniform ISM, with the ISM number density fixed at $n_\mathrm{ISM}=10\ \mathrm{cm^{-3}}$. The light curves and posterior distributions are obtained using Posterior-3 from Table~\ref{tab:VegasAfterglow}.}
    \label{fig:afterglow_model3}
\end{figure*}

\begin{figure*}[htbp]
    \centering

    \begin{subfigure}{0.7\linewidth}
        \centering
        \includegraphics[width=\linewidth]{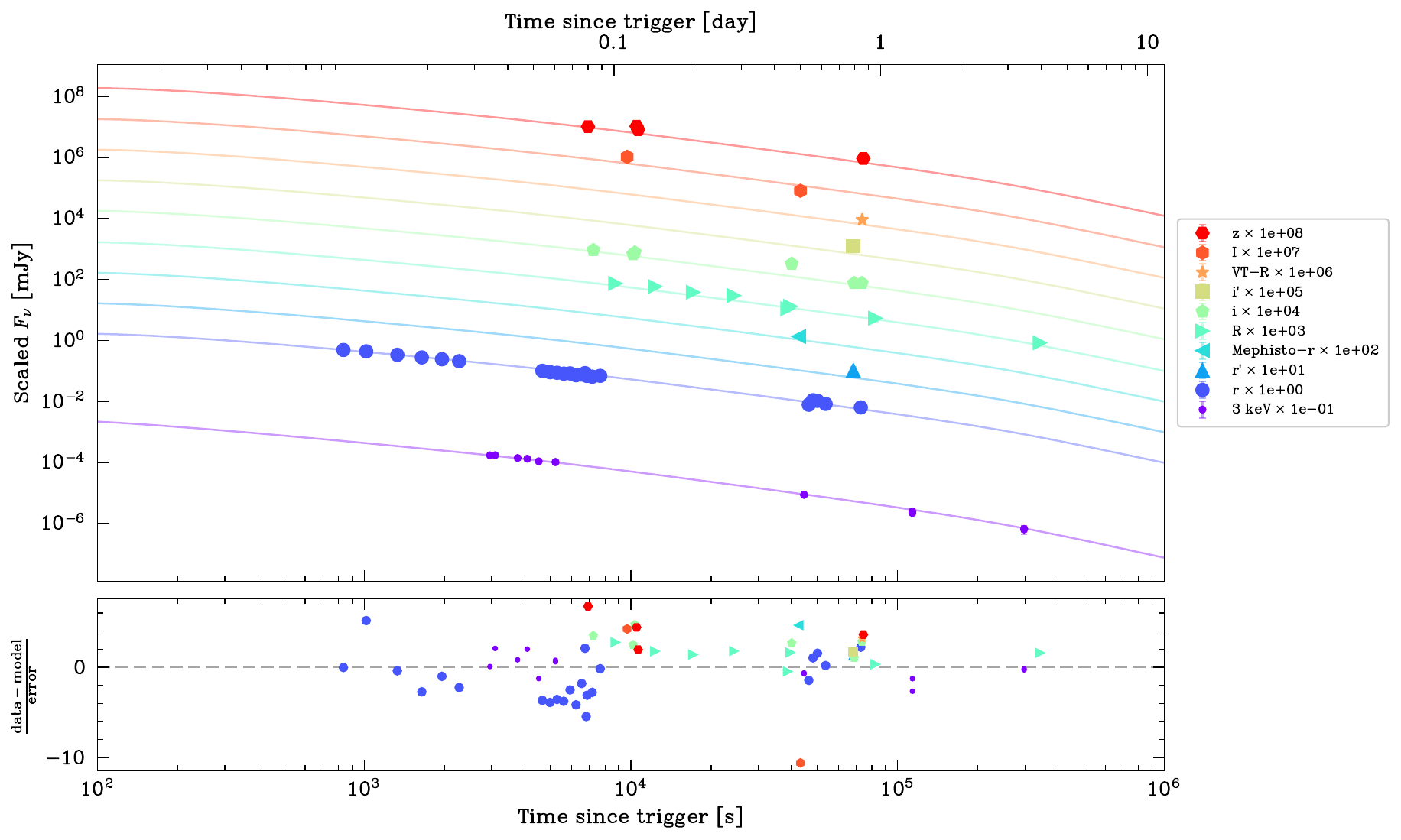}
    \end{subfigure}

    \vspace{0.5cm}

    \begin{subfigure}{0.7\linewidth}
        \centering
        \includegraphics[width=\linewidth]{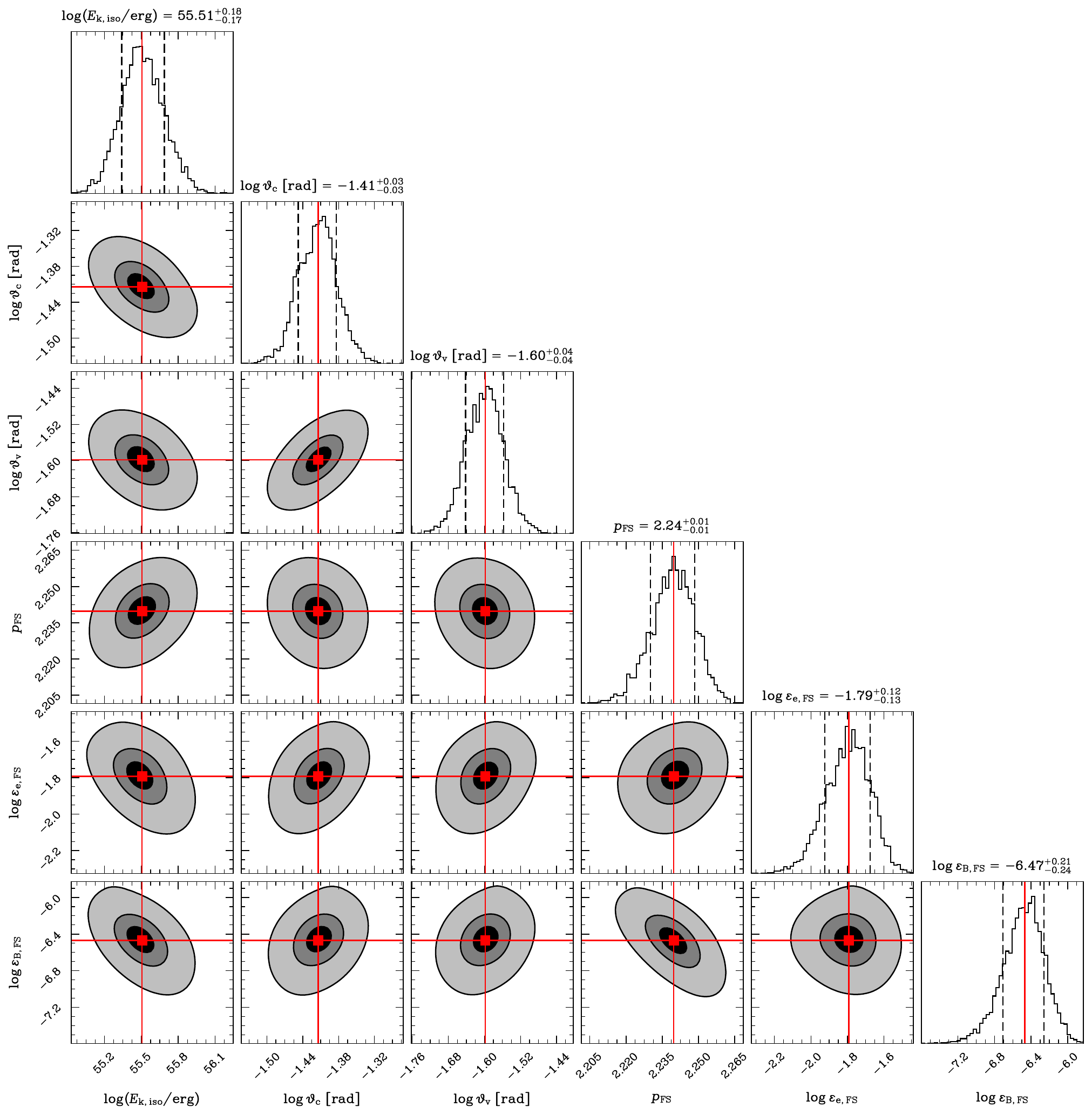}
    \end{subfigure}

    \caption{Multiwavelength light curves of the afterglow (top panel) and the corresponding marginalized posterior distribution (bottom panel). The afterglow is modeled as a top-hat jet propagating through a uniform ISM, with the initial bulk Lorentz factor fixed at $\Gamma_0=500$ and the ISM number density fixed at $n_\mathrm{ISM}=10\ \mathrm{cm^{-3}}$. The light curves and posterior distributions are obtained using Posterior-4 from Table~\ref{tab:VegasAfterglow}.}
    \label{fig:afterglow_model4}
\end{figure*}

The Posterior-1, Posterior-2, Posterior-3, and Posterior-4 distributions are reported in Table~\ref{tab:VegasAfterglow} and shown in Figures~\ref{fig:afterglow_model1}, \ref{fig:afterglow_model2}, \ref{fig:afterglow_model3}, and \ref{fig:afterglow_model4}, respectively. A detailed description of these posterior distributions is provided in Section~\ref{sec:afterglow_modeling}.

\begin{table*}[ht!]
\caption{Bayesian inference results of afterglow model parameters, with uniform priors adopted for all parameters.}
\label{tab:VegasAfterglow}
\centering
\begin{tabular}{cccccc}
\hline\hline
Parameter & Uniform Prior & Posterior-1 & Posterior-2 & Posterior-3 & Posterior-4 \\
\hline
$\log (E_{\mathrm{k,iso}}/\mathrm{erg})$ 
& $\mathcal{U}(50, 60)$
& $55.68_{-0.80}^{+0.85}$ 
& $55.29_{-0.65}^{+0.73}$ 
& $55.93_{-0.63}^{+0.83}$ 
& $55.51_{-0.17}^{+0.18}$ \\

$\log \Gamma_0$
& $\mathcal{U}(1, 4)$
& $3.36_{-0.56}^{+0.44}$
& 2.70 (fixed)
& $3.66_{-0.28}^{+0.23}$
& 2.70 (fixed) \\

$\log \theta_{\mathrm{c}}$ [rad]
& $\mathcal{U}(-3, 0.2)$
& $-1.08_{-0.39}^{+0.26}$
& $-0.87\pm0.13$
& $-1.48_{-0.12}^{+0.09}$
& $-1.41\pm0.03$ \\

$\log \theta_{\mathrm{v}}$ [rad]
& $\mathcal{U}(-3, 0.2)$
& $-1.32_{-0.40}^{+0.27}$
& $-1.10_{-0.13}^{+0.14}$
& $-1.73_{-0.14}^{+0.11}$
& $-1.60\pm0.04$ \\

$\log (n_{\mathrm{ISM}}/\mathrm{cm^{-3}})$
& $\mathcal{U}(-6, 6)$
& $3.80_{-2.58}^{+1.50}$
& $4.96_{-0.84}^{+0.71}$
& 1 (fixed)
& 1 (fixed) \\

$p_{\mathrm{FS}}$
& $\mathcal{U}(2.01, 3)$
& $2.19\pm0.02$
& $2.19_{-0.01}^{+0.02}$
& $2.20\pm0.01$
& $2.24\pm0.01$ \\

$\log \epsilon_{\mathrm{e,FS}}$
& $\mathcal{U}(-10, 0)$
& $-2.32_{-0.73}^{+0.69}$
& $-1.90_{-0.58}^{+0.53}$
& $-2.55_{-0.70}^{+0.55}$
& $-1.79_{-0.13}^{+0.12}$ \\

$\log \epsilon_{\mathrm{B,FS}}$
& $\mathcal{U}(-10, 0)$
& $-7.75_{-0.97}^{+1.50}$
& $-8.45_{-0.61}^{+0.53}$
& $-6.13_{-0.34}^{+0.24}$
& $-6.47_{-0.24}^{+0.21}$ \\

$\chi$
& 1 (fixed)
& 1 (fixed)
& 1 (fixed)
& 1 (fixed)
& 1 (fixed) \\

\hline
$\ln \mathcal{Z}$
& ---
& $-271.45 \pm 0.15$
& $-271.72 \pm 0.16$
& $-272.10 \pm 0.16$
& $-293.64 \pm 0.16$ \\

Figure
& ---
& \ref{fig:afterglow_model1}
& \ref{fig:afterglow_model2}
& \ref{fig:afterglow_model3}
& \ref{fig:afterglow_model4} \\

\hline
\end{tabular}

\tablefoot{The uncertainties of the above parameters are reported as the $1\sigma$ credible intervals. The additional subscript $\mathrm{FS}$ represents the contribution from the forward-shock component.}
\end{table*}

\clearpage
\section{VLT/X-shooter spectrum}
\begin{figure*}[ht!]
    \centering

    \begin{subfigure}{0.9\linewidth}
        \centering
        \includegraphics[width=\linewidth]{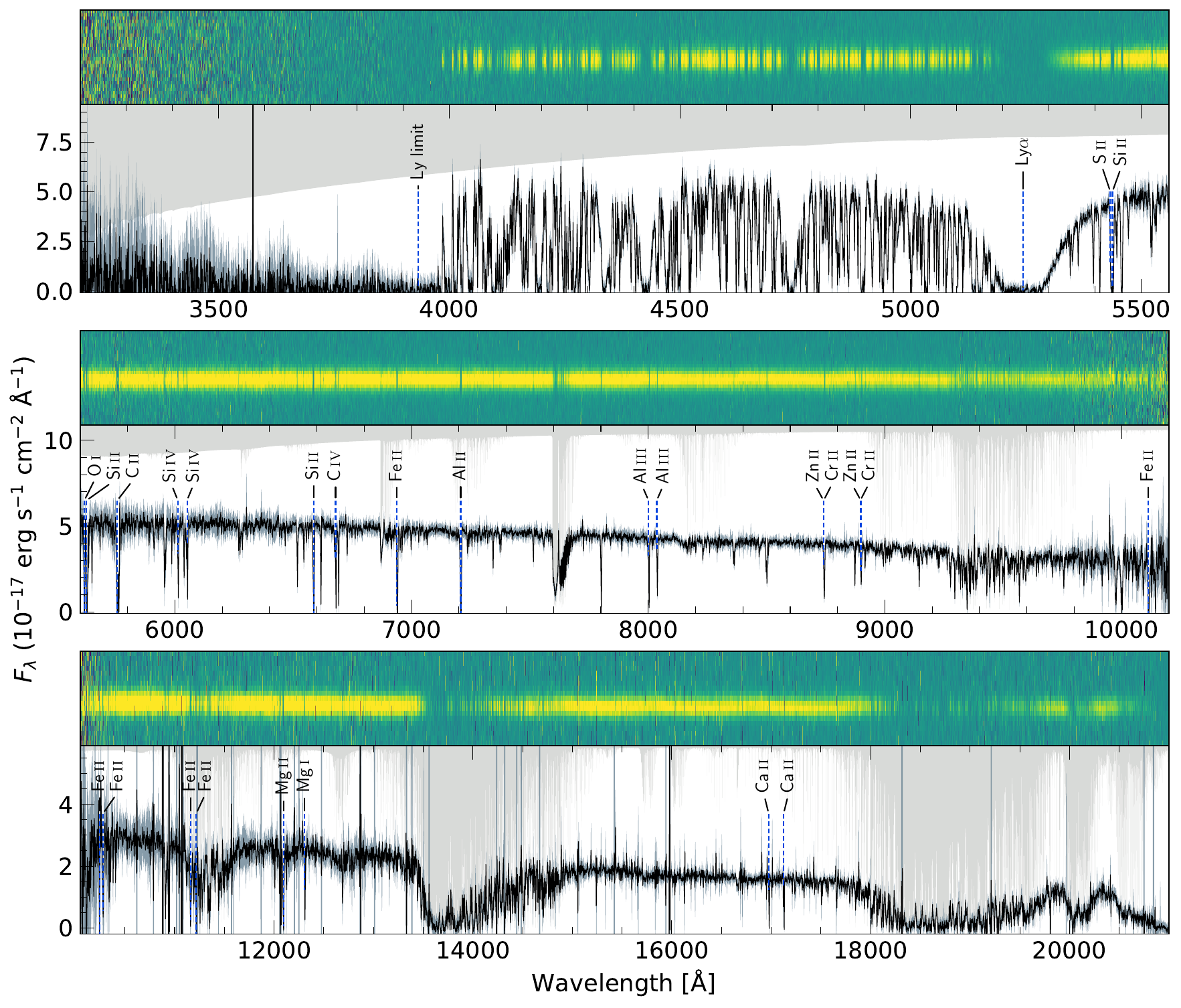}
        \caption{Spectrum obtained by VLT/X-shooter.}
    \end{subfigure}

    \vspace{0.5cm}

    \begin{subfigure}{\linewidth}
        \centering
        \includegraphics[width=\linewidth]{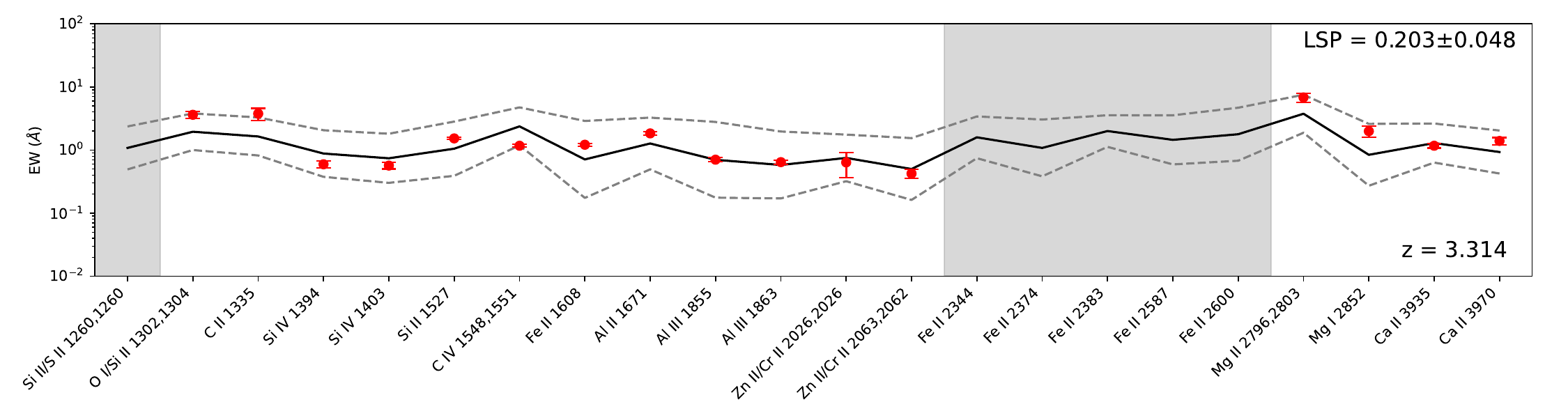}
        \caption{Equivalent widths measured from the afterglow spectrum.}
    \end{subfigure}

    \caption{
    Spectrum and absorption-line analysis of the EP250226a afterglow.
    The redshift of EP250226a was determined to be $z=3.314$.
    In panel (a), from top to bottom, the panels correspond to the UVB,
    VIS, and NIR arms of VLT/X-shooter. In each panel, the upper section
    presents the two-dimensional spectrum, while the lower section shows
    the corresponding one-dimensional extracted spectrum. Identified
    absorption features are indicated by blue dashed vertical lines.
    For clarity, the one-dimensional spectra are smoothed using a
    Savitzky-Golay filter, with the associated uncertainties shown.
    Wavelength regions significantly affected by telluric absorption are
    indicated by gray shaded bands. 
    In panel (b), the equivalent widths measured from the afterglow
    spectrum are shown in red and are reported in Table~\ref{tab:EW}.
    The solid black line represents the average value derived from the GRB
    sample \citep{2012A&A...548A..11D}, while the dashed lines indicate
    the corresponding upper and lower $1\sigma$ standard deviations.
    The gray shaded regions mark wavelength intervals that lack spectral coverage or are significantly contaminated by strong telluric absorption.} 
    \label{fig:VLTspec}
\end{figure*}

Several absorption lines were identified in the VLT/X-shooter spectrum (see Section~\ref{sec:Spectroscopic_Observations} for details). The EWs of the identified absorption lines are listed in Table~\ref{tab:EW}, while their positions in the spectrum are marked in Figure~\ref{fig:VLTspec}.

\begin{table}[ht!]
\caption{Rest-frame equivalent widths of typical absorption lines in GRB afterglows.}
\label{tab:EW}
\centering
\begin{tabular}{lc}
\hline\hline
Feature & $\mathrm{EW}_{\rm rest}$ (\AA) \\
\hline
Si\,II/S\,II 1260,1260 & ---\\
O\,I/Si\,II 1302,1304 & $3.62^{+0.95}_{-0.45}$\\
C\,II 1335 & $3.77^{+1.23}_{-0.81}$\\
Si\,IV 1394 & $0.59^{+0.09}_{-0.08}$\\
Si\,IV 1403 & $0.57^{+0.08}_{-0.07}$\\
Si\,II 1527 & $1.53^{+0.07}_{-0.07}$\\
C\,IV 1548,1551 & $1.17^{+0.07}_{-0.06}$\\
Fe\,II 1608 & $1.21^{+0.06}_{-0.07}$\\
Al\,II 1671 & $1.84^{+0.15}_{-0.13}$\\
Al\,III 1855 & $0.70^{+0.06}_{-0.05}$\\
Al\,III 1863 & $0.64^{+0.05}_{-0.05}$\\
Zn\,II/Cr\,II 2026,2026 & $0.64^{+0.13}_{-0.27}$\\
Zn\,II/Cr\,II 2063,2062 & $0.42^{+0.04}_{-0.07}$\\
Fe\,II 2344 & ---\\
Fe\,II 2374 & ---\\
Fe\,II 2383 & ---\\
Fe\,II 2587 & ---\\
Fe\,II 2600 & ---\\
Mg\,II 2796,2803 & $6.82^{+1.44}_{-1.10}$\\
Mg\,I 2852 & $1.99^{+0.49}_{-0.40}$\\
Ca\,II 3935 & $1.17^{+0.08}_{-0.09}$\\
Ca\,II 3970 & $1.39^{+0.24}_{-0.18}$\\
\hline
\end{tabular}
\tablefoot{
The $\mathrm{EW}_{\rm rest}$ entries for several absorption lines are left blank in the table because these lines are either significantly affected by telluric absorption or lie near the boundaries between the X-shooter spectral arms, making their EW measurements potentially unreliable.}
\end{table}

\FloatBarrier 
\twocolumn

\end{appendix}
\end{document}